\documentclass[11pt]{article}

\usepackage{jheppub}
\usepackage{amsmath,amssymb,graphicx} 
\usepackage{epsf}
\usepackage{epsfig}

\newcommand{\ssw}{s^{2}_W}
\newcommand{\cw}{c_W}

\newcommand{\re}{{\rm Re}}

\title{Interplay of vector-like top partner multiplets in a realistic 
  mixing set-up} 

\author[a]{Giacomo Cacciapaglia,} 
\author[a,1]{Aldo Deandrea,
   \note{also Institut Universitaire de France, 103 boulevard
     Saint-Michel, 75005 Paris, France}} 
\author[b]{Naveen Gaur,}
\author[c]{Daisuke Harada,}
\author[d,e]{Yasuhiro Okada,}
\author[f]{Luca Panizzi}

\affiliation[a]{Universit\'e de Lyon, France; Universit\'e Lyon 1,
  CNRS/IN2P3, UMR5822 IPNL,\\ F-69622 Villeurbanne Cedex, France.} 
\affiliation[b]{Department of Physics, Dyal Singh College (University of Delhi), Lodi Road,
  New Delhi - 110003, India} 
\affiliation[c]{Centre for High Energy Physics, Indian Institute of
  Science, Bangalore 560012, India} 
\affiliation[d]{KEK Theory Center, Institute of Particle and Nuclear
  Studies, KEK, 1-1 Oho, Tsukuba, Ibaraki 305-0801, Japan} 
\affiliation[e]{Department of Particle and Nuclear Physics, Graduate
  University for Advanced Studies (Sokendai), 1-1 Oho,  
Tsukuba, Ibaraki 305-0801, Japan}
\affiliation[f]{School of Physics and Astronomy, University of
  Southampton, Highfield,\\ Southampton SO17 1BJ, UK} 

\emailAdd{g.cacciapaglia@ipnl.in2p3.fr}
\emailAdd{deandrea@ipnl.in2p3.fr}
\emailAdd{gaur.nav@gmail.com}
\emailAdd{harada@cts.iisc.ernet.in}
\emailAdd{yasuhiro.okada@kek.jp}
\emailAdd{L.Panizzi@soton.ac.uk}

\abstract{The ATLAS and CMS collaborations at the LHC have performed 
  analyses on the existing data sets,
  studying the case of one vector-like fermion or multiplet coupling
  to the standard model Yukawa sector. In the near future, with more
  data available, these experimental collaborations will start to
  investigate more realistic cases. The presence of more than one
  extra vector-like multiplet is indeed a common  situation in many
  extensions of the standard model. The interplay of these vector-like
  multiplet between precision electroweak bounds, flavour and collider
  phenomenology is  a important question in view of establishing
  bounds or for the discovery of  physics beyond the standard
  model. In this work we study the phenomenological consequences of
  the presence of two vector-like multiplets. We analyse the
  constraints on such scenarios from tree-level data and oblique
  corrections for the case of mixing to each of the SM generations. 
  In the present work, we limit to scenarios with two top-like partners and no mixing in the down-sector.}

\begin{document}

\begin{flushright}
  KEK-TH-1729     \\
  LYCEN-2014-09
\end{flushright}
\maketitle
\flushbottom


\section{Introduction}                          \label{sec:introduction} 

The Large Hadron Collider (LHC) has confirmed the effective description of the electroweak
sector given by the Standard Model (SM) Lagrangian with the discovery of the Higgs boson and the analysis of its properties. It also features a
very strong potential for the discovery or exclusion of new physics/particles, thus opening the possibility of investigating both
strongly and weakly coupled extensions of the Standard Model. New vector-like (VL) fermions are often present in many of the extensions
of the SM, especially in relation with the top sector (top partners,
as for example in composite Higgs models \cite{Agashe:2004rs,Contino:2006qr,Giudice:2007fh,Matsedonskyi:2012ym},
extra-dimensional models \cite{Antoniadis:1990ew,Antoniadis:2001cv,Csaki:2003sh,Hosotani:2004wv,Cacciapaglia:2009pa},   
little Higgs models \cite{ArkaniHamed:2002qx,ArkaniHamed:2002qy,Schmaltz:2005ky}, gauge-Higgs models \cite{Hosotani:1983xw}, gauge coupling 
unification \cite{Choudhury:2001hs,Panico:2008bx} and models with an extended custodial symmetry \cite{Agashe:2006at,Chivukula:2011jh}). 
Both CMS \cite{twikiCMS} and ATLAS \cite{twikiATLAS} have recently devoted a considerable effort in the analyses apt to setting
bounds on this type of new particles. Initially, simplifying assumptions were considered (mixing only with the third generation of SM quark
family or specific decay modes) \cite{delAguila:2000rc,AguilarSaavedra:2005pv,Anastasiou:2009rv,AguilarSaavedra:2009es,Cacciapaglia:2010vn,Marzocca:2012zn,DeSimone:2012fs,Falkowski:2013jya,Aguilar-Saavedra:2013wba,Ellis:2014dza}. 
However the most recent analyses, due to larger data samples, allow exploring more general situations with mixing of VL quarks with the first two 
generation of SM quarks \cite{Cacciapaglia:2011fx,Okada:2012gy,Buchkremer:2013bha,Delaunay:2013pwa,Barducci:2014ila}. 
Considering the presence of a complete multiplet of the symmetries of the Standard Model is however not enough in some realistic scenarios: in fact, 
theoretically justified models often contain a multiplet of larger global symmetries which can be described in terms of several multiplets which are close 
in mass. The various multiplets then mix with each other via the Higgs interactions. 
The presence of general mixing structures and the interplay of different multiplets typically affects the tree-level and loop-level
bounds, thereby modifying the results expected by performing simplified analyses based on a single particle or a single multiplet.
This work is devoted to the detailed exploration of general structures and mixing of more than one VL quark multiplet and, specifically, we study in
detail the implications of the presence of \textit{two} VL quark multiplets mixings with \textit{any} of the 3 SM quark generation. 
We also focus on a specific sub-set of scenarios where both VL multiplets contain a top partner ({\sl i.e.} with electric charge $+2/3\ e$), and where 
eventual bottom partners (i.e. with electric charge $-1/3\ e$) do not mix with the SM down sector.
This choice is done to minimise the constraints from flavour, which are very severe on mixing in the down sector only. In the scenarios we 
selected, larger mixing angles are allowed thus providing larger single production cross sections at the LHC.
These scenarios are also theoretically justified in models where the new physics couples dominantly to the top quark.
Of course, depending upon the multiplet considered, non-SM quarks, {\sl i.e.} quarks having non-SM electric charge, may be present in the 
considered multiplets. We have estimated the constraints on such scenarios from electroweak precision (EWP) data (oblique and non-oblique)
and current LHC data.    

The paper is organized as follows: in Section \ref{sec:model} we classify (through their $(SU(2)_L, U(1)_Y)$ quantum numbers) all the 
possible pairs of VL multiplets that can interact with SM quarks via a SM Higgs doublet. In Section \ref{sec:yukawa} the Yukawa couplings 
of the VL multiplets with SM quark generations are described. In the same section we have also identified three kind of scenarios depending 
on the multiplet content, namely: top-type multiplets, bottom-type multiplets and mixed multiplets. The mass matrices  of the cases we 
considered in our analysis are provided in Section \ref{sec:massmatrices}. 
Section \ref{sec:bounds} is devoted to a discussion of the bounds
and constraints on masses and mixing parameters we considered for the numerical analysis: in Section \ref{sec:bounds:1} we describe the 
tree-level bounds, while the bounds  coming from oblique corrections, {\sl i.e.} $S$ and $T$ parameters
\cite{Peskin:1990zt,Peskin:1991sw}, are discussed in Section \ref{sec:bounds:2}. 
In Section \ref{sec:results} we present the results of our numerical analysis: in Sections \ref{sec:results:1}-\ref{sec:results:4} the bounds
on parameter space from tree-level constraints and EWP tests are described, while in Section \ref{sec:results:5} we provide a comparison 
between the bounds from single VL top production at LHC14 with flavour and EWP bounds. We finally conclude in Section \ref{sec:concl}.   

\section{Vector-like multiplets}                  
\label{sec:model}
The minimal set of VL multiplets that can mix with SM quarks and a SM (or SM-like) Higgs
boson have been extensively studied in literature \cite{delAguila:1982fs,delAguila:2000rc,Aguilar-Saavedra:2013wba,Cacciapaglia:2010vn,Cacciapaglia:2011fx,Okada:2012gy,Buchkremer:2013bha}.
%
\begin{table}[htp]
\begin{center}
\begin{tabular}{| l | c | c | c | c | c |}
\hline
Multiplet   & $\psi$ & $(SU(2)_{L},U(1)_{Y})$  & $T_{3}$ & $Q_{EM}$ & Yukawa to SM \\ \hline
Singlet 2/3 &  $U$   & $({\bf 1},2/3)$         & 0       & +2/3    & Yes  \\ \hline
Doublet 7/6 & $\left( \begin{array}{c}  X^{5/3} \\  U   \end{array}
     \right)$   & $({\bf 2},7/6)$     
     &  $\begin{array}{c} +1/2 \\ -1/2 \end{array}$     
     &  $\begin{array}{c}+5/3 \\ +2/3 \end{array} $          & Yes       \\ \hline
Triplet 5/3 & $\left( \begin{array}{c} X^{8/3} \\ X^{5/3} \\  U
  \end{array} \right)$ & $({\bf 3},5/3)$ & $\begin{array}{c} +2 \\ +1
  \\ 0 \end{array}$    & $\begin{array}{c} +8/3 \\ +5/3 \\ +2/3
\end{array}$           & No                 \\ \hline 
\end{tabular}
\caption{Quantum numbers for the top--type VL multiplets (up to triplets), explicitly indicating weak isospin, hypercharge, electric charge ($Q_{EM}$) and if a direct Yukawa coupling to SM quarks is allowed.}
\label{QN_vec-top}
\end{center}
\end{table} 
%
In Tables \ref{QN_vec-top}-\ref{QN_vec-mixed} we list the  $SU(2)_{L}\times U(1)_{Y}$ quantum numbers of the VL quark multiplets that can
have interactions -- when taken alone or in pairs -- with SM quark generations and Higgs boson doublet under $SU(2)_{L} \times U(1)_{Y}$ symmetry. 
The tables are organized as follows:    
\begin{itemize}
\item{} top-type multiplets (Table \ref{QN_vec-top}): multiplets
  containing one VL top partner but no bottom partners ({{\sl i.e.}} no VL
  quark  with electric charge $-1/3\ e$). In addition to a
  top partner, these multiplets may contain quarks with exotic charges
  $5/3\ e$ and $8/3\ e$.   
\item{} bottom-type multiplets (Table \ref{QN_vec-bottom}): multiplets
  containing one VL bottom partner but no top partners ({{\sl i.e.}} no VL
  quark  with charge  $2/3\ e$). In addition to a
  bottom partner these multiplets may contain quarks with exotic
  charges $-4/3\ e$ and $-7/3\ e$.  
\item{} mixed multiplets (Table \ref{QN_vec-mixed}): multiplets
  containing both VL top and bottom partners. In addition these multiplets
  may contain all of the exotic charged VL quarks.  
\end{itemize}
The multiplets in these Tables constitute the building blocks we will use to construct scenarios with 2 VL multiplets~\footnote{A model 
where quarks and leptons were taken as a part of quadruplet is given in \cite{Uschersohn:1984cg}}.

\begin{table}[htp]
\begin{center}
\begin{tabular}{| l | c | c | c | c | c |}\hline
Multiplet & $\psi$    &     $(SU(2)_{L},U(1)_{Y})$     &      $T_{3}$
& $Q_{EM}$    & Yukawa to SM   \\ \hline 
Singlet-1/3 & $D$      &     $({\bf 1},-1/3)$                   &
0         &  -1/3    & Yes             \\ \hline 
Doublet -5/6 & $\left(\begin{array}{c}D \\ Y^{-4/3}\end{array}\right)$ 
    & $({\bf 2},-5/6)$      &  $\begin{array}{c}+1/2 \\ -1/2 \end{array}$ 
    & $\begin{array}{c} -1/3 \\ -4/3 \end{array}$ & Yes
    \\ \hline 
Triplet -4/3 & $\left(\begin{array}{c}D \\ Y^{-4/3} \\ Y^{-7/3}
  \end{array} \right)$ & $({\bf 3},-4/3)$ & $\begin{array}{c}0 \\ -1\\
  -2 \end{array}$ & $\begin{array}{c}-1/3 \\ -4/3 \\ -7/3 \end{array}$ & No
    \\ \hline 
\end{tabular}
\caption{Quantum numbers for the bottom--type VL multiplets (up to
  triplets), explicitly indicating weak isospin, hypercharge, electric charge ($Q_{EM}$) and if a direct Yukawa coupling to SM quarks is allowed.} 
\label{QN_vec-bottom}
\end{center}
\end{table} 


\begin{table}[tp]
\begin{center}
\begin{tabular}{| l | c | c | c | c | c |}       \hline
Multiplet    & $\psi$   &    $(SU(2)_{L},U(1)_{Y})$      &
$T_{3}$   & $Q_{EM}$   & Yukawa to SM    \\ \hline 
Doublet 1/6 & $\left(\begin{array}{c}U \\ D \end{array}\right)$     &
$({\bf 2},1/6)$      
    &  $\begin{array}{c}+ 1/2 \\ -1/2 \end{array}$ &
    $\begin{array}{c}+ 2/3 \\ - 1/3 \end{array}$ & Yes$^\ast$
    \\ \hline 
Triplet 2/3 & $\left(\begin{array}{c}X^{5/3} \\ U \\ D \end{array}
\right)$   &  $({\bf 3},2/3)$ & $\begin{array}{c}+1 \\ 0 \\ -1
\end{array}$ & $\begin{array}{c}+5/3 \\ +2/3 \\ -1/3 \end{array} $ & Yes
\\ \hline 
Triplet -1/3 & 
$\left(\begin{array}{c}U \\ D \\ Y^{-4/3} \end{array}\right)$   &
$({\bf 3},-1/3)$ & $\begin{array}{c}+1 \\ 0 \\ -1 \end{array}$ &
$\begin{array}{c} +2/3 \\ -1/3 \\ -4/3 \end{array} $ & Yes
\\ \hline 
Quadruplet 7/6 & $\left(\begin{array}{c}X^{8/3} \\ X^{5/3} \\ U \\ D
  \end{array}\right)$ & $({\bf 4},7/6)$ &  $\begin{array}{c}+3/2  \\
  +1/2 \\ -1/2 \\ -3/2 \end{array}$   & $\begin{array}{c} +8/3 \\ +5/3
  \\ +2/3 \\ -1/3 \end{array} $     & No              \\  \hline 
Quadruplet 1/6 &$ \left(\begin{array}{c}X^{5/3} \\ U \\ D \\ Y^{-4/3}
  \end{array} \right)$ & $({\bf 4},1/6)$ &   $\begin{array}{c}+3/2 \\
  +1/2 \\ -1/2 \\ -3/2 \end{array}$ & $\begin{array}{c}+5/3 \\ +2/3 \\
  -1/3 \\ -4/3 \end{array}$      & No               \\  \hline  
Quadruplet - 5/6 &$ \left( \begin{array}{c}U \\D
    \\Y^{-4/3}\\Y^{-7/3}\end{array}\right)$ & $({\bf 4},-5/6)$    &
$\begin{array}{c}+3/2 \\ +1/2 \\ -1/2 \\ -3/2 \end{array}$ &
$\begin{array}{c}+2/3 \\ -1/3\\-4/3\\-7/3\end{array}$ & No \\  \hline  
\end{tabular}
\caption{Quantum numbers for the mixed--type VL fermion multiplets (up
  to quadruplets), explicitly indicating weak isospin, hypercharge,
  electric charge ($Q_{EM}$) and if a direct Yukawa coupling to SM
  quarks is allowed. For the Doublet 1/6, one can write an independent
  Yukawa coupling with the right-handed up and down quarks.}   
\label{QN_vec-mixed}
\end{center}
\end{table} 

\section{New Yukawa couplings}         
\label{sec:yukawa} 
The SM Yukawa couplings are the coefficients $y_u^{i,j}$ and
$y_d^{i,j}$, where $u$ and $d$ refer to the coupling of the up-type  
and down-type quarks respectively and the indices $i$ and $j$ label
the three SM generations ($i,j = 1,2,3$).  These couplings allow 
the interactions of the SM quarks with the Higgs bosons according to
the following Lagrangian terms:  
\begin{equation}
\mathcal{L}_{SM} = - y_u^{i,j}\, \bar{Q}_L^i \tilde{H} u_R^j -
y_d^{i,j}\, \bar{Q}_L^i H d_R^j + h.c. \,,
\label{eq:3:1}
\end{equation}
where $H=({\bf 2},1/2)$ is the Higgs boson doublet coupling to down-type quarks, $\tilde{H}=  i \tau^2 H^*$ is the same Higgs
multiplet coupling to up-type quarks, $Q_L=({\bf 2},\frac{1}{6})$ is the SM quark doublet, $u_R = ({\bf 1}, \frac{2}{3})$ and 
$d_R = ({\bf 1}, -\frac{1}{3})$ are the SM singlets. After the Higgs boson gets its Vacuum Expectation Value (VEV) we obtain:  
\begin{equation}
\mathcal{L}_{SM} = - \left(\tilde{m}^{up}\right)^{ij} \, \bar{u}_L^i
u_R^j  - \left(\tilde{m}^{down}\right)^{ij} \, \bar{d}_L^i d_R^j +
h.c. \,;
\label{eq:3:2}
\end{equation}
where $\tilde{m}^{up}$ and $\tilde{m}^{down}$ are the SM up-type and down-type $3 \times 3$ mass matrices for quarks.
In the following of the paper, we will work in the basis where the SM Yukawas are diagonal and the eigenvalues are real and positive.
This implies that the phases of the SM quark fields are fixed (up to an overall phase -- the Baryon number), and that a mixing matrix 
$\tilde{V}_{\rm CKM}$ appears in the couplings of the SM quark fields to the $W$ boson. We anticipate that this mixing matrix is not the 
measured CKM matrix, because of the effect of mixing to the VL quarks.

The presence of VL multiplets allows us to add Yukawa interactions between the VL multiplets and the SM quarks.
Due to $SU(2)$ products of representations, new quark doublets can couple with the SM right-handed singlets, 
while new quark singlets and triplets can couple to SM left-handed doublets. 
However, as we are considering a more general case in which more than one VL multiplet is present, new
Yukawa interactions between two VL quark multiplets and the SM Higgs doublet appear. \\

In the following we will not considered two multiplets of same type (same hypercharge). 
As we are interested in the interplay of VL quarks, we have considered cases that satisfy the 
following conditions:
\begin{itemize}
\item{} there must be two VL top quarks.
\item{} eventual VL bottom quarks do not mix with SM bottom sector, {\sl i.e.}
  we can take the mixing to be zero without affecting the top sector,
  to ensure that the model is not constrained too much by the stringent flavour
  physics and $Zbb$ coupling bounds from the bottom sector. 
\end{itemize}
The latter conditions tells us that the only multiplet containing a bottom quark that we allow is the Doublet--1/6, for which the Yukawa involving the down sector is independent form the Yukawa involving the up sector.
These conditions leave us with only four multiplet combinations, 
namely:
\begin{itemize} 
\item{} Singlet ($Y =  2/3$) + Doublet ($Y = 7/6$);
\item{} Doublet ($Y=7/6$) + Triplet ($Y=5/3$);
\item{} Singlet ($Y = 2/3$) + Doublet ($Y = 1/6$);
\item{} Doublet ($Y = 1/6$) + Doublet ($Y = 7/6$). 
\end{itemize} 
The analytical expressions of the mass matrices for the above combinations will be derived in the following. 
For the remaining cases with two different VL multiplets, the mass matrices are presented in Appendix \ref{appendix:A}. \\

The notation we will use to refer to the new Yukawa and mass terms in the Lagrangian is
the following: 
\begin{itemize}
\item{} $\mathcal{L}_{V-SM}$: Yukawa interactions between a VL multiplet and SM quarks; 
\item{} $\mathcal{L}_{V-V}$: Yukawa interactions between two VL multiplets;
\item{} $\mathcal{L}_{\rm mass}$: mass terms after the Higgs boson acquires its VEV and pure VL mass terms.
\end{itemize}
We will also denote the non-SM Yukawa couplings as:
\begin{itemize}
\item{} $\lambda_I^k$: Yukawa between the VL quark $I=1,2$ with the SM quark of generation $k$;
\item{} $\lambda_{Id}^k$: Yukawa coupling of the Doublet--1/6 with the right handed bottom (this coupling will be assumed to be very small in our analysis);
\item{} $\xi_1$: Yukawa between two VL quarks, involving a left-handed doublet (or quadruplet) and a right-handed singlet (or triplet);
\item{} $\xi_2$: Yukawa between two VL quarks, involving a left-handed singlet (or triplet) and a right-handed doublet (or quadruplet).
\end{itemize}
After the Higgs develops its VEV, $\langle H \rangle = \frac{v}{\sqrt{2}} \left(0, 1\right)^T$, these terms will generate mass terms mixing the VL quarks among themselves and with the SM quarks.
In the mass matrices we will consistently use the following notation:
\begin{itemize}
\item{} $y_I^k = \lambda_I^k \frac{v}{\sqrt{2}}$, when the mixing involves a VL doublet (or quadruplet);
\item{} $x_I^k =  \lambda_I^k \frac{v}{\sqrt{2}}$, when the mixing involves a VL singlet (or triplet);
\item{} $y_{Id}^k = \lambda_{Id}^k \frac{v}{\sqrt{2}}$, when the mixing involves a VL doublet (or quadruplet) of down type quark;
\item{} $x_{Id}^k =  \lambda_{Id}^k \frac{v}{\sqrt{2}}$, when the mixing involves a VL singlet (or triplet) of down type quark;
\item{} $\omega = \xi_1 \frac{v}{\sqrt{2}}$ and $\omega' = \xi_2 \frac{v}{\sqrt{2}}$, for the mixing among VL multiplets.
\end{itemize}

Note that for VL multiplets with the same quantum numbers as the SM quarks, a direct mass mixing can be written down: however, this 
term is not physical, as it can be easily removed by redefining the fields corresponding to the SM and VL quarks. In the following, therefore, 
we will never consider this term in the Lagrangians. Finally, all the new Yukawa couplings are potentially complex couplings: for each case, we will 
specify the number of physical phases, recalling that we chose to work in the basis where the SM Yukwas are real, positive and diagonal (thus 3 
mixing angles and one phase are already accounted for in $\tilde{V}_{\rm CKM}$). We also chose the VL masses to be real and positive, thus fixing 
the relative phase between the left and right-handed components of the VL quarks.


\subsection{Singlet $Y=2/3$ and Doublet $Y=7/6$}    \label{sec:yukawa:1}

The Doublet $Y=7/6$ couples to Singlets $Y=2/3$ (both SM and VL): 
\begin{eqnarray}
\mathcal{L}_{V-SM} &=& - \lambda_{1}^k\, \bar{\psi}_{1L} H u_R^k  -
\lambda_2^k \bar{Q}_L \tilde{H} \psi_{2R} + h.c. \,,  \label{eq:3:4} \\
\mathcal{L}_{V-V} &=& - \xi_{1} \, \bar{\psi}_{1L} H \psi_{2R} -
\xi_{2}\, \bar{\psi}_{1R} H \psi_{2L}+ h.c. \,,  \label{eq:3:5}
\end{eqnarray}
where $\psi_{1}=({\bf 2},\frac{7}{6})=\left(X_1^{5/3} \ U_1\right)^T$ and $\psi_{2}=({\bf 1},\frac{2}{3})=U_2$. 
In this Lagrangian, we can use the relative phases between the two VL quarks to fix $\xi_1 > 0$, so that $\xi_{2}$ contains one physical phase.
The relative phase between the VL and the SM quarks can be used to fix one of the 6 phases contained in $\lambda_{1,2}^k$. In total, 
the model has 6 additional phases to the SM ones, when all new Yukawas are non-vanishing. The mass Lagrangian is: 
\begin{eqnarray}
\mathcal{L}_{\rm mass} &=& - y_1^k \bar{U}_{1L} u_R^k - x_2^k u_L^k
U_{2R} - \omega \bar{U}_{1L} U_{2R} - {\omega'} \bar{U}_{1R} U_{2L} 
- M_1 \, \bar{U}_{1L} U_{1R}           \nonumber \\  
&&   - M_2 \, \bar{U}_{2L} U_{2R} - M_1 \, \bar{X}_{1L}^{5/3} X_{1R}^{5/3} + h.c. \,.
 \label{eq:3:6}
\end{eqnarray}
This leads to the mass matrix:
\begin{equation}
M_u = \left( \begin{array}{ccc}
\left(\tilde{m}^{up}\right)_{3\times 3} & 0_{3 \times 1} &
\left(x_2^k\right)_{3 \times 1} \\ 
\left(y_1^k\right)_{1 \times 3} & M_1 & \omega \\
0_{1 \times 3} & \omega' & M_2 
\end{array} \right), \quad 
M_{X_1^{5/3}} = M_1 \,,
 \label{eq:3:7}
\end{equation}
where $\tilde{m}^{up}$ is the SM $3 \times 3$ mass matrix of up sector. The mass matrix can be diagonalised by two unitarity matrices:  
\begin{equation}
M_u = V_L^u \cdot M_u^{diag} \cdot (V^u_R)^\dagger\,.
 \label{eq:3:8}
\end{equation}
The general procedure for the diagonalisation of mass matrices is described in Section \ref{sec:massmatrices}. 



\subsection{Doublet $Y=7/6$ and Triplet $Y=5/3$ }
               \label{sec:yukawa:2}
The Triplet $Y=5/3$ couples to the Doublet $Y=7/6$, which in
turn couples to the SM singlet $u_R$: 
\begin{eqnarray}
\mathcal{L}_{V-SM} &=& - \lambda_1^k \, \bar{\psi}_{1L} H u^{k}_{R}
+ h.c. \,,     \label{eq:3:9} \\
\mathcal{L}_{V-V} &=& - \xi_1 \, \bar{\psi}_{1L} \tau^a \tilde{H}
(\psi_{2R})^a - \xi_2 \, \bar{\psi}_{1R} \tau^a  \tilde{H}
(\psi_{2L})^a + h.c. \,, 
 \label{eq:3:10}
\end{eqnarray}
where $\psi_1 =({\bf 2},\frac{7}{6})=\left(X^{5/3}_1 \ U_1\right)^T$ and $\psi_2
=({\bf 3},\frac{5}{3})=\left(X^{8/3}_2 \ X^{5/3}_2 \ U_2\right)^T$. 
We can again use the relative phase between VL quarks to fix $\xi_1>0$ (and leave $\xi_2$ complex), and remove one of the 3 phases of $\lambda_1^k$, thus the model contains 3 additional physical phases.
The mass
contributions from the Yukawa interactions, including the VL masses, give the Lagrangian:
\begin{eqnarray}
\mathcal{L}_{\rm mass} &=& - y_1^k \bar{U}_{1L} u_R^k
- \sqrt{2} \, \omega \bar{U}_{1L} U_{2R} - \omega \bar{X}^{5/3}_{1L}
X^{5/3}_{2R} - \sqrt{2} \, {\omega'} \bar{U}_{1R} U_{2L} - {\omega'} 
\bar{X}^{5/3}_{1R} X^{5/3}_{2L} - M_1 \, \bar{U}_{1L} U_{1R}
\nonumber \\ 
&& - M_1 \, \bar{X}^{5/3}_{1L} X^{5/3}_{1R} - M_2 \, \bar{U}_{2L}
U_{2R} - M_2 \, \bar{X}^{5/3}_{2L} X^{5/3}_{2R} - M_2 \,
\bar{X}^{8/3}_{2L} X^{8/3}_{2R} + h.c. \,,      
\label{eq:3:11}
\end{eqnarray}
leading to the mass matrices:
\begin{equation}
M_u = \left( \begin{array}{ccc}
\left(\tilde{m}^{up}\right)_{3\times 3} & 0_{3 \times 1} & 0_{3 \times
  1} \\ 
\left(y_1^k\right)_{1 \times 3} & M_1 & \sqrt{2} \, \omega \\
0_{1 \times 3} & \sqrt{2} \, \omega' & M_2 
\end{array} \right), \quad 
M_{X^{5/3}} = \left( \begin{array}{cc}
M_1 & \omega \\
{\omega}' & M_2
\end{array} \right)
, \quad 
M_{X^{8/3}} = M_2\,.
 \label{eq:3:12}
\end{equation}
Note that the mass matrix in the up sector is the same as the previous case (with $x_2 = 0$), while now there are two exotic charged 
quarks which mix via the Yukawa couplings $\xi_{1,2}$.


\subsection{Singlet $Y=2/3$ and Doublet $Y=1/6$}
\label{sec:yukawa:3} 
In this case, the two VL multiplets have the same quantum numbers of the SM quarks, therefore one can replicate all the standard Yukawa 
couplings, including an independent coupling for the right-handed downs $d_R$:
\begin{eqnarray}
\mathcal{L}_{V-SM} &=& 
    - \lambda_{1}^k\, \bar{\psi}_{1L} \tilde{H} u_R^k -
    \lambda_{1d}^k\, \bar{\psi}_{1L} H d_R^k  - \lambda_2^k 
    \bar{Q}^k_L \tilde{H} \psi_{2R} + h.c. \,, \label{eq:3:13}\\
\mathcal{L}_{V-V} &=& - \xi_1\, \bar{\psi}_{1L} \tilde{H} \psi_{2R} 
     - \xi_{2}\,\bar{\psi}_{1R} \tilde{H} \psi_{2L} + h.c. \,, 
         \label{eq:3:14}
\end{eqnarray}
where $\psi_1 =({\bf 2},\frac{1}{6})=\left(U_1 \ D_1\right)^T$ and $\psi_2 =
({\bf 1},\frac{2}{3}) = U_2$. 
Like in the cases above, we can use the relative phases of the VL quarks to make $\xi_1>0$, and remove one of the 6 phases in $\lambda_{1,2}^k$. 
The bottom coupling $\lambda_{1d}^k$ are also complex, however as explained above we will set these couplings to zero in the following. The mass 
terms take the form (where we normalise the mass parameters to have coefficient one for the top partners): 
\begin{eqnarray}
\mathcal{L}_{\rm mass} &=& 
  - y_{1}^k \bar{U}_{1L} u_R^k 
  - x_2^k u_L^k
  U_{2R} - \omega \bar{U}_{1L} U_{2R} - \omega' \bar{U}_{2L} U_{1R} 
   \nonumber \\  
&&   - M_1 \, \bar{U}_{1L} U_{1R} - M_2 \, \bar{U}_{2L} U_{2R} - M_1
\, \bar{D}_{1L} D_{1R} + h.c. \,.
  \label{eq:3:15}
\end{eqnarray}
The mass matrices, therefore, read: 
\begin{equation}
M_u = \left( \begin{array}{c c c}
\left(\tilde{m}^{up}\right)_{3\times 3} &  0_{3 \times 1} &
(x_2^k)_{3\times 1} \\ 
(y_{1}^k)_{1\times 3} & M_1 & \omega \\
0_{1\times3} & \omega' & M_2
\end{array} \right), \quad
M_d = \left( \begin{array}{cc}
\left(\tilde{m}^{down}\right)_{3\times 3} & 0_{1 \times 3} \\
(0)_{3\times 1} & M_1
\end{array} \right)\,.
 \label{eq:3:16}
\end{equation}
Now, the mass matrix in the up sector is the same as the first case, while the down sector mass matrix is diagonal (as we set to zero 
the mixing in the down sector). No exotic charges are present in this case.


\subsection{Doublet $Y=1/6$ and Doublet $Y=7/6$}
\label{sec:yukawa:4} 
This is the only case where we consider two doublets, thus both VL multiplets only couple to the right-handed SM quarks:
\begin{eqnarray}
\mathcal{L}_{V-SM} &=& 
    - \lambda_{1}^k\, \bar{\psi}_{1L} \tilde{H} u_R^k -
    \lambda_{1d}^k\, \bar{\psi}_{1L} H d_R^k - \lambda_2^k \,
    \bar{\psi}_{2L} H u_R^k + h.c. \,,   \label{eq:3:17}
\end{eqnarray}
where $\psi_1 =({\bf 2},\frac{1}{6})=\left(U_1 \ D_1\right)^T$ and $\psi_2 =
({\bf 2},\frac{7}{6}) = \left(X_2^{5/3} \ U_2\right)^T$. 
In this case, no Yukawa coupling between the two VL multiplet is allowed, therefore one can use the two free phases to remove one phase in 
$\lambda_1^k$ and one in $\lambda_2^k$, so that only 4 new phases are present in this model. Once again, we will set $\lambda_{1d}^k = 0$.
The mass Lagrangian and mass matrices become:
\begin{eqnarray}
\mathcal{L}_{\rm mass} &=& 
  - y_{1u}^k \bar{U}_{1L} u_R^k -  y_{1d}^k \bar{D}_{1L} d_R^k - y_2^k
  U_{2L} u^k_R                                   \nonumber \\ 
&&  - M_1 \, \bar{U}_{1L} U_{1R} - M_1 \, \bar{D}_{1L} D_{1R} -  M_2
\, \bar{U}_{2L} U_{2R} -  M_2 \, \bar{X}^{5/3}_{2L} X^{5/3}_{2R} + h.c. \,, 
\label{eq:3:18} 
\end{eqnarray}
and
\begin{equation}
M_u = \left( \begin{array}{c c c}
\left(\tilde{m}^{up}\right)_{3\times 3} &  0_{3 \times 1} &
0_{3\times1} \\ 
 (y_{1}^k)_{1\times 3} & M_1 &  0 \\
(y_2^k)_{1 \times 3}  & 0 & M_2
\end{array} \right), \quad
M_d = \left( \begin{array}{c c }
\left(\tilde{m}^{down}\right)_{3\times 3} & 0_{3\times 1}  \\
(0)_{1\times 3} & M_2 \\
\end{array} \right), \quad 
M_{X^{5/3}} = M_2 \,.      \label{eq:3:19}
\end{equation}
The structure of the up-sector mass matrix is now different from the cases above.


\subsection{Diagonalisation of the mass matrices}             
\label{sec:massmatrices}
We have discussed so far four special cases, where two top partners mix with the up-sector giving rise to $5\times 5$ mass matrices, 
while the down sector is always diagonal. More general cases, and the form of the mixing matrices, can be found in Appendix \ref{appendix:A}.
In the up sector, the mass matrices can be diagonalised by two unitary $5\times 5$ matrices:
\begin{equation}
M_{u} = V_L \cdot M_{u}^{diag} \cdot V^{\dagger}_R\,,
\label{eq:4:7}
\end{equation}
with:
\begin{equation}
M_{u}^{diag} = \left( \begin{array}{ccccc}
m_{u} & & & & \\ & m_{c} & & & \\ & & m_{t} & & \\ & & &
m_{t'_1} & \\ & & & & m_{t'_2} 
\end{array} \right) \,.   \label{eq:4:8}
\end{equation}
Indeed when two VL multiplet are present at the same time, 
there are three types of mixing structures which can
arise with the SM up quarks: 
\begin{itemize}
\item{Case A:} two semi-integer isospin multiplets (as doublets, quadruplets, etc.). In this case the mass matrix becomes:
\begin{equation}
M_{u}^{(A)} =
\left(
\begin{array}{ccccc}
\tilde{m}_{u} & & & 0 & 0 \\
& \tilde{m}_{c} & & 0 & 0 \\
& & \tilde{m}_{t} & 0 & 0 \\
y_1^1 & y_1^2 & y_1^3 & M_{1} & 0 \\
y_2^1 & y_2^2 & y_2^3 & 0 & M_{2} \\
\end{array}      \label{eq:4:9}
\right) \,;
\end{equation}
\item{Case B:} two integer isospin multiplets (as singlets, triplets, etc.). The mass matrix is:
\begin{equation}
M_{u}^{(B)} =
\left(
\begin{array}{ccccc}
\tilde{m}_{u} & & & x^1_1 & x^1_2 \\
& \tilde{m}_{c} & & x^2_1 & x^2_2 \\
& & \tilde{m}_{t} & x^3_1 & x^3_2 \\
0 & 0 & 0 & M_{1} & 0 \\
0 & 0 & 0 & 0 & M_{2} \\ 
\end{array}
\right) \,;        \label{eq:4:10}
\end{equation}
\item{Case C:} one semi-integer isospin multiplet and one integer isospin multiplet. The mass matrix is:
\begin{equation}
M_{u}^{(C)} =
\left(
\begin{array}{ccccc}
\tilde{m}_{u} & & & 0 & x_2^1 \\
& \tilde{m}_{c} & & 0 & x_2^{2} \\
& & \tilde{m}_{t} & 0 & x_2^{3} \\
y_1^{1} & y_1^{2} & y_1^{3} &  M_{1} & \omega \\
0 & 0 & 0 & \omega' & M_{2} \\
\end{array}
\right) \,.     \label{eq:4:11}
\end{equation}
\end{itemize}
The VL multiplets considered in our analysis belong to the cases indicated in Table \ref{tab:ABC}, where the combinations we are
considering for our numerical analysis have been highlighted. These mass matrices cannot be diagonalised analytically. One can obtain 
approximate results in the limit where the VL masses $M_{1,2}$ are much larger than the contribution from the Yukawa couplings, and general 
results can be found in the Appendix of Ref.~\cite{Buchkremer:2013bha}. In our numerical results, however, we will use a numerical procedure 
to find the correct mass eigenstates and mixing angles, detailed in the next subsection.
\begin{table}[htp]
\begin{center}
\begin{tabular}{c|c|c|c|c}
%
2nd $\downarrow$  \hspace*{0.2cm} 1st $\rightarrow$ & Singlet $Y=\frac{2}{3}$ & Doublet
$Y=\frac{1}{6}$ & Doublet $Y=\frac{7}{6}$ & Triplet $Y=\frac{5}{3}$ \\  \hline  
%
Singlet $Y=\frac{2}{3}$ & case B & \bf case C & \bf case C & case B     \\ \hline
Doublet $Y=\frac{1}{6}$ &        & case A     & \bf case A & case C     \\ \hline
Doublet $Y=\frac{7}{6}$ &        &            & case A     & \bf case C \\ \hline
Triplet $Y=\frac{5}{3}$ &        &            &            & case B     \\
\end{tabular}
\caption{
Mixing structures for the VL quarks multiplets considered in our analysis. The combinations we have studied in detail are in bold.}
\label{tab:ABC}
\end{center}
\end{table} 

Some models also contain two exotic quarks which mix via a $2 \times 2$ matrix, like in Section~\ref{sec:yukawa:2}:
\begin{equation} \label{eq:massX2/3}
M_{X^{5/3}} = \left( \begin{array}{cc}
M_1 & \omega \\
{\omega}' & M_2
\end{array} \right)\,.
\end{equation}
This mass matrix can be diagonalised by two Unitary matrices
\begin{equation}
M_{X} = V_{XL} \cdot M_X^{diag} \cdot V^{\dagger}_{XR}\,,
\end{equation}
with eigenvalues
\begin{equation}
M_{X_{1,2}}^2 = \frac{1}{2} \left( M_1^2 + M_2^2 + \omega^2 + |\omega'|^2 \mp \sqrt{(M_1^2 + M_2^2 + \omega^2 + |\omega'|^2)^2 - 4 |M_1 M_2 - \omega \omega'|^2} \right)\,.
\end{equation}
Parametrising the mixing matrices as
\begin{equation}
V_{XL/R} = \left( \begin{array}{cc}
\cos \alpha_{L/R} & e^{i \delta_{L/R}} \sin \alpha_{L/R} \\
- e^{- i \delta_{L/R}} \sin \alpha_{L/R} & \cos \alpha_{L/R} 
\end{array} \right)\,;
\end{equation}
the mixing angles and phases can be expressed as
\begin{equation} \label{eq:anglesX2/3}
e^{- i \delta_L} \sin (2 \alpha_L) = 2 \frac{M_1 \omega' + M_2 \omega}{M_{X_2}^2 - M_{X_1}^2}\,, \quad e^{- i \delta_R} \sin (2 \alpha_R) = 2 \frac{M_1 \omega + M_2 \omega'}{M_{X_2}^2 - M_{X_1}^2}\,.
\end{equation}


\subsubsection{Numerical procedure}
\label{sec:numerics}  
In order to evaluate the constraints and estimate the production cross-sections of VL quarks at colliders we need to write the
Lagrangians presented in Section \ref{sec:yukawa:1}--\ref{sec:yukawa:4} in the mass basis. In the SM the physical masses of 
quarks are uniquely defined by Yukawa couplings. The introduction of VL quarks with couplings to
all the three SM quark generations enlarges the mass matrices and results in variation of the physical masses of SM quarks. 
In Section \ref{sec:massmatrices} we have presented the structure of the $5 \times 5$ mass matrices in the gauge basis. 
The procedure we have adopted for diagonalisation is the following: 
\begin{itemize}
\item{\bf Step 1:} using the new Yukawa Couplings ($x_i, y_i, \omega,
  \omega'$) and the masses $M_1, M_2$ as input parameters  
we write the five eigenvalue equations (one for each of the physical quarks): 
\begin{equation}
| M_{u}^\dagger M_{u} - \lambda_i I | = 0 \,,
\label{eqn:5:1}
\end{equation}
where $M_{u}$ are the mass matrices and $\lambda_i$ are the eigenvalues (square of physical masses).
\item{\bf Step 2:} the first three equations correspond to the three
known values of SM quark physical masses ($m_{u}, m_{c},
m_{t}$). We solve these equations for the three SM Yukawa couplings, namely
$\tilde{m}_{u}, \tilde{m}_{c}, \tilde{m}_{t}$.
\item{\bf Step 3:} we insert the SM Yukawa couplings obtained in Step 2
into the mass matrices and diagonalize them again to get the mixing matrices
($V_L^{u}$ and $V_R^{u}$) and the physical masses of the VL
quarks.  
\end{itemize} 
In this process we only consider positive solution for the SM Yukawa couplings (Step 2), to be consistent with our original choice.
For simplicity, we also set all the new physical phases to zero, allowing ourselves only a negative sign of the Yukawa couplings 
when physically inequivalent to the positive sign.


\section{Tree-level and Electroweak Precision bounds} 
\label{sec:bounds}
\subsection{Tree-level bounds on VL quarks}
\label{sec:bounds:1} 
In order to study the tree-level bounds we need to recall the couplings of the VL fermions to the gauge bosons and in
particular to the Z boson. The complete structure is given in the appendix \ref{appendix:B}. The mass matrices are diagonalised as follows: 
\begin{equation}
M_{u} = V_L \cdot M_{u}^{diag}\cdot V_R^\dagger \,,
\label{eq:6:1}
\end{equation}
so that the mass eigenstates are defined as:
\begin{equation}
\left( \begin{array}{c}
u \\ c \\ t \\ t'_1 \\ t'_2
\end{array} \right)_{L/R} = V^\dagger_{L/R} \cdot \left( \begin{array}{c}
u^1 \\ u^2 \\ u^3 \\ U_1 \\ U_2
\end{array} \right)_{L/R} \,.           \label{eq:6:2}
\end{equation}
In the mass eigenstate basis, the couplings of the $Z$ boson read (for the up-quarks):
\begin{eqnarray}
g_{ZL}^{IJ} &=& \frac{g}{2 \cos \theta_W} \left[ \left( 1 -
    \frac{4}{3} \sin^2 \theta_W \right) \delta^{IJ} +  
\sum_{K=4,5} (2 T_3^{(K)} - 1) V_{L}^{\ast,KI} V_L^{KJ} \right],
         \label{eq:6:3} \\
g_{ZR}^{IJ} &=& \frac{g}{2 \cos \theta_W} \left[ \left(- \frac{4}{3}
    \sin^2 \theta_W \right) \delta^{IJ} + \sum_{K=4,5} 2 T_3^{(K)}   
V_{R}^{\ast,KI} V_R^{KJ} \right]\,,  \label{eq:6:4}
\end{eqnarray}
where $T_3^{(K)}$ is the weak isospin of the VL quark $K$.
Note that the modifications to the couplings with respect to the SM values (including off-diagonal terms) are all proportional to
the $V_{L/R}^{4I}$ and the $V_{L/R}^{5I}$ elements of the mixing matrices. Let's now consider the bounds applied generation by
generation. 


\subsubsection{Bounds on the first generation}
\label{sec:bounds::1:1} 


\subsubsection*{Atomic Parity Violation} 
The weak charge of a nucleus can be, in general, written as
\cite{Deandrea:1997wk}:
\begin{equation}
Q_W = (2 Z + N) (\tilde{g}_{ZL}^u + \tilde{g}_{ZR}^u) +
(Z+2N)(\tilde{g}_{ZL}^d + \tilde{g}_{ZR}^d) \,, \label{eq:6:5}
\end{equation}
where $g_Z = \frac{g}{2 \cos \theta_W} \tilde{g}_Z$.
In our case:
\begin{equation}
\delta Q_W = (2 Z + N) \sum_{K=4,5} \left((2 T_3^{(K)} - 1)
  |V_L^{K1}|^2 + 2 T_3^{(K)}  |V_R^{K1} |^2\right) \,. \label{eq:6:6}
\end{equation}
The strongest bound is for Cesium, for which $2Z+N = 188$, and
\cite{Agashe:2014kda}: 
\begin{equation}
\left. Q_W\right|_{\rm exp.} = -73.20 \pm 0.35\,, \qquad \left. Q_W
\right|_{SM} = -73.15 \pm 0.02 \,. \label{eq:6:7}
\end{equation}
Neglecting the theoretical error on the SM value, which is rather small:
\begin{equation}
\delta Q_W = - 0.05 \pm 0.35\,. \label{eq:6:8}
\end{equation}



\subsubsection{Bounds on the second generation}
          \label{sec:bounds:1:2} 


\subsubsection*{$Z$-couplings measured at LEP} 

The couplings of the charm have been well measured at
LEP~\cite{ALEPH:2005ab}: 
\begin{equation}
g_{ZL}^c = 0.3453 \pm 0.0036\,, \quad g_{ZR}^c = - 0.1580 \pm
0.0051\,, \quad \mbox{correlation} = 0.30 \,. \label{eq:6:9}
\end{equation}
One can use this input to reconstruct a $\chi^2$ distribution, and set bounds on the mixing angles: the most conservative approach is to 
assume that the central values correspond to the SM prediction, therefore any deviation must be smaller than the quoted errors. The
$\chi^2$ can be constructed as follows: 
\begin{equation}
\chi^2 = \sum_{i,j=1,2} \delta g^i (V^{-1})^{ij} \delta g^j\,,
\label{eq:6:10} 
\end{equation}
where $\delta g$ are the deviations in the two couplings (left and right handed) and:
\begin{equation}
V^{ij} = \rho^{ij} \sigma^i \sigma^j\,, \quad \mbox{where} \quad \rho 
= \left( \begin{array}{cc} 
1 &0.30 \\ 0.30 & 1 \end{array} \right) \,,       \label{eq:6:11}
\end{equation}
and $\sigma^j$ are the errors. In the plots below, we will draw confidence levels at 68\% ($\chi^2 = 2.30$ for 3 degrees of freedom),
95\% ($5.99$) and 99\% (9.21). 



\subsubsection{Bounds on the third generation}
     \label{sec:bounds:1:3} 


\subsubsection*{$W_{tb}$ couplings measured at TeVatron and LHC}

As there is no mixing in the bottom sector, the value of $V_{tb}$ is affected only by the mixing of the top with the VL quarks in the
left-handed sector: 
\begin{equation}
|V_{tb}|^2 = 1-\sum_{K=4,5} |V_L^{K3}|^2\,. \label{eq:6:12}
\end{equation}

A list of up-to-date direct measurements and lower bounds on $V_{tb}$ can be found here~\cite{Chiarelli:2013psr}. The strongest bound is
from a CMS measurements of single top cross sections at 7 TeV~\cite{Chatrchyan:2012ep}, which gives  
$|V_{tb}| > 0.92$ at 95\% CL. We will use this limit to define the allowed region, even though all
other searches, including Tevatron \cite{Aaltonen:2014ura}, CMS \cite{Khachatryan:2014iya} and ATLAS \cite{Friedrich:2014eqa} at 8 TeV, have  
bounds $|V_{tb}| > 0.80$ at 95\% CL (see also \cite{Lista:2014kza,Moon:2014cea} for summaries of the most recent results).




\subsection{Oblique corrections}           
    \label{sec:bounds:2} 
In the following we analyse the impact of the interplay of two VL multiplets with the complete three SM generations in order to
establish bounds from the electroweak precision tests (EPW) in term of the oblique corrections. These bounds will be compared with those
coming from tree-level observables. In order to parameterise the effect of the loop correction we will use the Peskin-Takeuchi
$S$, $T$ and $U$ parameters, defined as~\cite{Peskin:1990zt,Peskin:1991sw}:
\begin{eqnarray}
S &=& 16\pi \left[ \Pi_{33}^{\prime}(0) - \Pi_{3Q}^{\prime}(0) \right]
      \,,  \label{eq:7:1}        \\ 
T &=& \frac{4\pi}{s_{W}^{2}c_{W}^{2}m_{Z}^{2}} \left[ \Pi_{11}(0) -
  \Pi_{33}(0) \right] \,, \label{eq:7:2} \\ 
U &=& 16\pi \left[ \Pi_{11}^{\prime}(0) - \Pi_{33}^{\prime}(0) \right] \,.
\label{eq:7:3}
\end{eqnarray}
where $\Pi_{ij}$ are the scalar two point functions, related to the $W$, $Z$, and $A$ one--loop two point functions by: 
\begin{eqnarray}
\Pi_{AA} &=& e^{2}\Pi_{QQ} \,, \label{eq:7:4} \\
\Pi_{ZA} &=& \frac{e^{2}}{s_{W}c_{W}}\left( \Pi_{3Q} - s_{W}^{2}
  \Pi_{QQ} \right) \,,  \label{eq:7:5} \\ 
\Pi_{ZZ} &=& \frac{e^{2}}{s_{W}^{2}c_{W}^{2}} \left( \Pi_{33}
  -2s_{W}^{2}\Pi_{3Q} + s_{W}^{4}\Pi_{QQ} \right) \,,  \label{eq:7:6} \\ 
\Pi_{WW} &=& \frac{e^{2}}{s_{W}^{2}} \Pi_{11} \,.  \label{eq:7:7}
\end{eqnarray}
For this work we have taken the SM reference point masses to be $m_{h, {\rm ref}}=126$ GeV, $m_{t,{\rm ref}}=173$ GeV and  
$m_{b,{\rm ref}} = 4.2$ GeV.  If $\hat{S}_{VL}$ and $\hat{T}_{VL}$ is the contribution of the model (including VL quarks) to the S 
and T parameters then the deviations can be defined as \cite{Lavoura:1992np}:
\begin{eqnarray}
S = \hat{S}_{VL} - \hat{S}_{SM} \,, \quad
T = \hat{T}_{VL} - \hat{T}_{SM} \,,
\end{eqnarray}
where the SM reference values, $\hat{S}_{SM}$ and $\hat{T}_{SM}$, can be approximated as:
\begin{eqnarray}
\hat{S}_{SM} &\simeq& \frac{N_{c}}{6\pi} \left[ 3 - \frac{1}{3} \ln
  \left( \frac{m_{t}^{2}}{m_{b}^{2}} \right) \right] \,,
     \label{eq:7:8} \\ 
\hat{T}_{SM} &\simeq& \frac{N_{c}}{16\pi s_{W}^{2}c_{W}^{2}m_{Z}^{2}}
\left[ m_{t}^{2} + m_{b}^{2} - \frac{2m_{t}^{2}m_{b}^{2}} 
{m_{t}^{2}-m_{b}^{2}} \ln \left( \frac{m_{t}^{2}}{m_{b}^{2}} \right)
\right] \,.  \label{eq:7:9}
\end{eqnarray}
By fixing $U=0$, the experimental results for the $S$ and $T$ parameters are given by~\cite{Baak:2011ze,Baak:2012kk}:
\begin{equation}
S= 0.05 \pm 0.09 \,, \qquad T= 0.08 \pm 0.07 \,,  \label{eq:7:10}
\end{equation}
where the correlation between $S$ and $T$ in this fit is
$\rho_{ST}=0.91$.

If VL multiplets are present in the physical spectrum, the $\Pi_{ij}$ two-point functions get extra contributions
from the new particles circulating in the loops. Obviously if other particles than the VL quarks are present in a specific 
model of new physics, they may also contribute. Therefore the bounds obtained from the $S$ and $T$ parameters should be taken with this restriction in
mind. The detailed formulas for the contributions of the VL particles to the $S$ and $T$ parameters are given in Appendix \ref{appendix:C}. In the 
numerical study we have combined the tree-level and loop-level bounds from the $S$ and $T$ parameters, in the case of two multiplets of VL
quarks. This allows to study the effect of the interplay of the two multiplets and the effect of the extra Yukawa couplings among the two
VL multiplets ($\omega$ and $\omega'$ parameters). Even if in the previous sections we have calculated analytically the general mixing
structure under some approximations, for the numerical part of this analysis the mixing angles have been computed exactly by numerically 
diagonalising the mass matrices. 



\section{Results} 
      \label{sec:results}

For our numerical analysis we have considered the mass parameters $M_1$ and
$M_2$ of the VL quarks to be same, {\sl i.e.} $M_1 = M_2 = M$. 
As the models have too many parameters to make a meaningful scan, we will show results in two limiting cases, when possible:
\begin{itemize}
\item the VL quarks can mix with a single SM generation, but not with each other (i.e. $\omega \sim \omega' \ll M$);
\item the VL quarks mix with each other (wherever possible), but the mixing with SM quarks is very small. 
\end{itemize}
The results in these simple limits can give a general idea on the allowed value of the mixing parameters, even though the case where all of them 
are non-zero is more realistic. We will focus on the benchmark value for the VL mass of $800$ GeV as a recent CMS analysis 
\cite{CMS-PAS-B2G-12-017} sets a bound of $788$ GeV under the assumption of strong pair production of VL quarks and 100\% branching 
fractions to $qW$. In the following we also consider cases in which the VL quark does not decay to  $qW$.
For these cases the bound does not apply directly, but these VL quarks are in doublets containing also other VL quarks for which the bound applies. 
As it is reasonable to assume that mass splittings inside multiplets are not large compared to the mass scale of the multiplet, we shall apply this 
$800$ GeV benchmark value to all cases.



\subsection{Singlet $Y=2/3$ and Doublet $Y=7/6$}
    \label{sec:results:1}    

\begin{figure}[bth]
\begin{center}
\hspace*{-0.7cm}
\epsfig{file=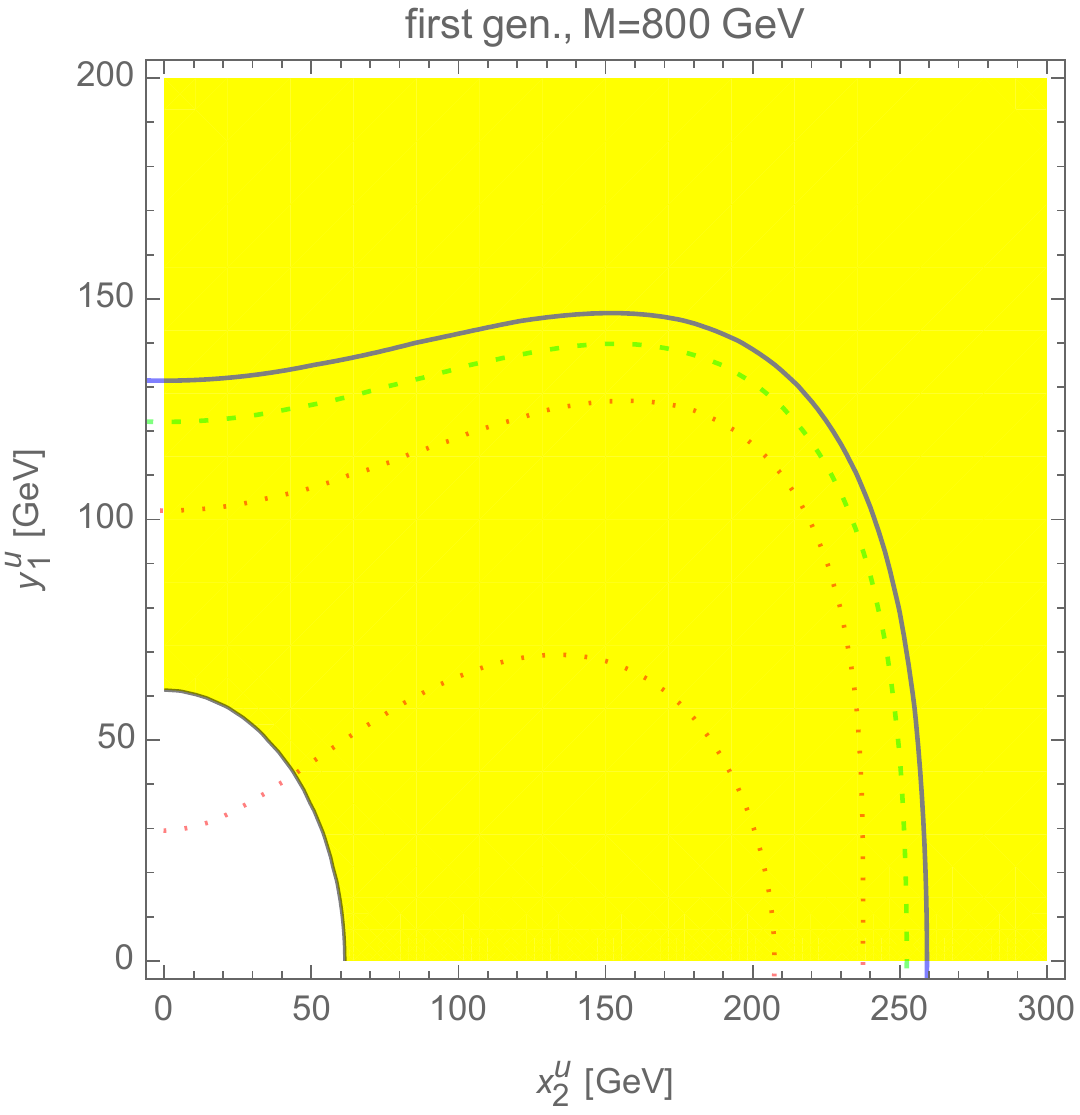,width=0.48\textwidth}
                 \hspace*{0.4cm}
\epsfig{file=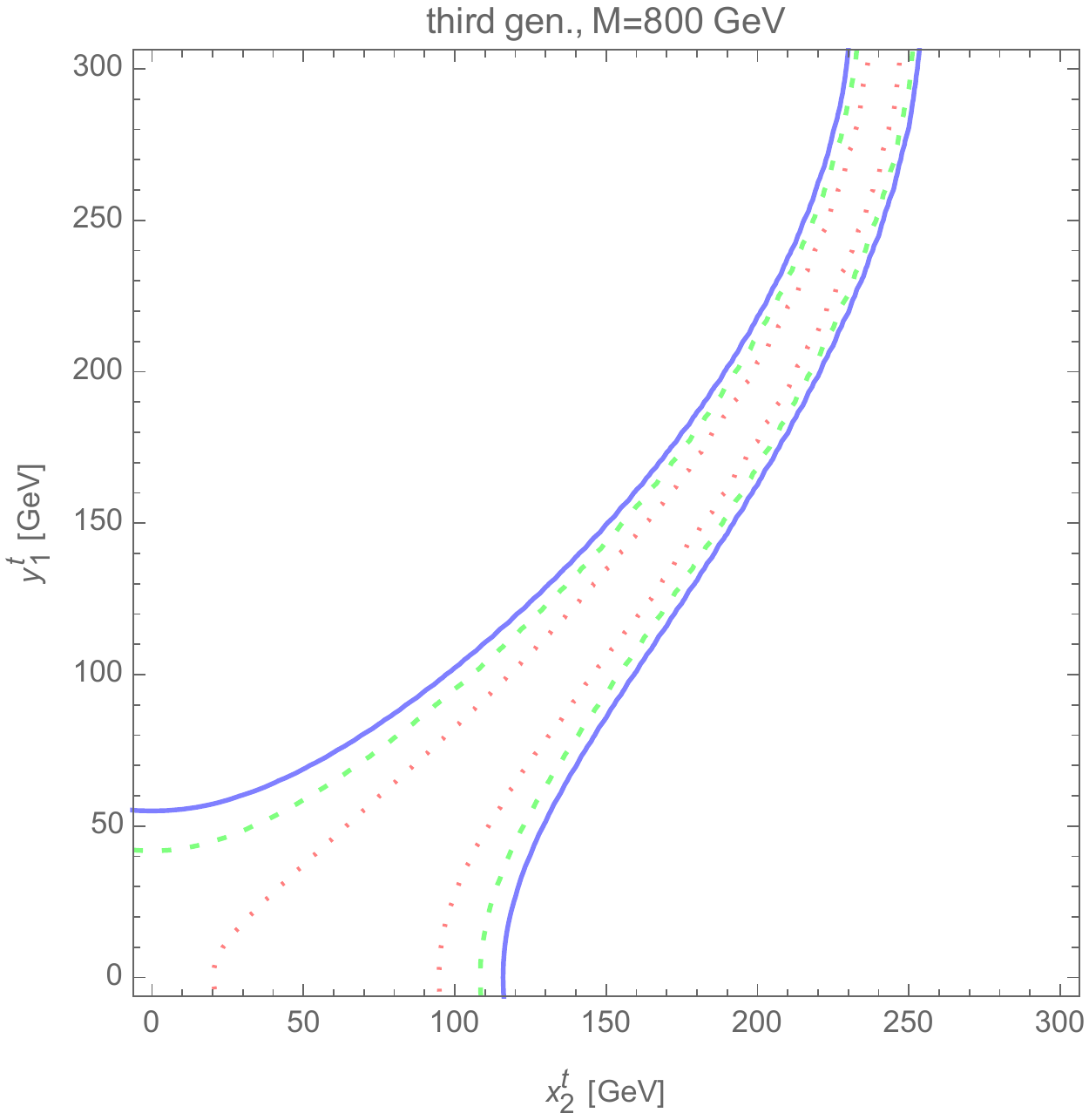,width=0.48\textwidth}
\end{center}
\caption{\underline{\bf \boldmath Singlet $Y=2/3$ and Doublet $Y=7/6$} (Section
  \ref{sec:yukawa:1}): EWP bounds at 1 $\sigma$ (red-dashed), 2  $\sigma$ (green-dashed) and 3 $\sigma$ (blue) for VL quarks coupling with
  the first (left panel) and third (right panel) SM generations, compared with the region excluded at 3$\sigma$ by tree-level bounds (yellow
  region). Here, $M = 800$ GeV, and $\omega = \omega'=0$. Only the first quadrant is shown as the figures are symmetric with respect to
  a sign change in the coordinates in the other 3 quadrants. Similar considerations apply to all the other figures of the same type.}     
\label{fig:snonsmd3}
\end{figure}
This scenario contains -- besides the SM particle spectrum --
two VL top quarks and one exotic quark with charge $5/3$. 
The Yukawa couplings and mass matrices for this scenario are given in 
Section \ref{sec:yukawa:1}. The additional parameters (apart from the SM ones) are:
$x_2^k$, $y_1^k$, $\omega$, $\omega'$ and $M$ with $k$ running on SM quark generations.  
We first study the case where $\omega' \sim \omega \sim 0$, and the VL quarks couple to a single generation: in this case, setting $\omega'$ to zero allows us to set both Yukawa couplings $x_2$ and $y_1$ to be real and positive.
The allowed regions in the parameter space, given the constraints from tree-level and EWP tests
discussed in Section \ref{sec:bounds}, are presented in Figure \ref{fig:snonsmd3}. 
We see that for couplings to the light generations (see Appendix \ref{appendix:D} for the plot with second generation), the tree level bounds always dominate, and require the mixing of VL quarks to be rather small. The case of the third generation is very different: the tree level bounds are very weak as they only come from $V_{tb}$, while EWP tests allow for large mixings, especially via a compensation between the doublet and singlet (in particular, $y_1^3$ can assume very large values). This situation can be very interesting in the single-production channel, where for instance the top partner may be produced via couplings to the first generation (the smaller coupling is easily compensated by the valence quark in the initial state~\cite{Buchkremer:2013bha}) and then decay into a third generation quark~\cite{Brooijmans:2014eja,Beauceron:2014ila,Basso:2014apa}.


\begin{figure}[tb]
\begin{center}
\hspace*{-0.7cm}
\epsfig{file=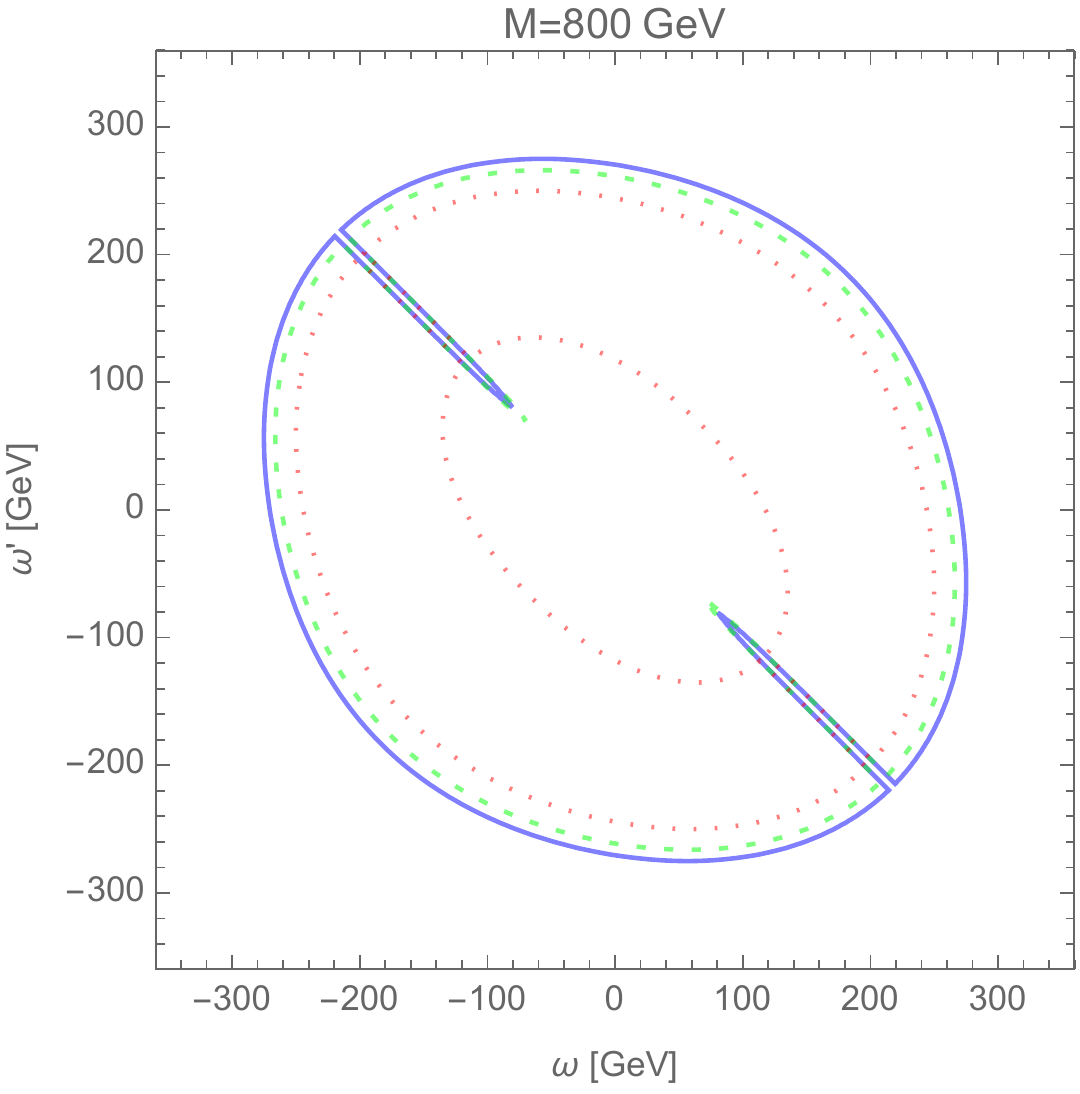,width=0.48\textwidth} 
\end{center}
\caption{\underline{\bf \boldmath Singlet $Y=2/3$ and Doublet $Y=7/6$} (Section
  \ref{sec:yukawa:1}):
  EWP bounds at 1$\sigma$ (red-dashed), 2$\sigma$ (green-dashed) and 3
  $\sigma$ 
  (blue) as a function of the new Yukawa couplings
  $\omega$ and $\omega'$ with $M = 800$ 
  GeV. We have assumed that there is
  no mixing of VL quarks with the SM quark generations {\sl i.e.} $x_2^k =
  y_1^k = 0$.}  
\label{fig:dd1}
\end{figure}

In Figure \ref{fig:dd1}  we show the EWP bounds in the plane 
of the Yukawa couplings between VL quarks, $\omega$ and $\omega'$, assuming that the other couplings are small. This plot gives a general idea on the allowed size of $\omega$ and $\omega'$: the bound is indeed not very strong, and values up to $300$ GeV are allowed. The plot is clearly symmetric under change of sign of either $\omega$ or $\omega'$, reflecting the one arbitrary phase in this sector.
As we are approximately decoupling the two $t'$s from the SM quarks, the mixing is dominated by the $2\times 2$ block of the VL quarks, similar to the matrix in Eq.~(\ref{eq:massX2/3}).
We then see that the mixing angles (Eq.(\ref{eq:anglesX2/3})) vanish when $\omega = - \omega'$, thus explaining the sharp dents in the excluded region. This effect only appears in our limiting choice $M_1 = M_2$ and for negligible mixing to SM quarks.



\subsection{Doublet $Y=7/6$ and Triplet $Y=5/3$}
        \label{sec:results:2}    

%
\begin{figure}[tb]
\begin{center}
\hspace*{-0.7cm}
\epsfig{file=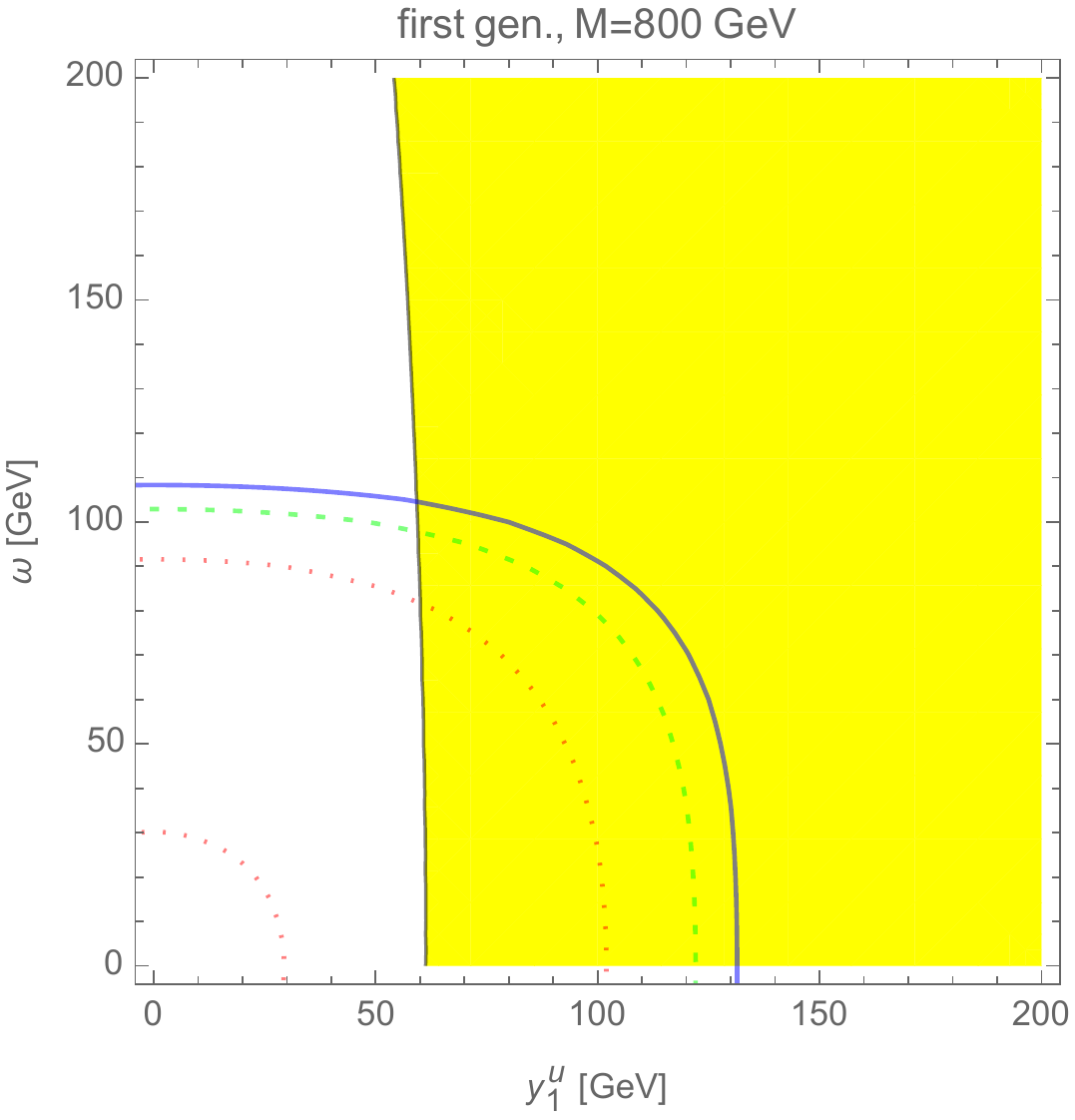,width=0.48\textwidth} 
     \hspace*{0.4cm}
\epsfig{file=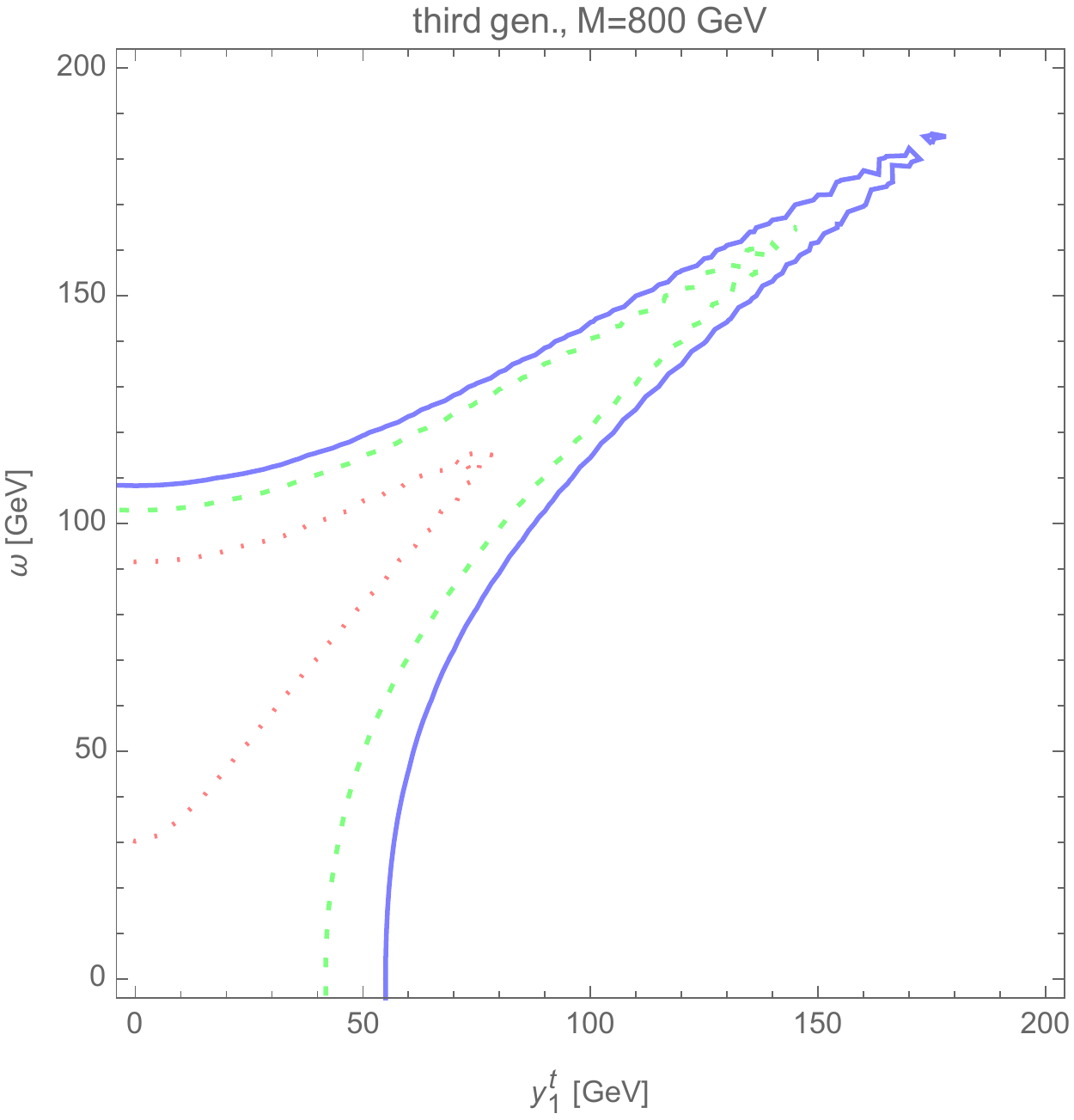,width=0.48\textwidth} 
\end{center}
\caption{\underline{\bf \boldmath Doublet $Y=7/6$ and Triplet $Y=5/3$} (Section
  \ref{sec:yukawa:2}):
  EWP bounds at 1$\sigma$ (red-dashed), 2$\sigma$ (green-dashed) and 3 $\sigma$ (blue) for VL quarks coupling with
  the first (left panel) and third (right panel) SM generations, compared with the region excluded at 3$\sigma$ by tree-level bounds (yellow
  region in the left panel). $M = 800$ GeV and $\omega = \omega'$.}   
\label{fig:tnonsmd3}
\end{figure}
This scenario contains two VL top quarks and three exotic quarks: two with charge $5/3$ and one with charge $8/3$. All of these states
contribute to the corrections to the EWP tests. The Yukawa couplings and mass matrices for this scenario are given in Section
\ref{sec:yukawa:2}. The additional Yukawa couplings in the model are: $y_1^k$, $\omega$, $\omega'$ and $M$ with $k$ running on 
SM quark generations. 

As there is a single Yukawa mixing involving the SM quarks, we decided to add a non-vanishing $\omega$ in the scan, setting $\omega' = \omega$. 
The combined bounds from tree-level and EWP tests are given in Figure \ref{fig:tnonsmd3} for scenarios where the VL quarks mix with either one of
the SM quark generation (case for second generation in Appendix \ref{appendix:D}). While three level bounds only exclude a large mixing to the SM 
quarks in the case of light generations, a bound on the VL Yukawa $\omega$ arises from the EWP bounds.
For the light generations, both Yukawas are constrained to be small.
For third generation, the EWP bounds give similar value near the axes, however there is a cancellation absent in the case of light generation which opens the parameter space for $\omega = \omega' \sim y_1^3$, so that larger mixing angles are allowed.

\begin{figure}[tb]
\begin{center}
\hspace*{-0.7cm}
\epsfig{file=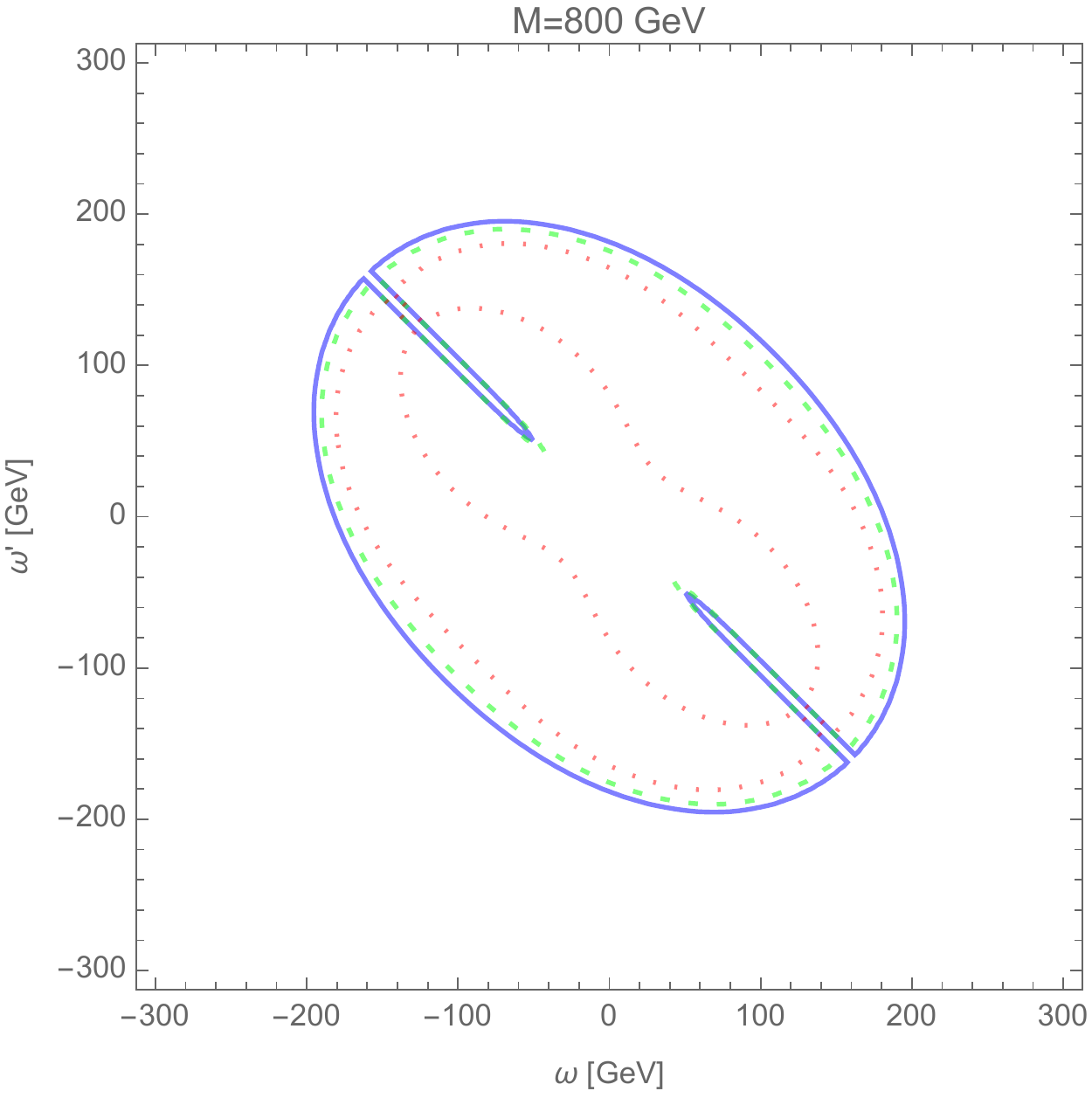,width=0.48\textwidth} 
\end{center}
\caption{\underline{\bf \boldmath Doublet $Y=7/6$ and Triplet $Y=5/3$} (Section
  \ref{sec:yukawa:2}):
  EWP bounds at 1$\sigma$ (red-dashed), 2$\sigma$ (green-dashed) and 3  $\sigma$ 
  (blue) as a function of the new Yukawa couplings $\omega$ and $\omega'$ with $M = 500/800$ GeV (left/right panel
  respectively). In addition the mixing of VL quarks with SM quarks is taken to be zero {\sl i.e.} $y_1^k = 0$. }   
\label{fig:tnonsmd1}
\end{figure}

In Figure \ref{fig:tnonsmd1} we show the EWP bounds in the plane of the new Yukawa couplings $\omega$ and $\omega'$. The bounds are not very strong, allowing values up to $200$ GeV. We also observe, like in Figure \ref{fig:dd1}, two dents for $\omega = - \omega'$ due to the vanishing of the mixing angles (for $M_1 = M_2$).
This is again an artifact of our choice of equal VL masses and vanishing couplings to the SM quarks.

\subsection{Singlet $Y=2/3$ and Doublet $Y=1/6$}
    \label{sec:results:3}    

This scenario contains a VL copy of the SM quarks: two VL top quarks and one VL bottom quark. The additional Yukawa couplings in the model, 
described in Section \ref{sec:yukawa:3}, are: $y_{1}^k$,  $x_2^k$, and $M$ with $k$ running on SM quark generations. Here we set the Yukawa in the down sector $y_{1d}^k=0$ to minimise bounds from flavour physics: this can be done independently on the up sector.

\begin{figure}[tb]
\begin{center}
\hspace*{-0.7cm}
\epsfig{file=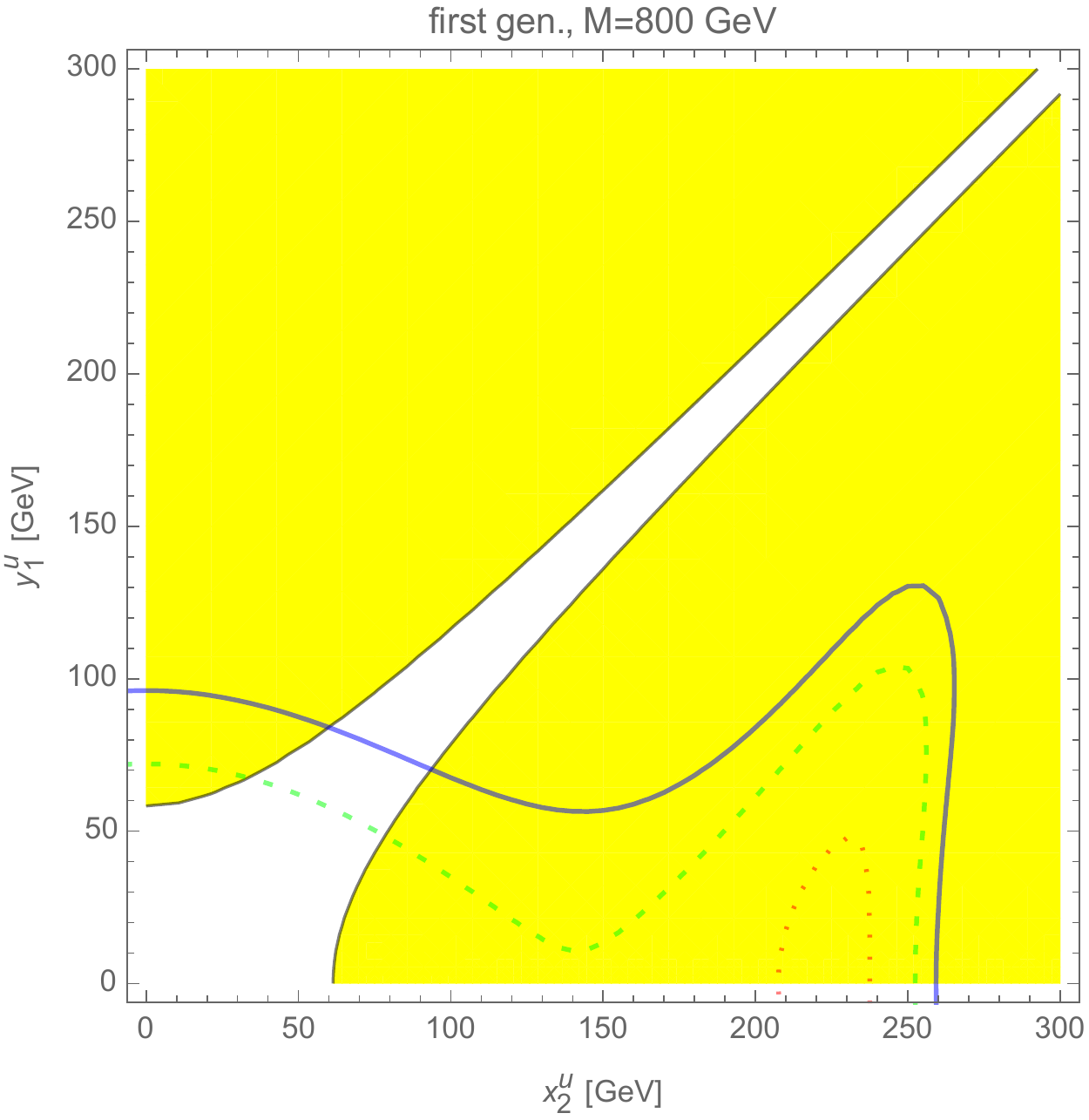,width=0.48\textwidth}
              \hspace*{.4cm} 
\epsfig{file=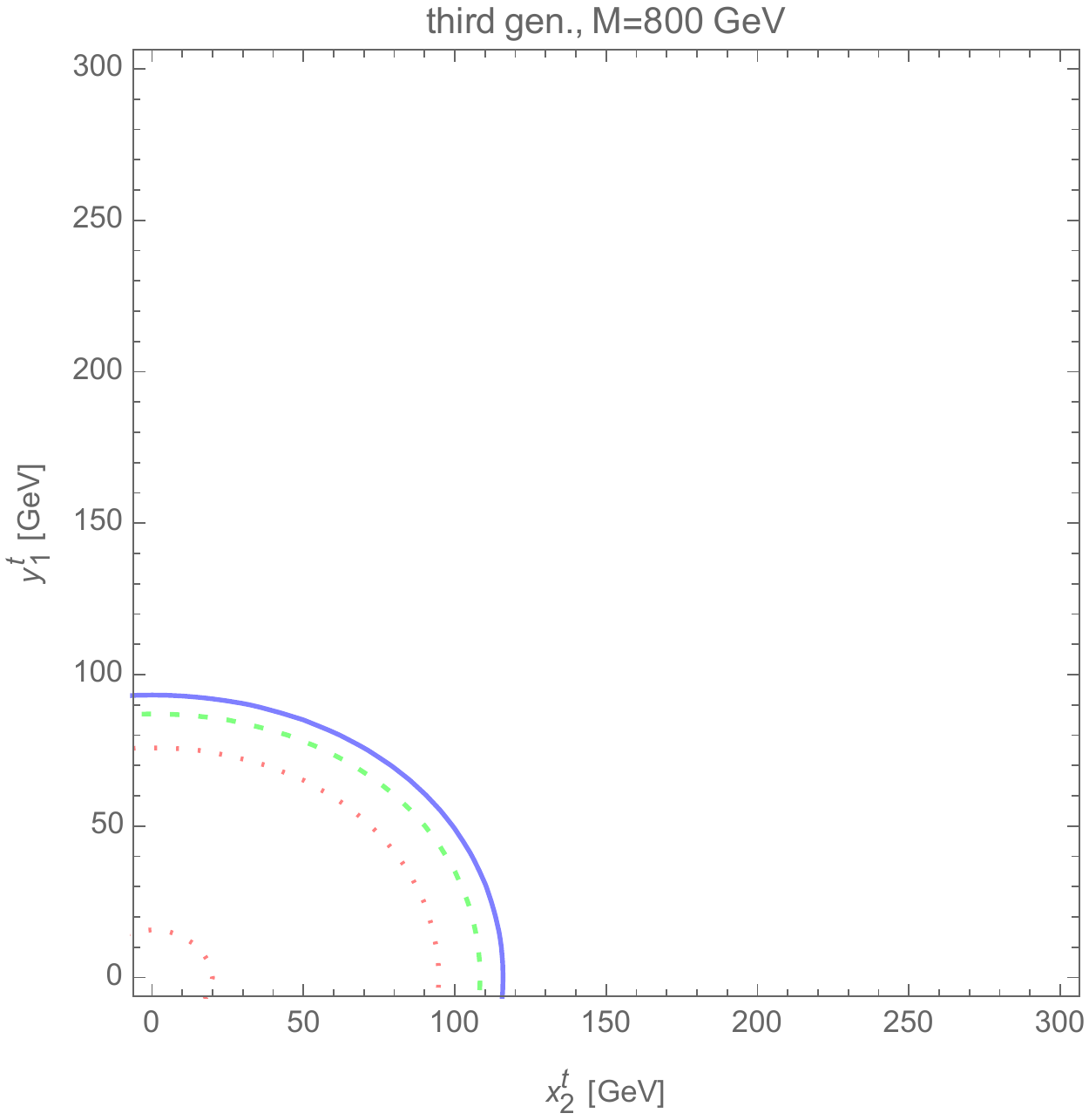,width=0.48\textwidth}
\end{center}
\caption{\underline{\bf \boldmath Singlet $Y=2/3$ and Doublet $Y=1/6$} (Section
  \ref{sec:yukawa:3}):
  EWP bounds at 1$\sigma$ (red-dashed), 2$\sigma$ (green-dashed) and 3 $\sigma$ (blue) for VL quarks coupling with
  the first (left panel) and third (right panel) SM generations, compared with the region excluded at 3$\sigma$ by tree-level bounds (yellow
  region in the left panel). Here, $M = 800$ GeV, and $\omega = \omega'=0$.} 
\label{fig:ssmd3}
\end{figure}

In Figure  \ref{fig:ssmd3} we show the combined bounds from EWP tests for VL quarks mixing with individual SM
quark generations. As in the previous scenarios, the mixing between VL quarks is taken to be zero, {\sl i.e.} $\omega = \omega' = 0$. 
This scenario exhibits quite distinctive features depending on the VL mixings and masses. 
In the case of mixing with the third generation, EWP bounds constrain quite tightly the allowed mixing parameters. 
In the case of mixing with first generation only, tree level bounds are quite tight too up to a cancellation for $y_1^1 \sim x_2^1$ where the mixing can be arbitrarily large. This throat may suggest the possibility of large compositeness in the light quark sector.
However, we see that EWP bounds exclude this region and point back to small mixing. For the case of mixing with second generation (see Appendix \ref{appendix:D}), EWP bounds dominate and again force the scenario to small mixing parameters. 

The bounds on $\omega$ and $\omega'$ are very similar to the previous cases, and we do not show them here.



\subsection{Doublet $Y=1/6$ and Doublet $Y=7/6$ }
    \label{sec:results:4}    
\begin{figure}[th]
\begin{center}
\hspace*{-0.7cm}
\epsfig{file=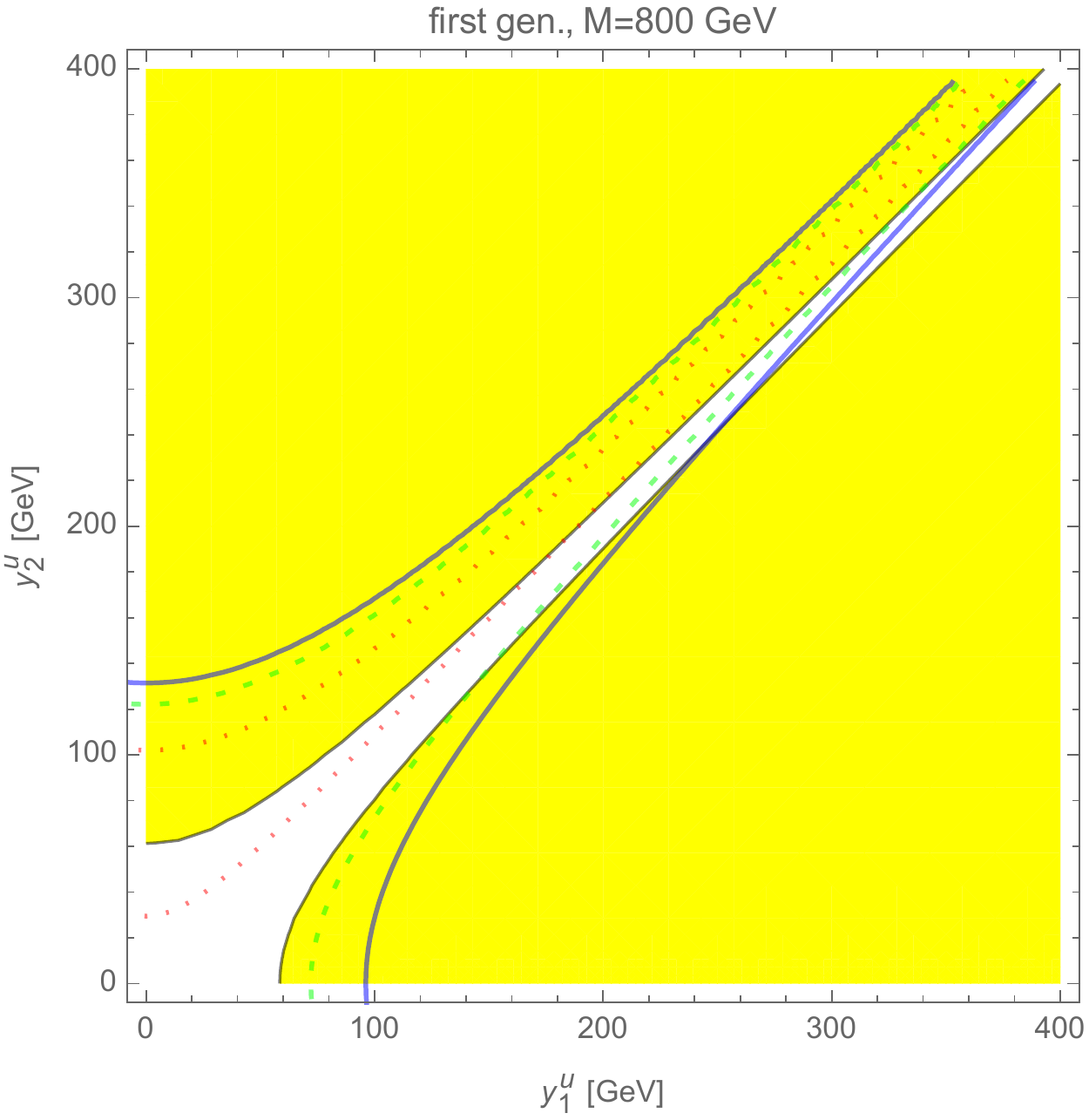,width=0.48\textwidth}
     \vspace*{0.4cm} 
\epsfig{file=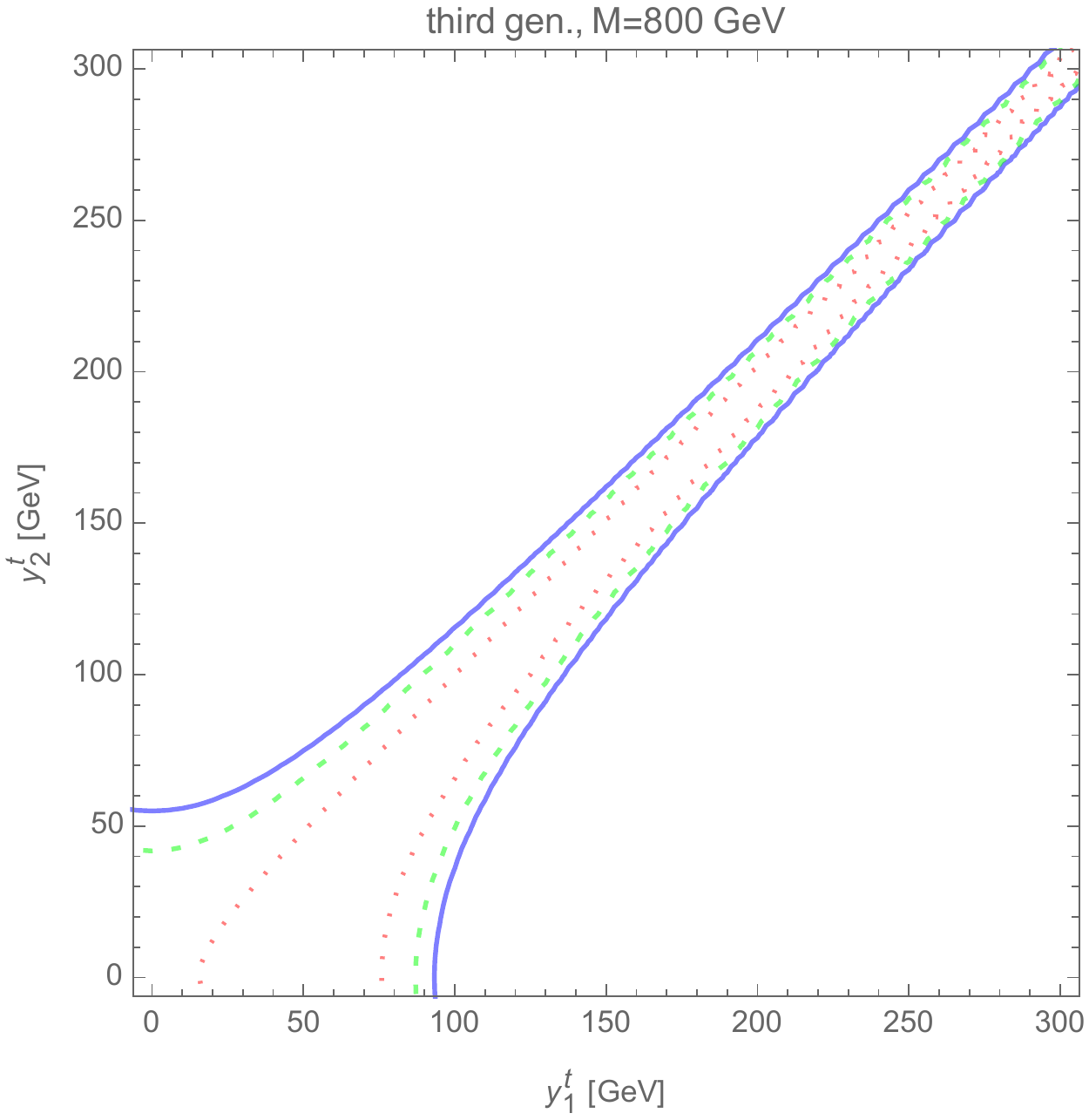,width=0.48\textwidth} 
 \end{center}
\caption{\underline{\bf\boldmath Doublet $Y=1/6$ and Doublet $Y=7/6$} (Section
  \ref{sec:yukawa:4}): EWP bounds at 1$\sigma$ (red-dashed), 2$\sigma$ (green-dashed) and 3 $\sigma$ (blue) for VL quarks coupling with
  the first (left panel) and third (right panel) SM generations, compared with the region excluded at 3$\sigma$ by tree-level bounds (yellow
  region in the left panel). $M = 800$ GeV,  $\omega = \omega'=0$.}  
\label{fig:dd3}
\end{figure}

This scenario is particularly interesting as it corresponds to a bi-doublet of the custodial SO(4) symmetry, which is often a basic ingredient for top partial compositeness in models of composite Higgs (see for instance~\cite{DeSimone:2012fs}).
This scenario contains two VL top quarks, one VL bottom quark and one exotic quark with charge
$5/3$. The additional parameters, described in Section
\ref{sec:yukawa:4}, are: $y_{1}^k$, $y_2^k$ and $M$ with $k$ running on SM quark generations. 
Note that for this scenario mixing between VL quarks ({\sl i.e.} $\omega$ and $\omega'$) is not allowed. 
Analogously to the model discussed in the previous Section \ref{sec:results:3}, 
this model also introduces an additional mixing in the bottom sector, which is again independent
from the the mixing in the top sector. Hence it is possible to impose the condition $y_{1d}^k = 0$ without affecting the top sector. 

The results for the combined tree-level and EWP bounds are given in Figure \ref{fig:dd3}. For the first generation (and the second, see Appendix \ref{appendix:D}), there is an 
interesting cancellation in the tree-level bounds for $|y_{1}^k| = |y_2^k|$: this is a 
consequence of an enhanced custodial symmetry, and this fact has been used in the
literature to justify $\mathcal{O}(1)$ mixings of VL quarks with light generations~\cite{Barcelo:2011wu}. 
EWP bounds show a similar cancellation, however along an axes which is a bit off compared to $|y_{1}^k| = |y_2^k|$, therefore a tension between the two allowed regions develops for large mixings.
Similar behaviour in the EWP bounds can be seen in the case of mixing to the third generation only.



\subsection{Single production cross sections}       
   \label{sec:results:5}
In this section a comparison between the tree-level and loop-level bounds and 
the bounds from single production processes at the LHC is provided for the scenarios above. 
The relevance of single production is given by the fact that its cross-section depends on both the masses
of the VL quarks and their couplings to the SM quarks; moreover, it is well known that 
single production becomes the dominant channel at the LHC, 
overcoming QCD pair production, when quark masses are higher than a certain (model-dependent) value.
For typical scenarios where VL quarks mix predominantly with third generation and mixing parameters are not too
constrained by flavour physics and EWP tests, the mass bounds from QCD pair production are already in the
region where the single production channel is relevant or even dominant \cite{Cacciapaglia:2011fx}.
So far, few experimental searches for single production of VL quarks 
have been performed. The ATLAS experiment has performed two searches including single production of VL quarks:
in \cite{ATLAS:2012apa} a search for singly-produced VL quarks coupling only with first generation
is performed, while \cite{ATLAS-CONF-2014-036} is a search for for pair+single production of 
VL quarks mixing with third generation only. The search \cite{ATLAS:2012apa} has already been 
considered in a previous analysis \cite{Buchkremer:2013bha} 
for comparing bounds from LHC and flavour physics, and we will consider
in this analysis part of the results obtained in that study.
To be specific, in the following we will consider the single production of a VL top partner in association with a light jet, 
and the mass of the VL quark will be fixed to 800 GeV. We consider exclusive coupling to each of the three SM generations for each scenario.

\begin{figure}[tb]
\begin{center}
\hspace*{-0.7cm} 
\epsfig{file=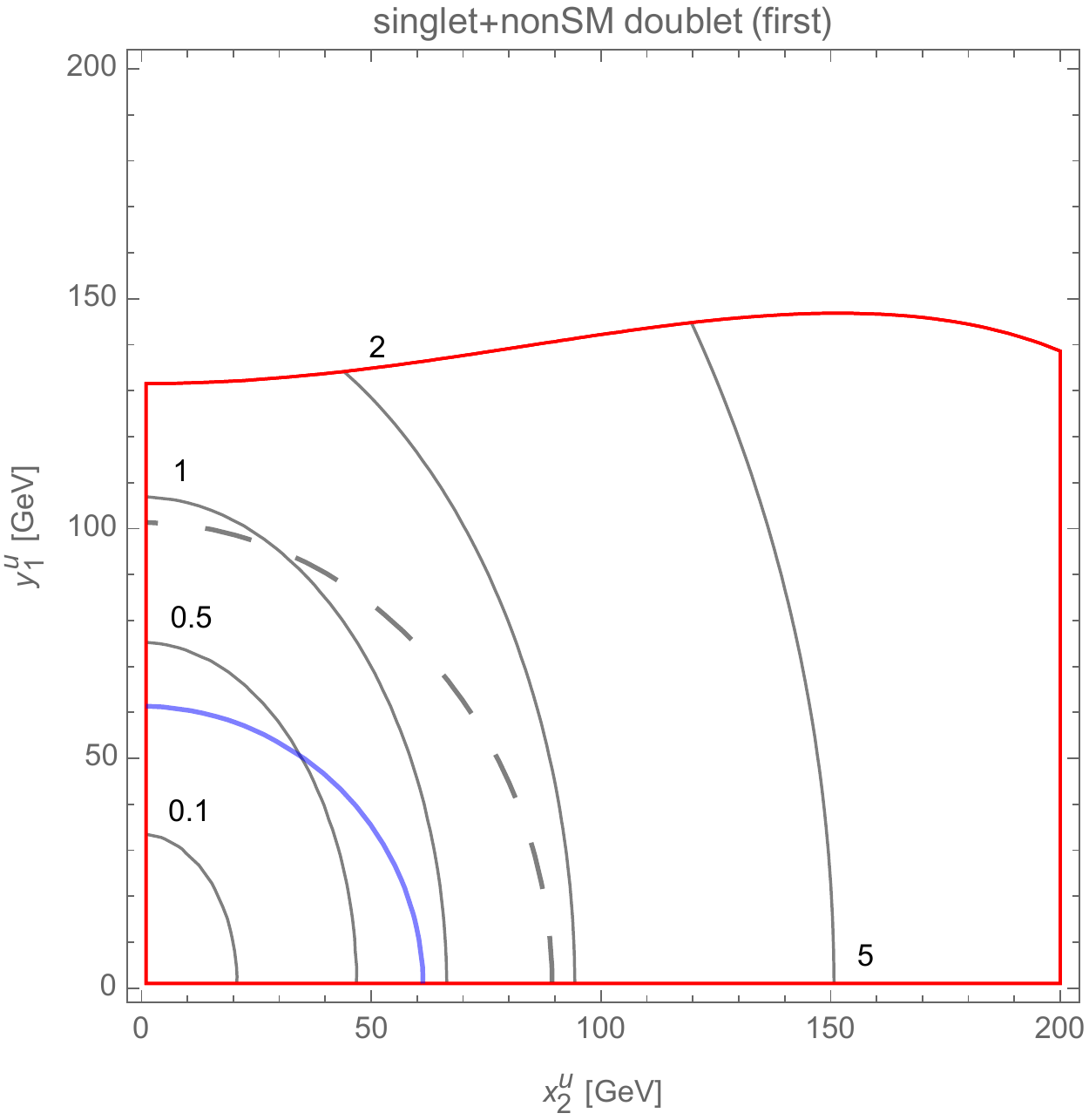,width=0.3\textwidth}
  \hspace*{0.4cm} 
\epsfig{file=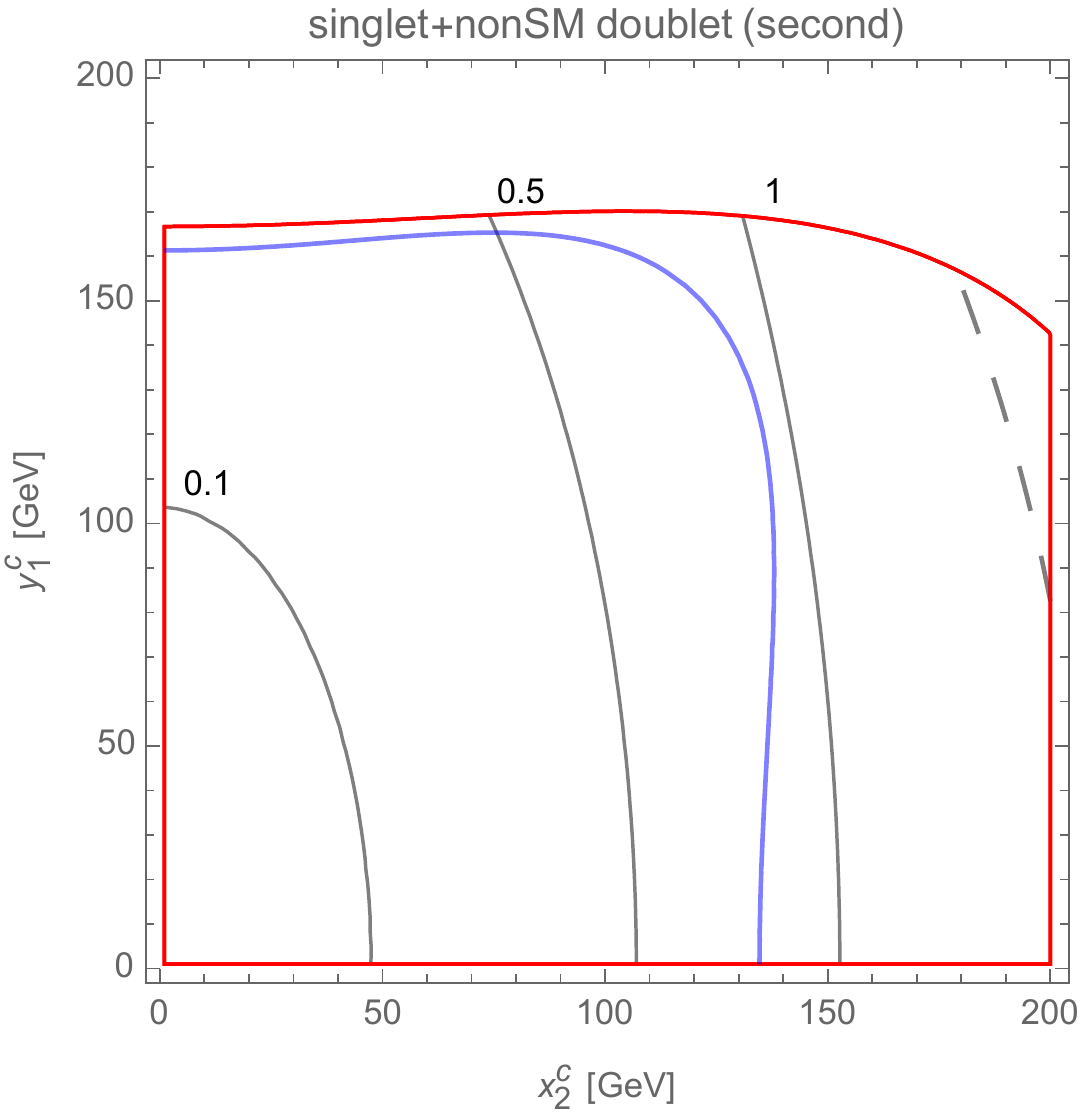,width=0.3\textwidth}
  \hspace*{0.4cm} 
\epsfig{file=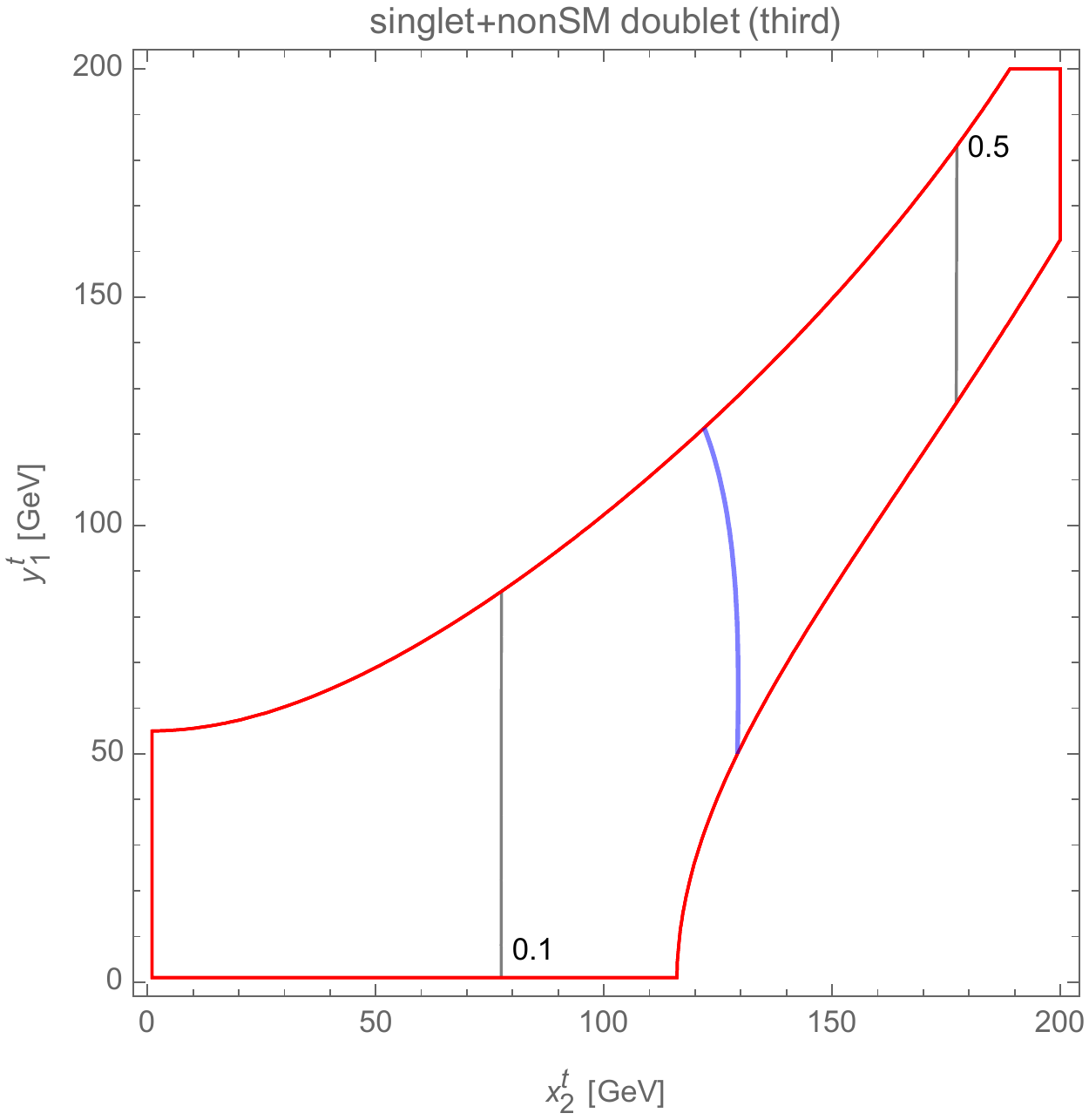,width=0.3\textwidth}
\end{center}
\caption{{\bf \boldmath Singlet $Y=2/3$ and Doublet $Y=7/6$}
  for mixing with first generation only (left), second generation only (middle), third generation (right), and for a mass of the VL quarks of 800
  GeV. The channel is T+jet. The grey contour lines correspond to
  cross-section values in picobarns at 14 TeV. The region inside the
  red line is allowed by the S and T parameters. The region inside the
  blue line is allowed by the tree-level bounds. The dashed black lines
  are the bounds from the ATLAS search \cite{ATLAS:2012apa}.} 
\label{fig:xsnonsmd} 
\end{figure}

The LHC bounds have been obtained by applying the model-independent
parametrisation described in \cite{Buchkremer:2013bha}. Considering
the observed cross-section reported in the ATLAS analysis
\cite{ATLAS:2012apa} and the universal coefficients computed in
\cite{Buchkremer:2013bha} (Tables 12-17) it is possible to set bounds on
the overall coupling strengths of the singly-produced VL quarks by using
the relations in Sections 3 and 4 of \cite{Buchkremer:2013bha}. 

\begin{figure}[tb]
\begin{center}
\hspace*{-0.7cm} 
\epsfig{file=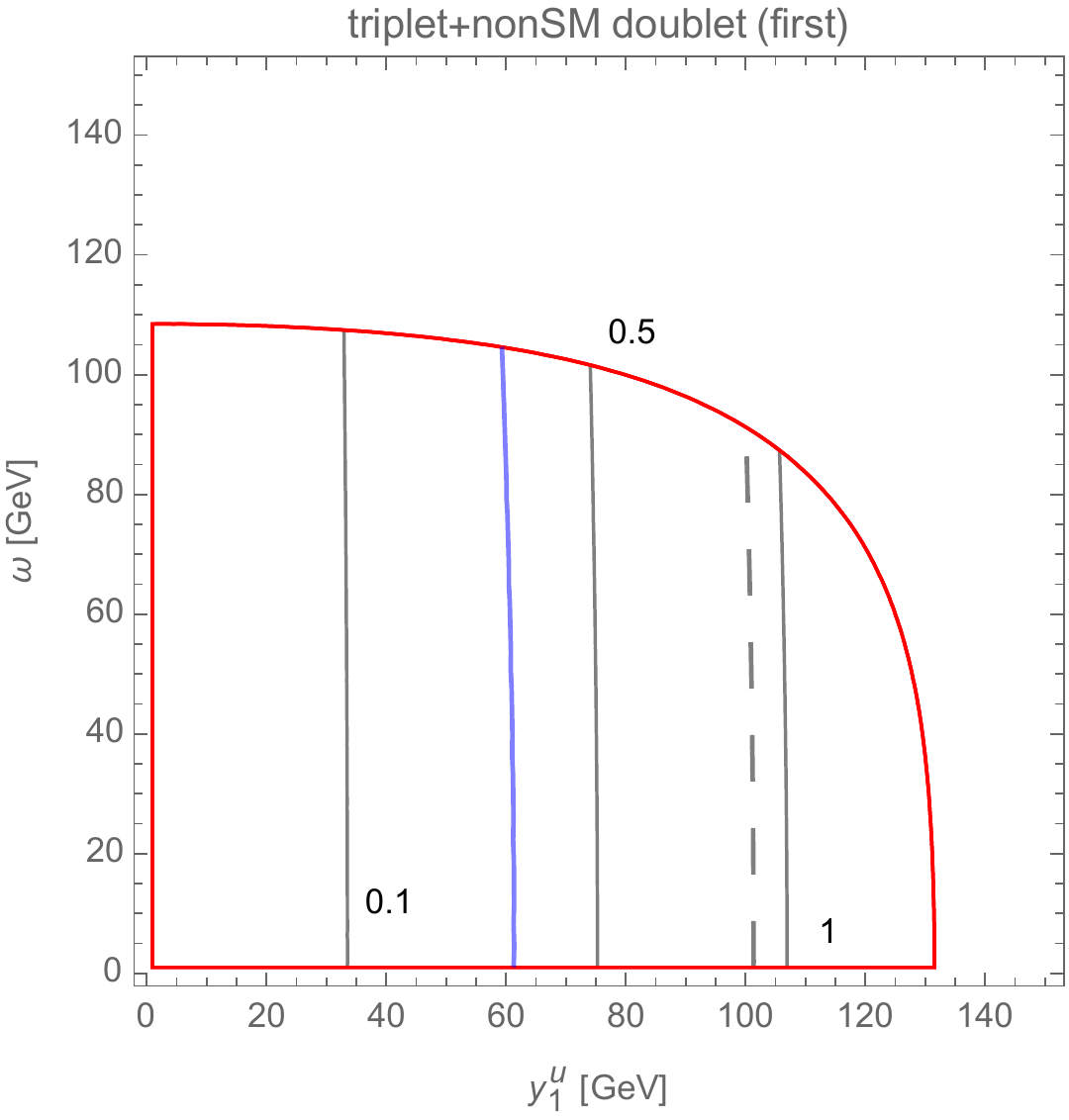,width=0.3\textwidth}
 \hspace*{0.4cm}
\epsfig{file=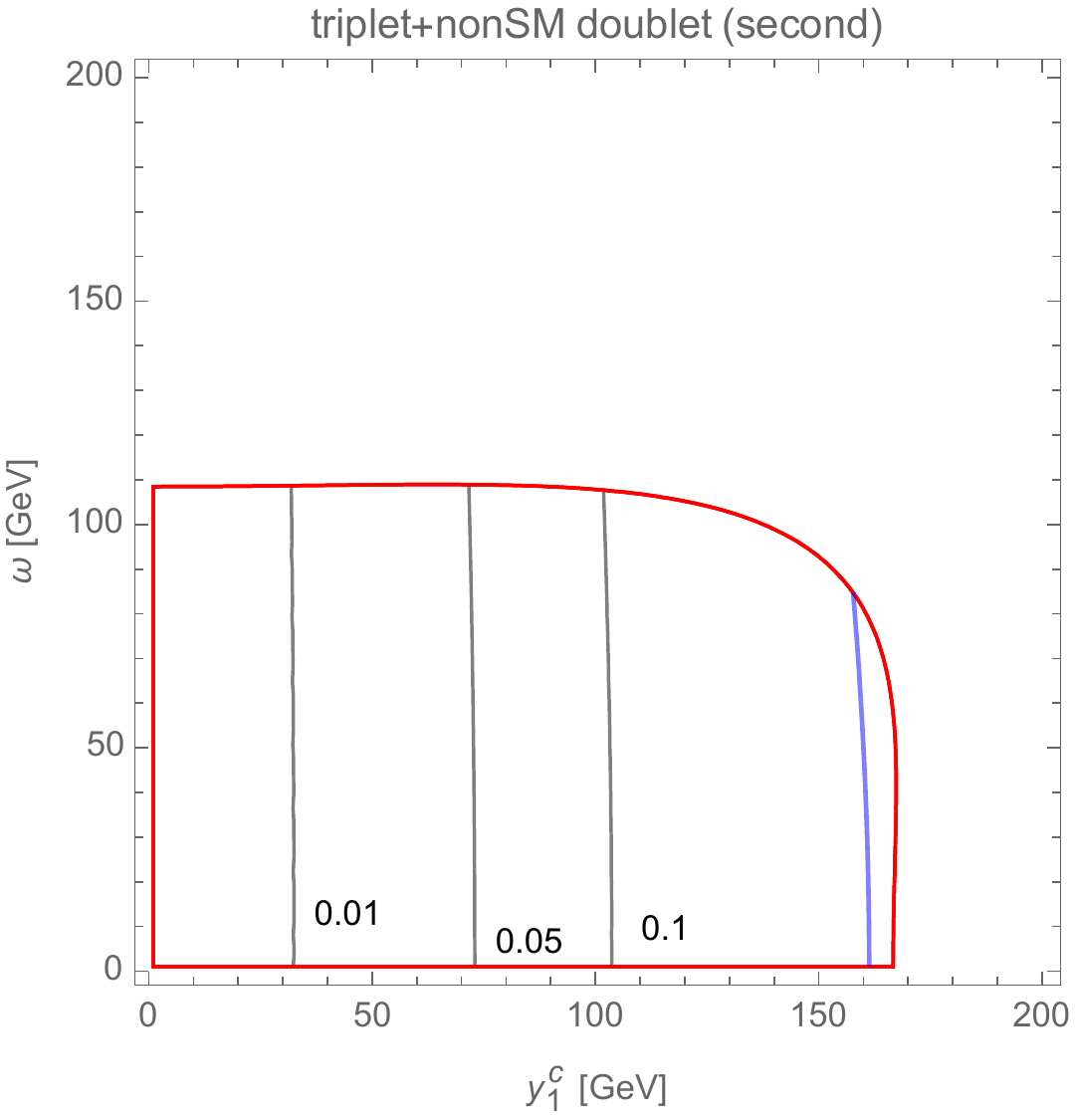,width=0.3\textwidth} 
 \hspace*{0.4cm}
\epsfig{file=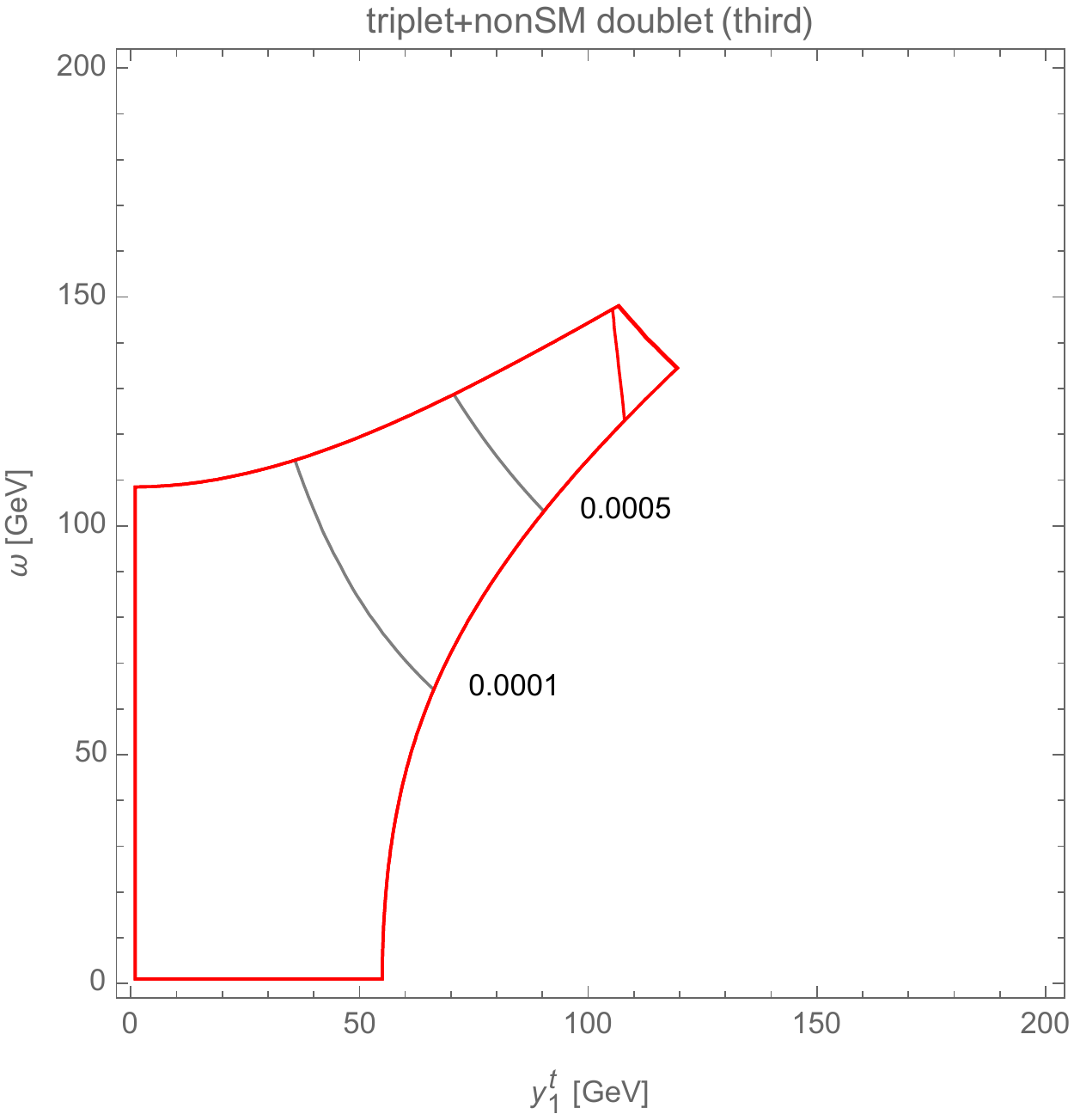,width=0.3\textwidth} 
\end{center}
\caption{{\bf \boldmath Doublet $Y = 7/6$ and Triplet $Y = 5/3$}
  for mixing with first generation only (left) ,second generation only (middle), third generation (right),  
  and for a mass of the VL quarks of 800 GeV. The channel is T+jet. The grey contour lines correspond
  to cross-section values in picobarns at 14 TeV. The region inside
  the red line is allowed by the S and T parameters. The region inside
  the blue line is allowed by the tree-level bounds. The dashed black lines
  are the bounds from the ATLAS search \cite{ATLAS:2012apa}.} 
\label{fig:xtripnonsmd} 
\end{figure}

\begin{figure}[tb]
\begin{center}
\hspace*{-0.7cm} 
\epsfig{file=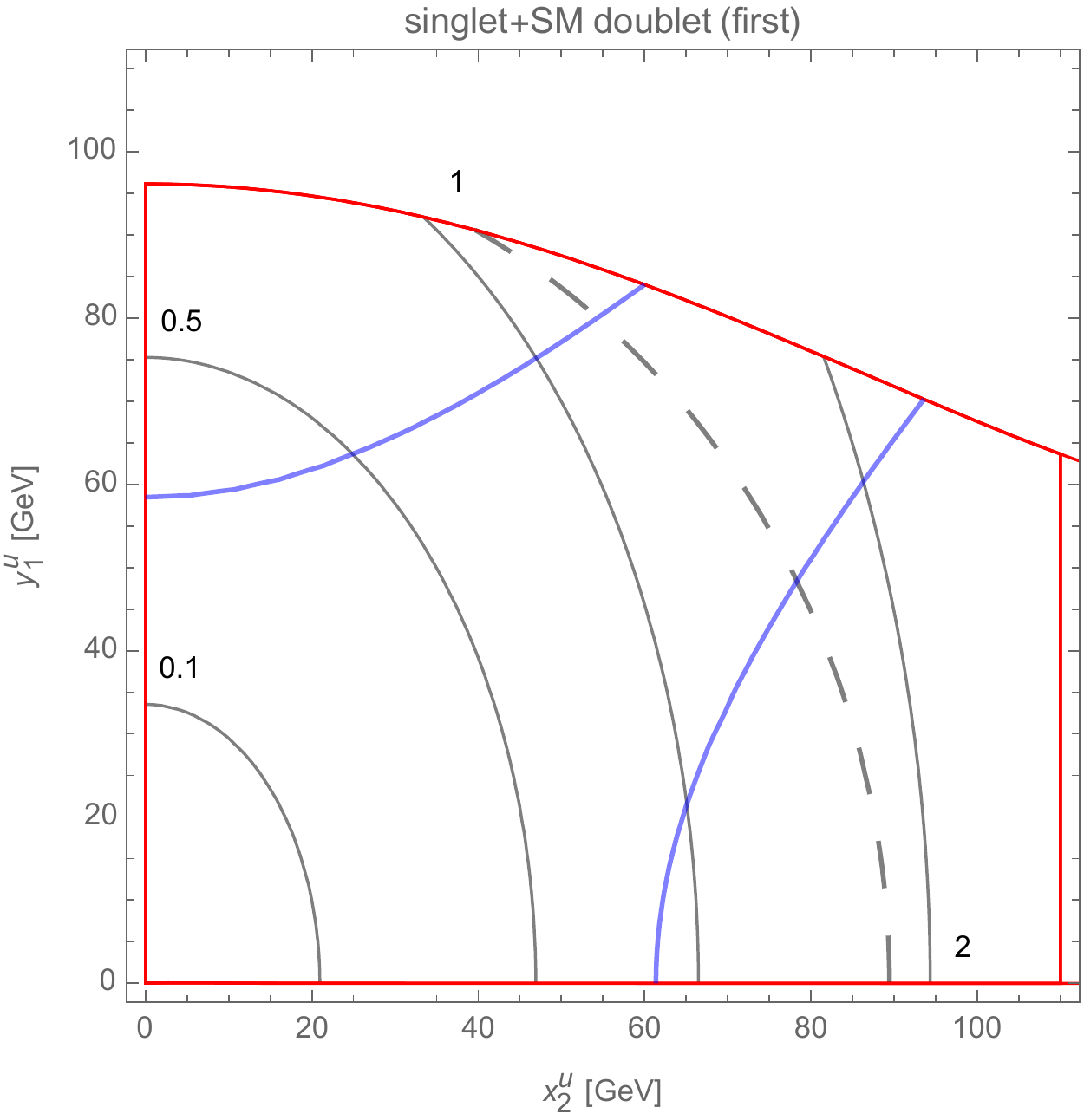,width=0.30\textwidth}
  \hspace*{0.4cm} 
\epsfig{file=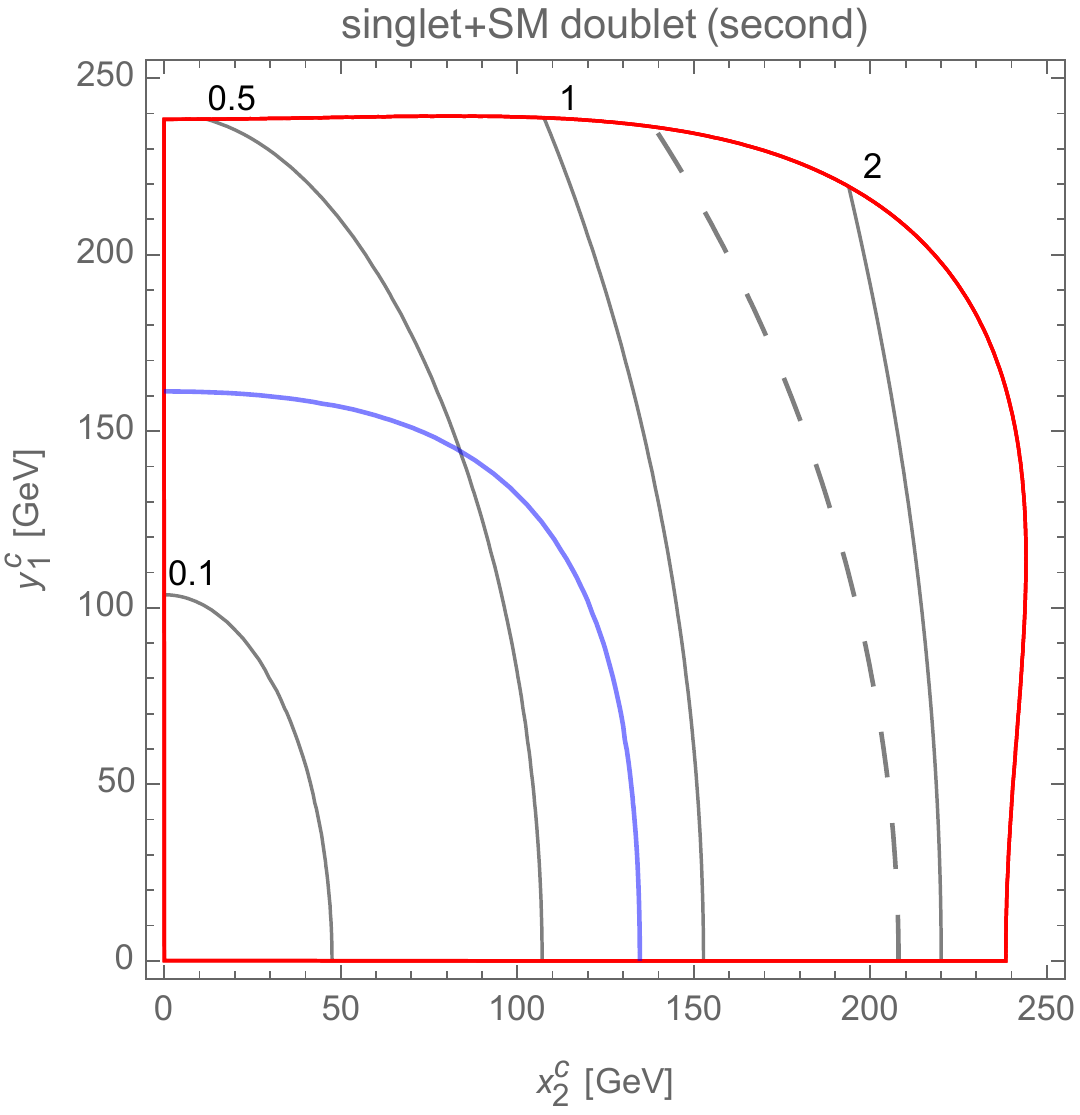,width=0.30\textwidth} 
  \hspace*{0.4cm} 
\epsfig{file=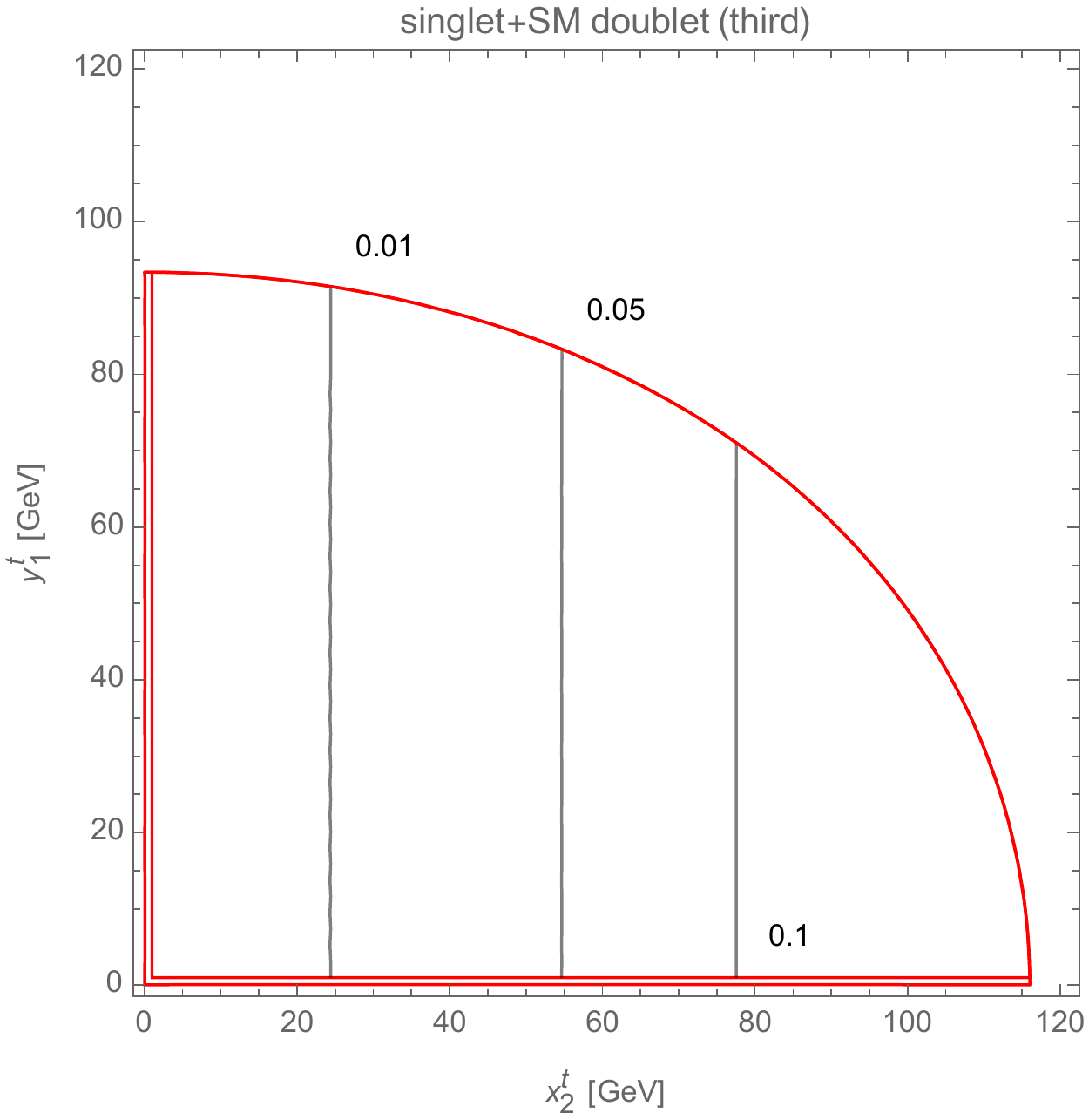,width=0.30\textwidth} 
\end{center}
\caption{{\bf \boldmath Singlet $Y=2/3$ and Doublet $Y=1/6$}
  for mixing with first generation only
  (left), second generation only (middle), third generation (right), and for a mass of the VL quarks
  of 800 GeV. The channel is T+jet. The grey contour lines correspond
  to cross-section values in picobarns at 14 TeV. The region inside
  the red line is allowed by the S and T parameters. The region inside
  the blue line is allowed by the tree-level bounds. The dashed black lines
  are the bounds from the ATLAS search \cite{ATLAS:2012apa}.} 
\label{fig:xssmd} 
\end{figure}

\begin{figure}[htb]
\begin{center}
\hspace*{-0.7cm} 
\epsfig{file=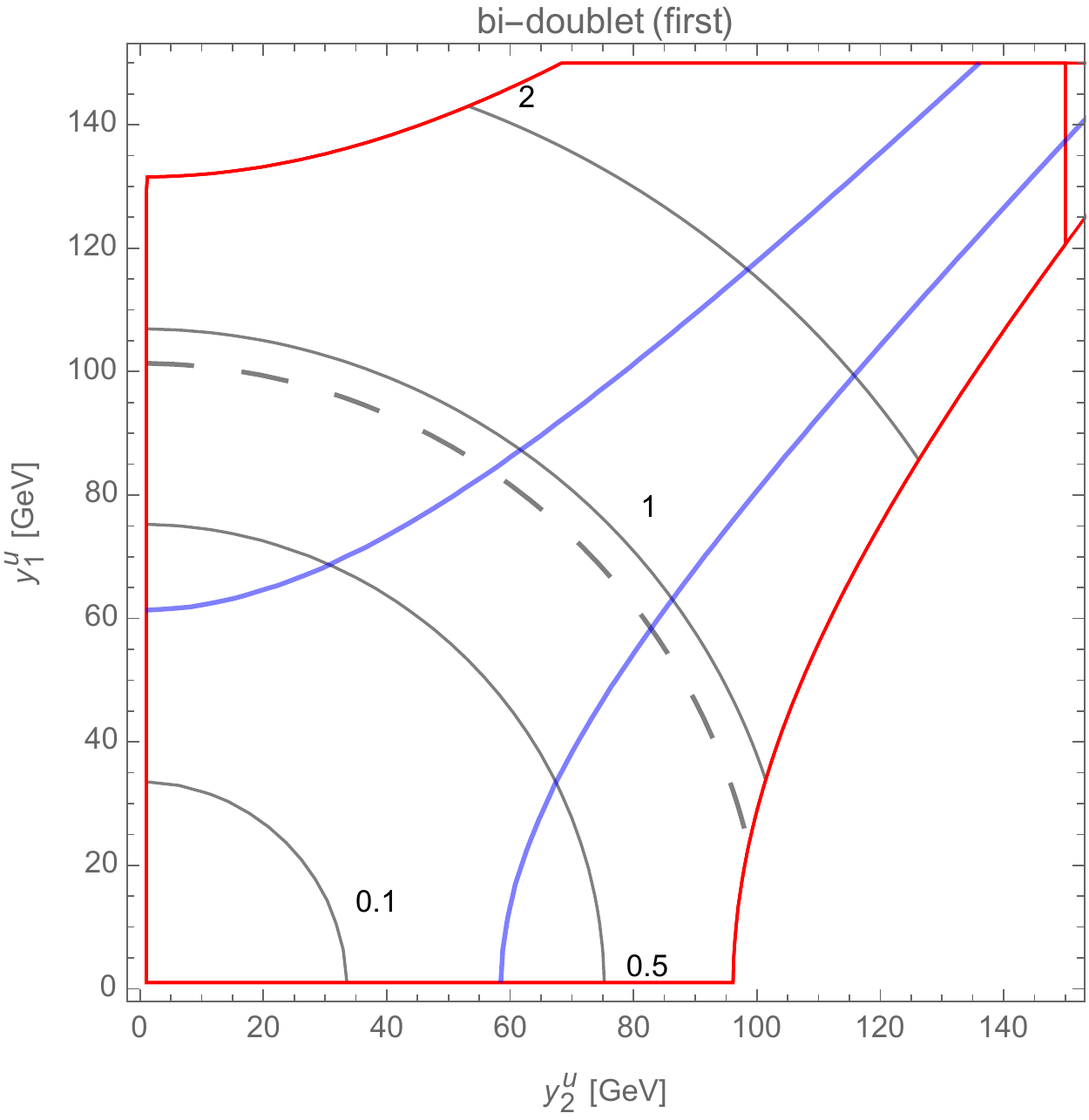,width=0.3\textwidth}
 \hspace*{0.4cm}
\epsfig{file=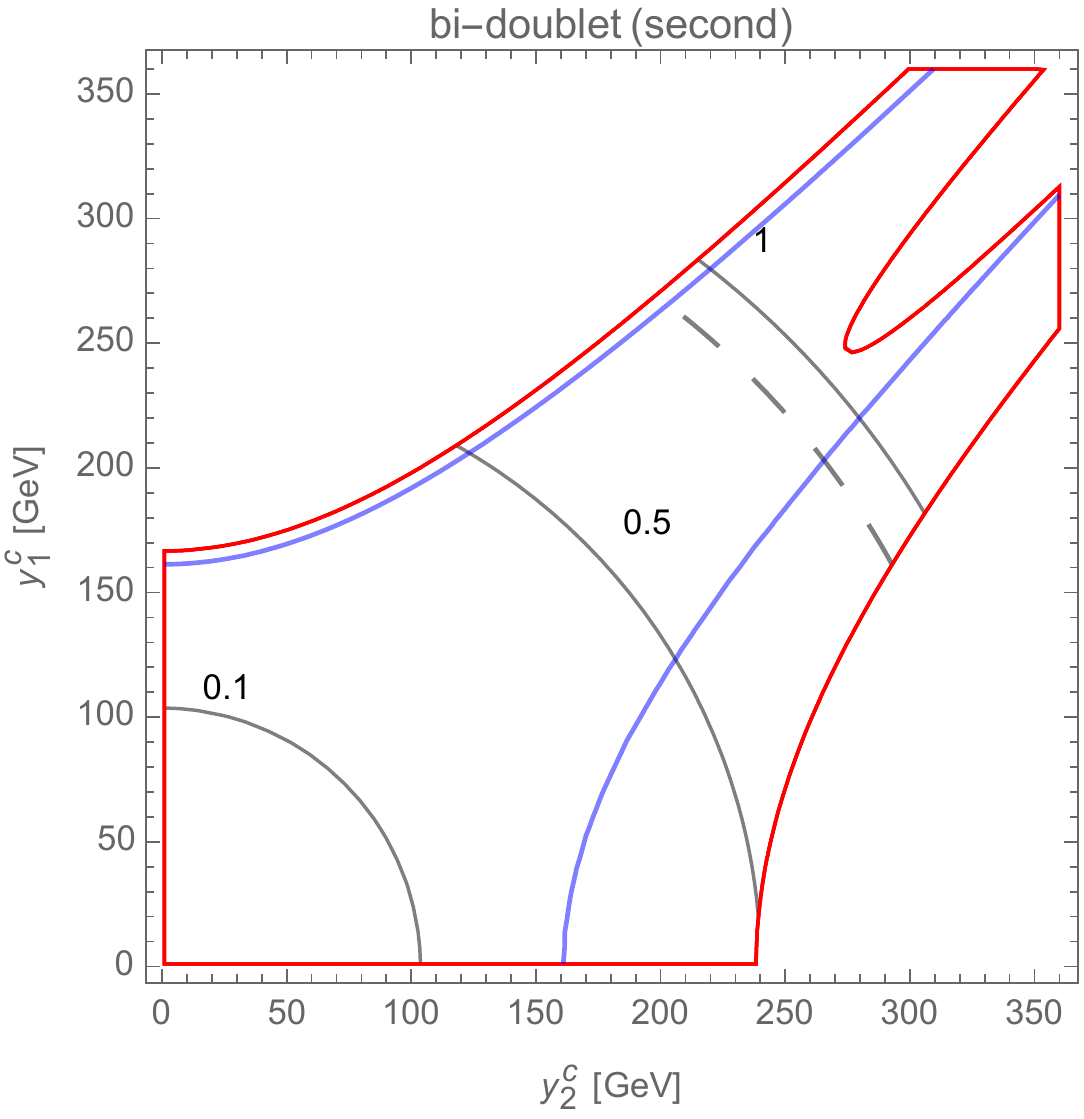,width=0.3\textwidth} 
 \hspace*{0.4cm}
\epsfig{file=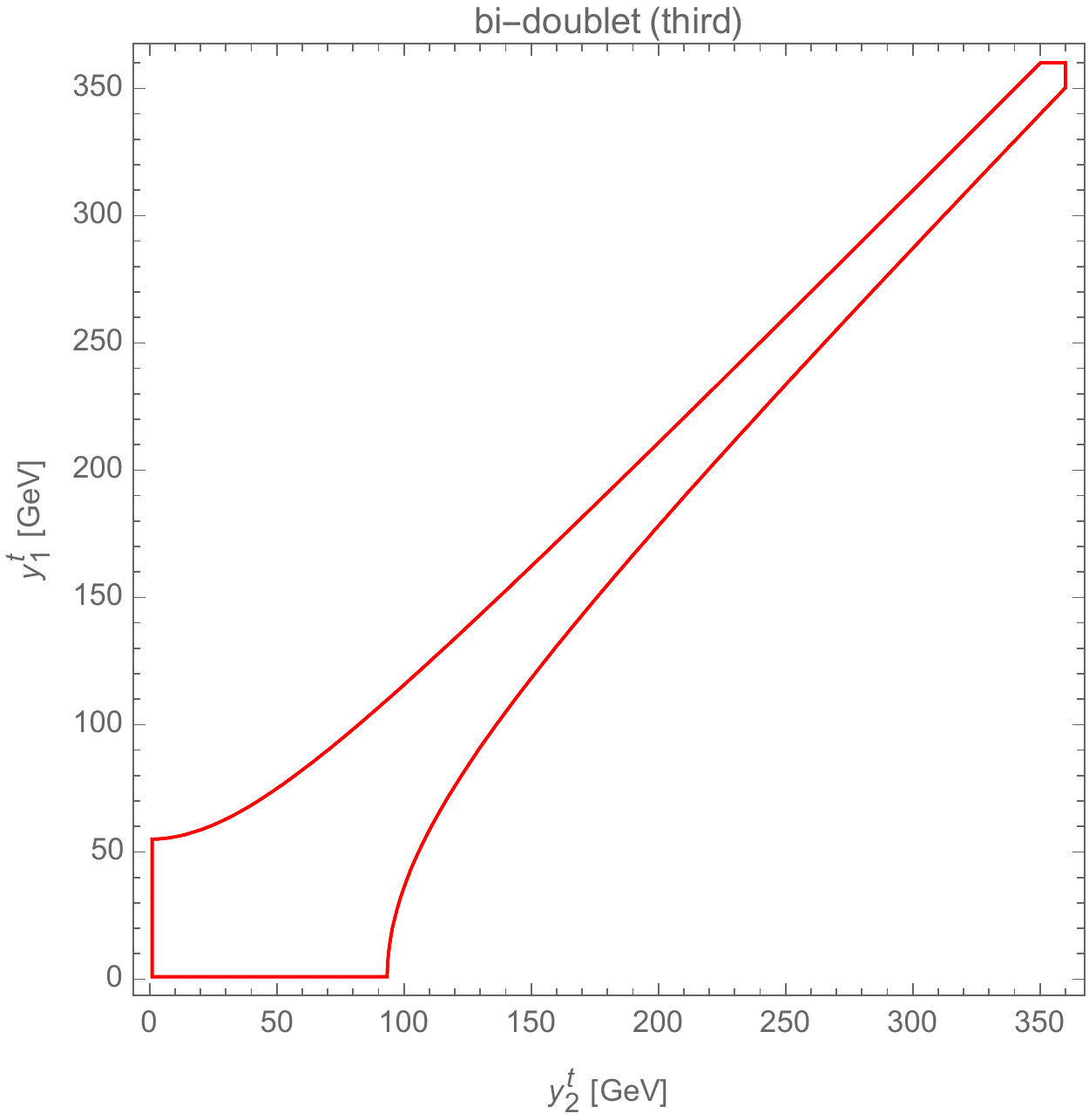,width=0.3\textwidth} 
\end{center}
\caption{{\bf \boldmath Doublet $Y = 1/6$ and Doublet $Y = 7/6$}
  for mixing with first generation only
  (left), second generation only (middle), third generation (right), and for a mass of the VL quarks
  of 800 GeV. The channel is T+jet. The grey contour lines correspond
  to cross-section values in picobarns at 14 TeV (this channel is not allowed in the case of the plot on the right). The region inside
  the red line is allowed by the S and T parameters. The region inside
  the blue line is allowed by the tree-level bounds. The dashed black lines
  are the bounds from the ATLAS search \cite{ATLAS:2012apa}.} 
\label{fig:xdnonsmd} 
\end{figure}

In the plots presented in 
Figs \ref{fig:xsnonsmd}, \ref{fig:xtripnonsmd}, \ref{fig:xssmd}, 
\ref{fig:xdnonsmd} the LHC bounds are directly compared with the tree-level and EWPT bounds.
The region inside the red line is the one allowed by the S
and T parameters (oblique corrections) whereas the blue
line marks the constraint from tree-level bounds. The EWPT bound should be taken only as an  
indication are the explicit assumption that no other extra
states contribute to the corrections is imposed; this simplification is not true in general,
e.g. in a complete model containing other new particles besides VL multiplets. 
The dashed black lines are the bounds at $3 \sigma$ derived by reinterpreting the results of the ATLAS search \cite{ATLAS:2012apa}. 
The grey lines represent the contours of the LHC production cross-section (in pb) 
for the process of single production of a VL top quark in association with a light jet ($p p \to T j $).  

The results of the ``Singlet ($Y=2/3$) and Doublet ($Y=7/6$)" scenario are summarised in 
Figure \ref{fig:xsnonsmd}. For this scenario, the tree-level bounds are the most 
stringent ones if VL quarks mix to the light generations, and they are stronger than the current bound form the ATLAS search. In both cases, the largest $T jet$ cross section allowed is between $0.5$ and $1$ pb.
In the case of mixing to the third generation only EWP and tree level bounds conspire to select small mixing, and the single production at 14 TeV is limited to small values around $100$ fb.
Figure \ref{fig:xtripnonsmd} shows the results for the ``Doublet ($Y=7/6$) 
and Triplet ($Y=5/3$)" scenario.  
In this case the oblique parameters 
constrain a much larger region of parameter space and this is due to the much 
richer exotic quark (quarks with charges $5/3$ and $8/3$) spectrum, that contributes 
to the corrections to the oblique parameters. 
The cross sections for the $t jet$ channel are much smaller than in the previous case, with value around $100$ fb for mixing to light generations: this is due to the suppression in the coupling of the $t'$ (belonging to a doublet) to the $W$ boson and a SM down-type quark. For the same reason, in the case of coupling to third generation only, the channel is nearly absent.
Figure \ref{fig:xssmd} refers to the ``Singlet ($Y=2/3$) and Doublet ($Y=1/6$)'' scenario. 
Due to the presence of a SM type VL doublet, this scenario contains right 
handed charged gauge boson couplings which give additional contributions to the
oblique parameters. However, the tree-level constraints are the most stringent for the parameter space of this scenario. 
We see that in the case of mixing to the first generation, large production rates are allowed, with cross sections above 1 pb and a region already probed by the ATLAS search. Smaller cross sections are attained in the case of mixing to the second generation, while the maximum values drop to  about $100$ fb for mixing to the third generation.
This is the scenario than offers the largest single-production cross sections, and it is a golden case to be studied at the Run 2 of the LHC.
Finally, in Figure \ref{fig:xdnonsmd} the results for the ``Doublet 
($Y=1/6$) and Doublet ($Y=7/6$)" scenario are presented. Again, though the presence of an exotic quark with charge 5/3 and of 
right-handed charged currents which contribute to the corrections to the oblique parameters, the tree-level constraints are
stronger for the most part of the parameter space. 
The large mixings allowed in the cancellation region produce very large $T jet$ cross sections, with the largest mixing already excluded by the ATLAS search. This is indeed a case where single production can be the most promising channel for the observation of the VL quarks.
In the case of mixing the third generation, the single production vanishes: the reason for this is that $t'$s belonging to doublets have very suppressed couplings to the $W b$, thus they cannot be produced in association with a light jet but only in association with tops.

An interesting common feature of all the scenario considered is that the
LHC bounds can be competitive if not stronger than the tree- and loop-level
bounds. Indeed the current LHC data we have considered 
are already able to constrain region of parameter space
otherwise allowed by other observables.

Another type of constraints can be obtained exploiting tools for the recasting of experimental searches
for pair production of VL quarks.
Considering the bounds on masses and couplings of the VL multiplets it is possible
to compute their branching ratios into SM states, and through the
recently developed software XQCAT
\cite{Barducci:2014ila,Barducci:2014gna,XQCATweb}, one can determine
the exclusion regions  by considering results from dedicated searches
in pair production and other searches not specifically designed for
VL quarks (such as SUSY analyses). This study can be performed systematically
for different combinations of VL multiplets, and we postpone this
analysis to a subsequent paper, where we will compare the bounds
obtained by XQCAT with dedicated simulations for specific scenarios. 




\section{Conclusions}          
\label{sec:concl}
Vector-like quarks are predicted by many theoretically motivated models 
of new physics. In most of these models VL quarks appear in complete multiplets 
and, usually, more than one multiplet is predicted. 
In this analysis we have considered scenarios with multiple VL quarks both from
the point of view of the general mixing structure with the three
Standard Model generations and considering the mixing pattern of these multiplets
for the determination of mixing effects and precision electroweak observables
both at tree-level and at loop-level. The specific case of two
different vector-like quark multiplets has been studied in detail,
with a special focus on multiplets containing a top partner. The main
result of our analysis is that tree-level and loop-level constraints
provide complementary information. 
Moreover the interplay of the vector-like multiplets
among themselves and with the Standard model quarks have important
consequences for phenomenology as in some cases large single production
cross-sections are possible and coupling with light generations is not
necessarily suppressed. These results have phenomenological
implications for LHC searches as the bounds we have extracted pinpoint
particular regions of the parameter space and suggest that in
realistic cases containing multiple multiplets of vector-like quarks,
cancellations are possible from tree-level bounds which allow large
values of the mixing parameters. Even if the EWP tests partially
allow to limit these regions where cancellations occur, one has to
keep in mind that these loop-level constraints are valid under the
assumption that no other states apart from the vector-like multiplets
contribute to the S and T parameters. Therefore it is clear that
direct searches by the LHC experimental collaborations in the next run
of the LHC will play a mayor role in constraining or discovering
physics beyond the Standard Model which contains vector-like
multiplets. 



\section*{Acknowledgments}
AD is partially supported by Institut Universitaire de France. AD and
GC also acknowledge partial support from the Labex-LIO (Lyon Institute
of Origins) under grant ANR-10-LABX-66, FRAMA (FR3127, F\'ed\'eration 
de Recherche ``Andr\'e Marie Amp\`ere ``) and IN2P3 Theory-LHC France
funding. The research of YO is supported in part by the Grant-in Aid 
for Scientific Research,Japan Society for the Promotion of Science 
(JSPS), No. 20244037 and No.22244031. The research of NG is partially
supported by Department of Science \& Technology (DST) project
no. SR/S2/HEP-09/2010. NG and DH are thankful to RECAPP, HRI and
organizers of WHEPP-13 where part of the work was done. The work of DH
was supported by Centenary post-doctoral fellowship of IISc,
Bengaluru.  


\appendix
\section*{Appendices}


\section{Lagrangian and mass matrices with two VL multiples }
   \label{appendix:A} 

The cases are classified into following four categories :
\begin{itemize}
\item{} {\bf Top type multiples:} Only have 2 top VL quarks and no
  bottom VL quark. 
\item{} {\bf Bottom type multiplets:} Only have 2 bottom VL quarks
  and no top VL quark. 
\item{} {\bf Hybrid multiplets:} mixing of top and bottom VL quarks
  with SM quarks are independent. Hence one can safely take mixing of
  bottom VL quarks with SM bottom sector to be zero to satisfy all the
  flavour physics bounds without affecting the top sector.  
\item{} {\bf Mixed multiplets:} Remaining cases. 
\end{itemize}


\subsection{Top multiplets}        \label{Appen:A:1}

In this appendix we consider the multiplets that are presented in the list
of Tables \ref{QN_vec-top}, \ref{QN_vec-bottom} and \ref{QN_vec-mixed}. 
The multiplets containing a top type partner but without any down type
partner, in terms of the  $(SU(2)_{L},U(1)_{Y})$ quantum numbers, are
the singlet $({\bf 1},2/3)$, the doublet $({\bf 2},7/6)$, and the
triplet $({\bf 3},5/3)$, all listed in Table \ref{QN_vec-top}.


\subsubsection{Singlet $Y=2/3$ and Doublet $Y=7/6$}   \label{Appen:A:1:1}  
Details are given in Section \ref{sec:yukawa:1}. 



\subsubsection{Doublet $Y=7/6$ and Triplet $Y=5/3$}   \label{Appen:A:1:2}
Details are given in Section \ref{sec:yukawa:2}. 



\subsubsection{Singlet $Y=2/3$ and Triplet $Y=5/3$}    \label{Appen:A:1:3}
\begin{equation}
\mathcal{L}_{V-SM} = - \lambda_2^k \bar{Q}_L^k \tilde{H} \psi_{2R} + h.c. \,,
  \label{eq:A:1}
\end{equation}
where $\psi_1 =({\bf 3},\frac{5}{3})=\left(X^{8/3}_1, X^{5/3}_1,
U_1\right)^{T}$ and $\psi_{2}=({\bf 1},\frac{2}{3})=U_2$. The mass lagrangian and mass matrices are:
\begin{equation}
\mathcal{L}_{mass} = - x_2^k u_L^k U_{2R} - M_1 \bar{X}^{8/3}_{1L} X^{8/3}_{1R}
- M_1 \bar{X}^{5/3}_{1L} X^{5/3}_{1R} - M_1 \bar{U}_{1L} U_{1R} -
M_2 \bar{U}_{2L} U_{2R} + h.c. \,,
\label{eq:A:2}
\end{equation}
\begin{equation}
M_u = \left( \begin{array}{ccc}
\left(\tilde{m}^{up}\right)_{3\times 3} & 0_{3 \times 1} &
\left(x_2^k\right)_{3 \times 1} \\   
0_{1 \times 3} & M_1 & 0 \\
0_{1 \times 3} & 0 & M_2
\end{array} \right), \quad   
M_{X^{5/3}} = M_1, \quad
M_{X^{8/3}} = M_1 \,,
\label{eq:A:3}
\end{equation}
where $\tilde{m}^{up}$ is the SM $3 \times 3$ mass matrix of the up
sector.  




\subsection{Bottom multiplets}  \label{Appen:A:2}
The multiplets  which do not contain a top type partner, but do
contain a down type partner, in terms of the  $(SU(2)_{L},U(1)_{Y})$
quantum numbers, are the singlet $({\bf 1},-\frac{1}{3})$ and the
doublet $({\bf 2},-\frac{5}{6})$ and triplet $({\bf
  3},-\frac{4}{3})$. All these multiplets are listed in Table
\ref{QN_vec-bottom}.  


\subsubsection{Singlet $Y=-1/3$ and Doublet $Y=-5/6$ }   \label{Appen:A:2:1}

\begin{eqnarray}
\mathcal{L}_{V-SM} &=& -  \lambda_{1d}^k\, \bar{\psi}_{1L} \tilde{H}
d_R^k  + h.c. \,,  \label{eq:A:4}  \\ 
\mathcal{L}_{V-V} &=& - \xi_1 \, \bar{\psi}_{1L} \tilde{H} \psi_{2R}  
- \xi_{2}\, \bar{\psi}_{1R} \tilde{H} \psi_{2L} + h.c. \,,
\label{eq:A:5}
\end{eqnarray}
where $\psi_1 =({\bf 2},-\frac{5}{6})=\left(D_1, Y_1^{-4/3}\right)^{T}$ and
$\psi_2 = ({\bf 1},-\frac{1}{3}) = D_2$. The mass lagrangian and mass matrices are:
\begin{eqnarray}
\mathcal{L}_{\rm mass} &=&  -  y_{1d}^k \bar{D}_{1L} d_R^k -
\omega \bar{D}_{1L} D_{2R}  - { \omega'} \bar{D}_{1R} D_{2L}\,
 - M_1 \, \bar{D}_{1L} D_{1R} - M_1 \, \bar{Y}_{1L} Y_{1R}   \nonumber \\
&& - M_2 \, \bar{D}_{2L} D_{2R} + h.c. \,,
\label{eq:A:6}
\end{eqnarray}
\begin{equation}
M_u = \left(\tilde{m}^{up} \right)_{3\times 3}, \quad 
M_d = \left( \begin{array}{ccc}
\left(\tilde{m}^{down}\right)_{3\times 3} & 0_{3\times 1} & 0_{3\times 1} \\
(y_{1d}^k)_{1\times 3} & M_1 &  \omega \\
0_{1\times 3} &  \omega' & M_2 
\end{array} \right), \quad
M_{Y^{-4/3}} = M_1 \,, 
\label{eq:A:7}
\end{equation}
where $\tilde{m}^{up}$ and $\tilde{m}^{down}$ are the SM $3 \times 3$ mass matrices of
the up and down sectors respectively.  



\subsubsection{Doublet $Y=-5/6$ and triplet $Y=-4/3$ }   \label{Appen:A:2:2}
\begin{eqnarray}
\mathcal{L}_{V-SM} &=& - \lambda_{1d}^k \, \bar{\psi}_{1L} \tilde{H} d^k_{R}
+ h.c. \,,      \label{eq:A:8} \\
\mathcal{L}_{V-V} &=& - \xi_1 \, \bar{\psi}_{1L} \tau^a H  (\psi_{2R})^a 
- \xi_2 \, \bar{\psi}_{1R} \tau^a  H (\psi_{2L})^a + h.c. \,,
\label{eq:A:9}
\end{eqnarray}
where $\psi_1 =({\bf 2},-\frac{5}{6})=\left(D_1,Y_1^{-4/3}\right)^{T}$ and 
$\psi_2 =({\bf 3},-\frac{4}{3})=\left(D_2, Y_2^{-4/3}, Y_2^{-7/3}\right)^{T}$. The mass 
lagrangian and mass matrices are:
\begin{eqnarray}
\mathcal{L}_{\rm mass} &=& - y_{1d}^k \bar{D}_{1L} d_R^k
- \sqrt{2} \, \omega \bar{D}_{1L} D_{2R} - \omega \bar{Y}^{-4/3}_{1L} Y^{-4/3}_{2R}
- \sqrt{2} \, {\omega'} \bar{D}_{1R} D_{2L} - {\omega'} \bar{Y}^{-4/3}_{1R} Y^{-4/3}_{2L} \nonumber \\
&& - M_1 \, \bar{D}_{1L} D_{1R} - M_1 \, \bar{Y}^{-4/3}_{1L} Y^{-4/3}_{1R}
   - M_2 \, \bar{D}_{2L} D_{2R} - M_2 \, \bar{Y}^{-4/3}_{2L} Y^{-4/3}_{2R} \nonumber \\
&& - M_2 \, \bar{Y}^{-7/3}_{2L} Y^{-7/3}_{2R} + h.c. \,,
\label{eq:A:10} 
\end{eqnarray}
\begin{equation}
M_d = \left( \begin{array}{ccc}
\left(\tilde{m}^{down}\right)_{3\times 3} & 0_{3\times 1} & 0_{3\times 1} \\
(y_{1d}^k)_{1\times 3} & M_1 &  \sqrt{2} \, \omega \\
0_{1\times 3} &  \sqrt{2} \, \omega' & M_2
\end{array} \right), \quad
M_{Y^{-4/3}} =
\left(\begin{array}{c c}
M_1     & \omega \\
\omega' & M_2
\end{array}  \right), \quad
M_{Y^{-7/3}} = M_2. 
\label{eq:A:11}
\end{equation}




\subsection{Hybrid multiplets}   \label{Appen:A:3}

Multiplets where the mixing parameters of top and bottom sectors are
independent. Hence one can evade the constraints coming from
b-sector by assuming mixing to be zero without effecting top sector.  


\subsubsection{SM Doublet $Y=1/6$ and Singlet $Y=2/3$}   \label{Appen:A:3:1}

Details are given in Section \ref{sec:yukawa:3}. 



\subsubsection{SM Doublet $Y=1/6$ and Doublet $Y=7/6$}   \label{Appen:A:3:2}

Details are given in Section \ref{sec:yukawa:4}. 



\subsubsection{SM Doublet $Y=1/6$ and Singlet $Y=-1/3$}   \label{Appen:A:3:3}
\begin{eqnarray}
\mathcal{L}_{V-SM} &=& 
    - \lambda_{1}^k\, \bar{\psi}_{1L} \tilde{H} u_R^k -
    \lambda_{1d}^k\, \bar{\psi}_{1L} H d_R^k  
     - \lambda_{2d}^k \bar{Q}^k_L H \psi_{2R} + h.c. \,,  \label{eq:A:12}
     \\
\mathcal{L}_{V-V} &=& 
      - \xi_1\, \bar{\psi}_{1L} H \psi_{2R} - \xi_{2}\,
      \bar{\psi}_{1R} H \psi_{2L} + h.c. \,,  \label{eq:A:13}
\end{eqnarray}
where $\psi_1 =({\bf 2},\frac{1}{6})=\left(U_1,D_1\right)^{T}$ and $\psi_2 =
({\bf 1},-\frac{1}{3}) = D_2$. The mass lagrangian and mass matrices are: 
\begin{eqnarray}
\mathcal{L}_{\rm mass} &=& 
   - y_1^k \bar{U}_{1L} u_R^k -  y_{1d}^k \bar{D}_{1L} d_R^k - x_{2d}^k d_L^k D_{2R}
   - \omega \bar{D}_{1L} D_{2R} - \omega' \bar{D}_{2L} D_{1R} \nonumber \\  
&& - M_1 \, \bar{D}_{1L} D_{1R}  
   - M_2 \, \bar{D}_{2L} D_{2R} - M_1 \, \bar{U}_{1L} U_{1R} + h.c. \,,
   \label{eq:A:14}
\end{eqnarray}
\begin{equation}
M_u = \left( \begin{array}{cc}
\left(\tilde{m}^{up}\right)_{3\times 3} & 0_{3 \times 1} \\
(y_1^k)_{1\times 3} & M_1
\end{array} \right)\,, \quad 
M_d = \left( \begin{array}{c c c}
\left(\tilde{m}^{down}\right)_{3\times 3} &  0_{3 \times 1} & (x_{2d}^k)_{3\times 1} \\
(y_{1d}^k)_{1\times 3} & M_1 & \omega \\
0_{1\times3} & \omega' & M_2
\end{array} \right)\,.  
\label{eq:A:15}
\end{equation}



\subsubsection{SM Doublet $Y=1/6$ and Doublet  $Y=-5/6$}   \label{Appen:A:3:4}
\begin{equation}
\mathcal{L}_{V-SM} = 
    - \lambda_{1}^k\, \bar{\psi}_{1L} \tilde{H} u_R^k -
    \lambda_{1d}^k\, \bar{\psi}_{1L} H d_R^k   - \lambda_{2d}^k
    \bar{\psi_2}_L \tilde{H} d^k_R + h.c. \,,   \label{eq:A:16}
\end{equation}
where $\psi_1 =({\bf 2},\frac{1}{6})=\left(U_1,D_1\right)^{T}$ and $\psi_2 =
({\bf 2},-\frac{5}{6}) = \left(D_2,Y_2^{-4/3}\right)^T$. The mass lagrangian and mass matrices are:
\begin{eqnarray}
\mathcal{L}_{\rm mass} &=& 
   - y_{1}^k \bar{U}_{1L} u_R^k -  y_{1d}^k \bar{D}_{1L} d_R^k - y_{2d}^k \bar{D}_{2L} d_R^k
   - M_1 \, \bar{U}_{1L} U_{1R} -  M_1 \, \bar{D}_{1L} D_{1R}      \nonumber \\ 
&& - M_2 \, \bar{D}_{2L} D_{2R} - M_2 \, \bar{Y}^{-4/3}_{2L} Y^{-4/3}_{2R} + h.c. \,,
 \label{eq:A:17}
\end{eqnarray}
\begin{equation}
M_u = \left( \begin{array}{cc}
\left(\tilde{m}^{up}\right)_{3\times 3} & 0_{3 \times 1} \\
(y_{1}^k)_{1\times 3} & M_1
\end{array} \right)\,, \quad 
M_d = \left( \begin{array}{c c c}
\left(\tilde{m}^{down}\right)_{3\times 3} &  0_{3 \times 1} & 0_{3\times1} \\
(y_{1d}^k)_{1\times 3} & M_1 & 0 \\
(y_{2d}^k)_{1\times 3}  & 0 & M_2
\end{array} \right)\,, \quad 
M_{Y^{-4/3}} = M_2 \,.  
\label{eq:A:18}
\end{equation}




\subsection{Mixed multiplets}        \label{Appen:A:4}
The remaining combinations contain multiplets with both a VL top partner and a VL bottom
partner but with non-independent mixing in the up and in the down sector. 
They are listed in Table \ref{QN_vec-mixed}. These combinations are not considered in our
numerical studies, however their mixing structure with the SM and the other VL multiplets
is described in the following. 


\subsubsection{SM Doublet $Y=1/6$ and Triplet $Y=2/3$ }   \label{Appen:A:4:1}
\begin{eqnarray}
\mathcal{L}_{V-SM} &=& - \lambda_{1}^k\, \bar{\psi}_{1L} \tilde{H} u_R^k 
  -  \lambda_{1d}^k\, \bar{\psi}_{1L} H d_R^k 
  - \lambda_2^k \, \bar{Q}^k_L \tilde{H} \tau^a \psi_{2R}^a + h.c. \,,
   \label{eq:A:19} \\
\mathcal{L}_{V-V} &=& 
      - \xi_1\, \bar{\psi}_{1L} \tilde{H} \tau^a \psi_{2R}^a  -
      \xi_{2}\, \bar{\psi}_{1R} \tilde{H} \tau^a \psi_{2L}^a  
      + h.c. \,,  \label{eq:A:20}
\end{eqnarray}
where $\psi_1 = ({\bf 2},1/6) = \left(U_1,D_1\right)^T$ and $\psi_2 = \left(X_2^{5/3},U_2,D_2\right)^T$. 
The mass lagrangian and mass matrices are:
\begin{eqnarray}
\mathcal{L}_{\rm mass} &=& 
   - y_{1}^k \bar{U}_{1L} u_R^k -  y_{1d}^k \bar{D}_{1L} d_R^k 
   -  x_2^k \left( \bar{u}^k_{L} U_{2R} + \sqrt{2} \, \bar{d}^k_{L} D_{2R}\right)
   - \omega \left(\bar{U}_{1L} U_{2R} + \sqrt{2} \bar{D}_{1L} D_{2R}\right)    \nonumber \\ 
&& - \omega' \left(\bar{U}_{2L} U_{1R} + \sqrt{2} \bar{D}_{2L} D_{1R}\right) 
   - M_1 \, \bar{U}_{1L} U_{1R} - M_1 \, \bar{D}_{1L} D_{1R} 
   - M_2 \, \bar{U}_{2L} U_{2R}                                           \nonumber \\
&& - M_2 \, \bar{D}_{2L} D_{2R} - M_2 \, \bar{X}^{5/3}_{2L}
X^{5/3}_{2R}  + h.c. \,, 
\label{eq:A:21}
\end{eqnarray}
\begin{equation}
M_u = \left( \begin{array}{c c c}
\left(\tilde{m}^{up}\right)_{3\times 3} &  0_{3 \times 1} & (x_2^k)_{3\times 1} \\
(y_{1}^k)_{1\times 3} & M_1 & \omega \\
0_{1\times3} & \omega' & M_2
\end{array} \right)\,, \quad 
M_d = \left( \begin{array}{c c c}
\left(\tilde{m}^{down}\right)_{3\times 3} &  0_{3 \times 1} & \sqrt{2}
\, (x_2^k)_{3\times 1} \\ 
(y_{1d}^k)_{1\times 3} & M_1 & \sqrt{2} \, \omega \\
0_{1\times3} & \sqrt{2} \, \omega' & M_2
\end{array} \right)\,, \quad                   
M_{X^{5/3}} = M_2 \,.
\label{eq:A:22}
\end{equation}



\subsubsection{SM Doublet $Y=1/6$ and Triplet $Y=-1/3$}  \label{Appen:A:4:2}
\begin{eqnarray}
\mathcal{L}_{V-SM} &=& - \lambda_{1}^k\, \bar{\psi}_{1L} \tilde{H} u_R^k
-  \lambda_{1d}^k\, \bar{\psi}_{1L} H d_R^k  
- \lambda_2^k \, \bar{Q}_L^k H \tau^a \psi_{2R}^a + h.c. \,,
\label{eq:A:23} \\
\mathcal{L}_{V-V} &=& 
- \xi_1\, \bar{\psi}_{1L} H \tau^a \psi_{2R}^a  - \xi_{2}\,
\bar{\psi}_{1R} H \tau^a \psi_{2L}^a    + h.c. \,,
\label{eq:A:24} 
\end{eqnarray}
where $\psi_1 = ({\bf 2},\frac{1}{6}) = \left(U_1,D_1\right)^T$ and $\psi_2 =
\left({\bf 3},-\frac{1}{3}\right) = \left(U_2,D_2,Y_2^{-4/3}\right)^T$. The mass lagrangian and mass matrices are:
\begin{eqnarray}
\mathcal{L}_{\rm mass} &=& 
   - y_{1}^k \bar{U}_{1L} u_R^k -  y_{1d}^k \bar{D}_{1L} d_R^k 
   - x_2^k \left(\sqrt{2} \, \bar{u}_{L}^k U_{2R} - \bar{d}_{L}^k D_{2R}\right)
   - \omega \left(\sqrt{2} \, \bar{U}_{1L} U_{2R} - \bar{D}_{1L} D_{2R}\right) \nonumber \\ 
&& - \omega' \left(\sqrt{2} \, \bar{U}_{2L} U_{1R} - \bar{D}_{2L} D_{1R}\right) 
   - M_1 \, \bar{U}_{1L} U_{1R} - M_1 \, \bar{D}_{1L} D_{1R} 
   - M_2 \, \bar{U}_{2L} U_{2R} \nonumber \\
&& - M_2 \, \bar{D}_{2L} D_{2R} - M_2 \, \bar{X}^{5/3}_{2L} X^{5/3}_{2R} + h.c. \,,   
\label{eq:A:25}
\end{eqnarray}
\begin{equation}
M_u = \left( \begin{array}{c c c}
\left(\tilde{m}^{up}\right)_{3\times 3} &  0_{3 \times 1} & \sqrt{2} \, (x_2^k)_{3\times 1} \\
 (y_{1}^k)_{1\times 3} & M_1 & \sqrt{2} \, \omega \\
0_{1\times3} & \sqrt{2} \, \omega' & M_2
\end{array} \right)\,, \quad
M_d = \left( \begin{array}{c c c}
\left(\tilde{m}^{down}\right)_{3\times 3} &  0_{3 \times 1} & - (x_2^k)_{3\times 1} \\
(y_{1d}^k)_{1\times 3} & M_1 & - \omega \\
0_{1\times3} & - \omega' & M_2
\end{array} \right)\,, \quad 
M_{X^{5/3}} = M_2 \,.
\label{eq:A:26}
\end{equation}



\subsubsection{Triplet $Y=2/3$ and Singlet $Y=2/3$ }   \label{Appen:A:4:3}
\begin{equation}
\mathcal{L}_{V-SM} = - \lambda_{1}^k\, \bar{Q}_{L} \tilde{H} \psi_{1R}  
  - \lambda_2^k \, \bar{Q}_L^k \tilde{H} \tau^a \psi_{2R}^a + h.c. \,,
\label{eq:A:27}
\end{equation}
where $\psi_1 = ({\bf 1},\frac{2}{3}) = U_1$ and $\psi_2 = ({\bf 3},\frac{2}{3}) = \left(U_2,D_2,Y_2^{-4/3}\right)^T$.
The mass lagrangian and mass matrices are:
\begin{eqnarray}
\mathcal{L}_{\rm mass} &=& 
   - x_1^k \bar{u}^k_{L} U_{1R} 
   -  x_2^i \left(  \bar{u}^i_{L} U_{2R} +  \sqrt{2} \, \bar{d}^i_{L} D_{2R}\right)   \nonumber \\
&& - M_1 \, \bar{U}_{1L} U_{1R} - M_2 \, \bar{U}_{2L} U_{2R}    
   - M_2 \, \bar{D}_{2L} D_{2R} - M_2 \, \bar{Y}^{-4/3}_{2L} Y^{-4/3}_{2R}  + h.c. \,,
   \label{eq:A:28}
\end{eqnarray}
\begin{equation}
M_u = \left( \begin{array}{c c c}
\left(\tilde{m}^{up}\right)_{3\times 3} &  (x_1^k)_{3 \times 1} &  (x_2^k)_{3\times 1} \\
0_{1\times 3} & M_1 & 0 \\
0_{1\times3} & 0 & M_2
\end{array} \right)\,, \quad
M_d = \left( \begin{array}{c c}
\left(\tilde{m}^{down}\right)_{3\times 3} & \sqrt{2} \, (x_2^k)_{3\times 1} \\
0_{1\times 3} & M_2 
\end{array} \right) \,, \quad 
M_{Y^{-4/3}} = M_2 \,.  
\label{eq:A:29}
\end{equation}



\subsubsection{Triplet $Y=2/3$ and Doublet $Y=7/6$ }   \label{Appen:A:4:4}
\begin{eqnarray}
\mathcal{L}_{V-SM} &=& - \lambda_{1}^k\, \bar{\psi}_{1L} H u_R^k  
         - \lambda_2^k \, \bar{Q}_L^k \tilde{H} \tau^a \psi_{2R}^a +
         h.c. \,,  \label{eq:A:30} \\ 
\mathcal{L}_{V-V} &=& 
      - \xi_1\, \bar{\psi}_{1L} H \tau^a \psi_{2R}^a  - \xi_{2}\,
      \bar{\psi}_{1R} H \tau^a \psi_{2L}^a + h.c. \,,  \label{eq:A:31}
\end{eqnarray}
where $\psi_1 = ({\bf 2},\frac{7}{6}) = \left(X^{5/3}_1,U_1\right)^T$ and
$\psi_2 = ({\bf 3},\frac{2}{3}) = \left(X^{5/3}_2,U_2,D_2\right)^T$. The mass lagrangian and mass matrices are:  
\begin{eqnarray}
\mathcal{L}_{\rm mass} &=& 
   - y_1^k \bar{U}_{1L} u_R^k  
   -  x_2^i \left( \bar{u}^i_{L} U_{2R} +  \sqrt{2}  \, \bar{d}^i_{L} D_{2R}\right)
   - \omega \left(\sqrt{2} \, \bar{X}^{5/3}_{1L} X_{2R}^{5/3} 
   - \bar{U}_{1L} U_{2R}\right)    \nonumber \\  
&& - \omega' \left(\sqrt{2} \, \bar{X}^{5/3}_{2L} X_{1R}^{5/3} -  \bar{U}_{2L} U_{1R}\right) 
   - M_1 \, \bar{U}_{1L} U_{1R}  - M_1 \, \bar{X}^{5/3}_{1L} X^{5/3}_{1R}  
   - M_2 \, \bar{X}^{5/3}_{2L} X^{5/3}_{2R} \nonumber \\
&& - M_2 \, \bar{U}_{2L} U_{2R} - M_2 \, \bar{D}_{2L} D_{2R}  + h.c. \,,
\label{eq:A:32}   
\end{eqnarray}
\begin{eqnarray}
M_u = \left( \begin{array}{c c c}
\left(\tilde{m}^{up}\right)_{3\times 3} &  0_{3 \times 1} &  (x_2^k)_{3\times 1} \\
 (y_1^k)_{1\times 3} & M_1 &  - \omega \\
0_{1\times3} & - \omega' & M_2
\end{array} \right)\, &,& \quad 
M_d = \left( \begin{array}{c c }
\left(\tilde{m}^{down}\right)_{3\times 3} & \sqrt{2} \, (x_2^k)_{3\times 1} \\
0_{1\times 3} & M_2 \\
\end{array} \right)\, \nonumber \\
M_{Y^{-4/3}} &=&
\left( \begin{array}{c c }
M_1 &  \sqrt{2} \, \omega  \\
\sqrt{2} \, \omega'  & M_2 \\
\end{array} \right) \,. 
\label{eq:A:33}
\end{eqnarray}



\subsubsection{Triplet $Y=2/3$ and Singlet $Y=-1/3$ }   \label{Appen:A:4:5}
\begin{equation}
\mathcal{L}_{V-SM} =  - \lambda_1^k \, \bar{Q}_L^k \tilde{H} \tau^a \psi_{1R}^a  
     -  \lambda_{2d}^k\, \bar{Q}_{L}^k H \psi_{2R}  +   h.c. \,,
     \label{eq:A:34} \\ 
\end{equation}
where $\psi_1 = ({\bf 3},\frac{2}{3}) = \left(X_1^{5/3},U_1,D_1\right)^T$ and
$\psi_2 = ({\bf 1},-\frac{1}{3}) = D_2$. The mass lagrangian and mass matrices are:
\begin{eqnarray}
\mathcal{L}_{\rm mass} &=& 
   - x_1^k \, \left(  \bar{u}^k_{L} U_{1R} +  \sqrt{2} \,  \bar{d}^k_{L} D_{1R}\right)
   - x_{2d}^k \, \bar{d}_L^k D_{2R} 
   - M_1 \, \bar{X}^{5/3}_{1L} X^{5/3}_{1R} - M_1 \, \bar{U}_{1L} U_{1R} \nonumber \\
&& - M_1 \, \bar{D}_{1L} D_{1R}  - M_2 \, \bar{D}_{2L} D_{2R}  + h.c. \,,
\label{eq:A:35} 
\end{eqnarray}
\begin{equation}
M_u = \left( \begin{array}{c c }
\left(\tilde{m}^{up}\right)_{3\times 3} & (x_1^k)_{3\times 1} \\
0_{1\times 3} & M_1 \\
\end{array} \right)\,, \quad 
M_d = \left( \begin{array}{c c c} 
\left(\tilde{m}^{down}\right)_{3\times 3} &  \sqrt{2} \, (x_1^k)_{3
\times 1} &  (x_{2d}^k)_{3\times 1} \\ 
0_{1\times 3} & M_1 &  0   \\
0_{1\times3} & 0 & M_2
\end{array} \right)\,, \quad
M_{X^{5/3}} = M_1 \,.
\label{eq:A:36}
\end{equation}



\subsubsection{Triplet $Y=2/3$ and Doublet $Y=-5/6$ }   \label{Appen:A:4:6}
\begin{equation}
\mathcal{L}_{V-SM} =  - \lambda_1^k \, \bar{Q}_L^k \tilde{H} \tau^a \psi_{1R}^a  
- \lambda_{2d}^k\, \bar{\psi}_{2L} \tilde{H} d_R^k + h.c. \,,  \label{eq:A:37}
\end{equation}
where $\psi_1 = ({\bf 3},2/3) = \left(X_1^{5/3},U_1,D_1\right)^T$ and $\psi_2 =
({\bf 2},-5/6) = \left(D_2,Y_2^{-4/3}\right)^T$. The mass lagrangian and mass matrices are:
\begin{eqnarray}
\mathcal{L}_{\rm mass} &=& 
   - x_1^k  \left(  \bar{u}^k_{L} U_{1R} +  \sqrt{2} \, \bar{d}^k_{L} D_{1R}\right)
   - y_{2d}^k \, \bar{D}_{2L} d_R^k 
   - M_1 \, \bar{X}^{5/3}_{1L} X^{5/3}_{1R} - M_1 \, \bar{U}_{1L} U_{1R} \nonumber \\
&& - M_1 \, \bar{D}_{1L} D_{1R}  - M_2 \, \bar{D}_{2L} D_{2R}  + h.c. \,,
\label{eq:A:38}
\end{eqnarray}
\begin{equation}
M_u = \left( \begin{array}{c c }
\left(\tilde{m}^{up}\right)_{3\times 3} & (x_1^k)_{3\times 1} \\
0_{1\times 3} & M_1 \\
\end{array} \right)\,, \quad 
M_d = \left( \begin{array}{c c c}
\left(\tilde{m}^{down}\right)_{3\times 3} &  \sqrt{2} \, (x_1^k)_{3
\times 1} &  0_{3\times 1} \\ 
0_{1\times 3} & M_1 &  0   \\
(y_{2d}^k)_{1\times 3} & 0 & M_2
\end{array} \right)\,, \quad
M_{X^{5/3}} = M_1 \,.
\label{eq:A:39}
\end{equation}



\subsubsection{Triplet $Y=-1/3$ and Singlet $Y=2/3$ }    \label{Appen:A:4:7}
\begin{equation}
\mathcal{L}_{V-SM} = - \lambda_1^k \, \bar{Q}_L^k H \tau^a \psi_{1L}^a 
- \lambda_{2}^k\, \bar{Q}_{L}^k \tilde{H} \psi_{2R}  + h.c. \,,
\label{eq:A:40}
\end{equation}
where $\psi_1 = ({\bf 3},-1/3) = \left(U_1,D_1,Y_1^{-4/3}\right)^T$ and 
$\psi_2 = ({\bf 1},2/3) = U_2$.
The mass lagrangian and mass matrices are:
\begin{eqnarray}
\mathcal{L}_{\rm mass} &=& 
   - x_1^k \, \left(\sqrt{2} \, \bar{u}^k_{L} U_{1R} -  \bar{d}^k_{L} D_{1R}\right)
   - x_2^k \bar{u}_L^k U_{2R} 
   - M_1 \, \bar{U}_{1L} U_{1R}   - M_1 \, \bar{D}_{1L} D_{1R} 
     \nonumber \\ 
&&  - M_1 \, \bar{Y}^{-4/3}_{1L} Y^{-4/3}_{1R}
- M_2 \, \bar{U}_{2L} U_{2R}  + h.c. \,,
\label{eq:A:41}
\end{eqnarray}
\begin{eqnarray}
M_u = \left( \begin{array}{c c c}
\left(\tilde{m}^{up}\right)_{3\times 3} &  \sqrt{2} \, (x_1^k)_{3
  \times 1} &  (x_2^k)_{3\times 1} \\ 
0_{1\times 3} & M_1 &  0 \\
0_{1\times3} & 0 & M_2
\end{array} \right)\, &,& \quad
M_d = \left( \begin{array}{c c }
\left(\tilde{m}^{down}\right)_{3\times 3} & - (x_1^k)_{3\times 1} \\
0_{1\times 3} & M_1 \\
\end{array} \right) \,, \nonumber \\ 
M_{Y^{-4/3}} &=& M_1 \,.
\label{eq:A:42}
\end{eqnarray}



\subsubsection{Triplet $Y=-1/3$ and Doublet $Y=7/6$ }   \label{Appen:A:4:8}
\begin{equation}
\mathcal{L}_{V-SM} = - \lambda_1^k \, \bar{Q}^k_L H \tau^a \psi_{1L}^a 
- \lambda_{2}^k\, \bar{Q}^k_{L} \tilde{H} \psi_{2R}  + h.c. \,,
\label{eq:A:43}
\end{equation}
where $\psi_1 = ({\bf 3},-1/3) = \left(U_1,D_1,Y_1^{-4/3}\right)^T$ and $\psi_2
= ({\bf 2},7/6) = \left(X_2^{5/3},U_2\right)^T$. The mass lagrangian and mass matrices are:
\begin{eqnarray}
\mathcal{L}_{\rm mass} &=& 
   - x_1^k \, \left(\sqrt{2} \, \bar{u}^k_{L} U_{1R} -  \bar{d}^k_{L} D_{1R}\right)
   - x_2^k \, \bar{u}^k_L U_{2R} 
   - M_1 \, \bar{U}_{1L} U_{1R}   - M_1 \, \bar{D}_{1L} D_{1R}  \nonumber \\ 
&&    - M_1 \, \bar{Y}^{-4/3}_{1L} Y^{-4/3}_{1R}
- M_2 \, \bar{U}_{2L} U_{2R}  + h.c. \,, 
\label{eq:A:44}
\end{eqnarray}
\begin{eqnarray}
M_u = \left( \begin{array}{c c c}
\left(\tilde{m}^{up}\right)_{3\times 3} &  \sqrt{2} \, (x_1^k)_{3\times 1} &  (x_2^k)_{3\times 1} \\ 
0_{1\times 3} & M_1 &  0 \\
0_{1\times3} & 0 & M_2
\end{array} \right)\, &,& \quad
M_d = \left( \begin{array}{c c }
\left(\tilde{m}^{down}\right)_{3\times 3} & - (x_1^k)_{3\times 1} \\
0_{1\times 3} & M_1 \\
\end{array} \right) \, , \nonumber \\  
M_{Y^{-4/3}} &=& M_1 \,. 
\label{eq:A:45}
\end{eqnarray}



\subsubsection{Triplet $Y=-1/3$ and Singlet $Y=-1/3$ }   \label{Appen:A:4:9}
\begin{equation}
\mathcal{L}_{V-SM} = - \lambda_1^k \, \bar{Q}_L^k H \tau^a \psi_{1L}^a 
- \lambda_{2d}^k \, \bar{Q}_L^k H \psi_{2R} + h.c. \,,
\label{eq:A:46}
\end{equation}
where $\psi_1 = ({\bf 3},-1/3) = \left(U_1,D_1,Y_1^{-4/3}\right)^T$ and 
$\psi_2 = ({\bf 1},-1/3) = D_2$. 
The mass lagrangian and mass matrices are:
\begin{eqnarray}
\mathcal{L}_{\rm mass} &=& 
   - x_1^k \, \left(\sqrt{2} \,  \bar{u}^k_{L} U_{1R} -  \, \bar{d}^k_{L} D_{1R}\right)
   - x_{2d}^k \, \bar{d}_L^k D_{2R} - M_1 \, \bar{U}_{1L} U_{1R} - M_1 \, \bar{D}_{1L} D_{1R} \nonumber \\
&& - M_1 \, \bar{Y}^{-4/3}_{1L} Y^{-4/3}_{1R}  - M_2 \, \bar{D}_{2L} D_{2R}  + h.c. \,,
\label{eq:A:47}
\end{eqnarray}
\begin{eqnarray}
M_u = \left( \begin{array}{c c }
\left(\tilde{m}^{up}\right)_{3\times 3} &  \sqrt{2} \, (x_1^k)_{3 \times 1}  \\
0_{1\times 3} & M_1 
\end{array} \right) \, &,& \quad
M_d = \left( \begin{array}{c c c}
\left(\tilde{m}^{down}\right)_{3\times 3} & - (x_1^k)_{3 \times 1} & (x_{2d}^k)_{3\times 1} \\
0_{1\times 3} & M_1 &  0 \\
0_{1\times3} & 0 & M_2
\end{array} \right) \,,  \nonumber \\ 
M_{Y^{-4/3}} &=& M_1 \,. 
\label{eq:A:48}
\end{eqnarray}



\subsubsection{Triplet $Y=-1/3$ and Doublet $Y=-5/6$ }   \label{Appen:A:4:10}
\begin{eqnarray}
\mathcal{L}_{V-SM} &=& - \lambda_{1d}^k\, \bar{\psi}_{1L} \tilde{H} d_R^k
- \lambda_2^k \, \bar{Q}_L^k H \tau^a \psi_{2L}^a + h.c. \,,
\label{eq:A:49}      \\ 
\mathcal{L}_{V-V} &=& - \xi_1\, \bar{\psi}_{1L} \tilde{H} \tau^a
\psi_{2R}^a  - \xi_{2}\, \bar{\psi}_{1R} \tilde{H} \tau^a \psi_{2L}^a
+  h.c. \,, \label{eq:A:50}    
\end{eqnarray} 
where $\psi_1 = ({\bf 2},-5/6) = \left(D_1,Y^{-4/3}_1\right)^T$ and $\psi_2 =
({\bf 3},-1/3) = \left(U_2,D_2,Y_2^{-4/3}\right)^T$. The mass lagrangian and mass matrices
 are:
\begin{eqnarray}
\mathcal{L}_{\rm mass} &=& 
   - y_{1d}^k \bar{D}_{1L} d_R^k  
   - x_2^k  \, \left( \sqrt{2} \, \bar{u}^k_{L} U_{2R} -   \bar{d}^k_{L} D_{2R}\right)
   - \omega \left(  \bar{D}_{1L} D_{2R} + \sqrt{2} \, \bar{Y}^{-4/3}_{1L} Y_{2R}^{-4/3} \right) \nonumber \\  
&& - \omega' \left( \bar{D}_{2L} D_{1R} + \sqrt{2} \, \bar{Y}^{-4/3}_{2L} Y_{1R}^{-4/3} \right) 
   - M_1 \, \bar{D}_{1L} D_{1R}  - M_1 \, \bar{Y}^{-4/3}_{1L} Y^{-4/3}_{1R} \nonumber \\
&& - M_2 \, \bar{U}_{2L} U_{2R} 
   - M_2 \, \bar{D}_{2L} D_{2R}  - M_2 \, \bar{Y}^{-4/3}_{2L} Y^{-4/3}_{2R}  + h.c. \,, 
\label{eq:A:51}
\end{eqnarray}
\begin{eqnarray}
M_u = \left( \begin{array}{c c }
\left(\tilde{m}^{up}\right)_{3\times 3} & \sqrt{2} \, (x_2^k)_{3\times 1} \\
0_{1\times 3} & M_2 \\
\end{array} \right)\, &,& \quad 
M_d = \left( \begin{array}{c c c}
\left(\tilde{m}^{down}\right)_{3\times 3} &  0_{3 \times 1} & - (x_2^k)_{3\times 1} \\
 \left(y_{1d}^k\right)_{1\times 3} & M_1 &   \omega \\
0_{1\times3} &  \omega' & M_2
\end{array} \right)\,,  \nonumber \\ 
M_{Y^{-4/3}} &=& 
\left( \begin{array}{c c }
M_1 &  \sqrt{2} \, \omega  \\
\sqrt{2} \, \omega'  & M_2 \\
\end{array} \right) \,. 
\label{eq:A:52}
\end{eqnarray}



\subsubsection{Triplet $Y=2/3$ and Triplet $Y=-1/3$ }    \label{Appen:A:4:11}
\begin{eqnarray}
\mathcal{L}_{V-SM} &=& - \lambda_1^k \, \bar{Q}^k_L H \tau^a \psi_{1R}^a 
            - \lambda_2^k \, \bar{Q}^k_L \tilde{H} \tau^a \psi_{2R}^a
            + h.c. \,,     
            \label{eq:A:53}
\end{eqnarray}
where $\psi_1 = ({\bf 3},-1/3) = \left(U_1,D_1,Y_1^{-4/3}\right)^T$ and $\psi_2 =
({\bf 3},2/3) = \left(X_2^{5/3},U_2,D_2\right)^T$. The mass lagrangian and mass matrices 
are:  
\begin{eqnarray}
\mathcal{L}_{\rm mass} &=& 
   - x_1^k \, \left( \sqrt{2} \, \bar{u}^k_{L} U_{1R} -  \bar{d}^k_{L} D_{1R}\right)
   - x_2^k \, \left( \bar{u}^k_{L} U_{2R} -  \sqrt{2} \, \bar{d}^k_{L} D_{2R}\right) \nonumber \\ 
&& - M_1 \, \bar{U}_{1L} U_{1R} 
   - M_1 \, \bar{D}_{1L} D_{1R}  - M_1 \, \bar{Y}^{-4/3}_{1L} Y^{-4/3}_{1R} \nonumber \\
&& - M_2 \, \bar{U}_{2L} U_{2R} 
   - M_2 \, \bar{X}^{5/3}_{2L} X^{5/3}_{2R} - M_2 \, \bar{D}_{2L} D_{2R} + h.c. \,,
   \label{eq:A:54}
\end{eqnarray}
\begin{eqnarray}
M_u = \left( \begin{array}{c c c}
\left(\tilde{m}^{up}\right)_{3\times 3} & \sqrt{2} \, (x_1^k)_{3\times 1} & (x_2^k)_{3\times1} \\
0_{1\times 3} & M_1 & 0 \\
0_{1\times 3} & 0     & M_2 
\end{array} \right)\,, &\quad&
M_d = \left( \begin{array}{c c c}
\left(\tilde{m}^{down}\right)_{3\times 3} & - (x_1^k)_{3 \times 1} & -
\sqrt{2} \, (x_2^k)_{3\times 1} \\ 
0_{1\times 3} & M_1 &   0  \\
0_{1\times3} &  0 & M_2
\end{array} \right)\,, \nonumber\\
M_{Y^{-4/3}} = M_1 \,, &\quad& 
M_{X^{5/3}} = M_2 \,. \label{eq:A:55}
\end{eqnarray}




\section{Couplings to gauge and Higgs bosons}  \label{appendix:B}

\begin{table}[htb]
\begin{center}
\begin{tabular}{|c|cc|cc|cc|cc|cc|}
\hline
Model & $\alpha_{1}$ & $\alpha_{2}$ & $\alpha_{1}^{X^{5/3}}$ & $\alpha_{2}^{X^{5/3}}$ & $\alpha_{1}^{X^{8/3}}$ & $\alpha_{2}^{X^{8/3}}$ & $\alpha_{1}^{Y^{-4/3}}$ & $\alpha_{2}^{Y^{-4/3}}$ & $\alpha_{1}^{Y^{-7/3}}$ & $\alpha_{2}^{Y^{-7/3}}$ \\ \hline
\ref{Appen:A:1:1} & $0$ & $0$ & $1$ & $0$ & $0$ & $0$ & $0$ & $0$ & $0$ & $0$ \\ \hline
\ref{Appen:A:1:2} & $0$ & $0$ & $1$ & $\sqrt{2}$ & $0$ & $-\sqrt{2}$ & $0$ & $0$ & $0$ & $0$ \\ \hline
\ref{Appen:A:1:3} & $0$ & $0$ & $\sqrt{2}$ & $0$ & $-\sqrt{2}$ & $0$ & $0$ & $0$ & $0$ & $0$ \\ \hline
\ref{Appen:A:2:1} & $0$ & $0$ & $0$ & $0$ & $0$ & $0$ & $1$ & $0$ & $0$ & $0$ \\ \hline
\ref{Appen:A:2:2} & $0$ & $0$ & $0$ & $0$ & $0$ & $0$ & $1$ & $-\sqrt{2}$ & $0$ & $\sqrt{2}$ \\ \hline
\ref{Appen:A:3:1} & $1$ & $0$ & $0$ & $0$ & $0$ & $0$ & $0$ & $0$ & $0$ & $0$ \\ \hline
\ref{Appen:A:3:2} & $1$ & $0$ & $0$ & $1$ & $0$ & $0$ & $0$ & $0$ & $0$ & $0$ \\ \hline
\ref{Appen:A:3:3} & $1$ & $0$ & $0$ & $0$ & $0$ & $0$ & $0$ & $0$ & $0$ & $0$ \\ \hline
\ref{Appen:A:3:4} & $1$ & $0$ & $0$ & $0$ & $0$ & $0$ & $0$ & $1$ & $0$ & $0$ \\ \hline
\ref{Appen:A:4:1} & $1$ & $\sqrt{2}$ & $0$ & $-\sqrt{2}$ & $0$ & $0$ & $0$ & $0$ & $0$ & $0$ \\ \hline
\ref{Appen:A:4:2} & $1$ & $-\sqrt{2}$ & $0$ & $0$ & $0$ & $0$ & $0$ & $\sqrt{2}$ & $0$ & $0$ \\ \hline
\ref{Appen:A:4:3} & $0$ & $\sqrt{2}$ & $0$ & $-\sqrt{2}$ & $0$ & $0$ & $0$ & $0$ & $0$ & $0$ \\ \hline
\ref{Appen:A:4:4} & $0$ & $\sqrt{2}$ & $1$ & $-\sqrt{2}$ & $0$ & $0$ & $0$ & $0$ & $0$ & $0$ \\ \hline
\ref{Appen:A:4:5} & $\sqrt{2}$ & $0$ & $-\sqrt{2}$ & $0$ & $0$ & $0$ & $0$ & $0$ & $0$ & $0$ \\ \hline
\ref{Appen:A:4:6} & $\sqrt{2}$ & $0$ & $-\sqrt{2}$ & $0$ & $0$ & $0$ & $0$ & $1$ & $0$ & $0$ \\ \hline
\ref{Appen:A:4:7} & $-\sqrt{2}$ & $0$ & $0$ & $0$ & $0$ & $0$ & $\sqrt{2}$ & $0$ & $0$ & $0$ \\ \hline
\ref{Appen:A:4:8} & $-\sqrt{2}$ & $0$ & $0$ & $1$ & $0$ & $0$ & $\sqrt{2}$ & $0$ & $0$ & $0$ \\ \hline
\ref{Appen:A:4:9} & $-\sqrt{2}$ & $0$ & $0$ & $0$ & $0$ & $0$ & $\sqrt{2}$ & $0$ & $0$ & $0$ \\ \hline
\ref{Appen:A:4:10} & $0$ & $-\sqrt{2}$ & $0$ & $0$ & $0$ & $0$ & $1$ & $\sqrt{2}$ & $0$ & $0$ \\ \hline
\ref{Appen:A:4:11} & $-\sqrt{2}$ & $\sqrt{2}$ & $0$ & $-\sqrt{2}$ & $0$ & $0$ & $\sqrt{2}$ & $0$ & $0$ & $0$ \\ \hline
\end{tabular}
\caption{
The coefficients $\alpha_{i}$, $\alpha_{i}^{X^{5/3}}$, $\alpha_{i}^{X^{8/3}}$, $\alpha_{i}^{Y^{-4/3}}$ and $\alpha_{i}^{Y^{-7/3}}$ ($i=1,2$) in the two vector-like multiplets models listed in Appendix~\ref{appendix:A}.
} \label{Tab:5}
\end{center}
\end{table}

The VL quarks couple to gauge bosons and the Higgs boson
according to their quantum numbers. In the following we give some
general formulas for the case of two VL quarks multiplets which
were used for our numerical results and which can be easily generalised 
for scenarios with more than two VL quark multiplets. 


\subsection{$W^{\pm}$ boson couplings}    \label{Appen:B:1}

In the gauge basis, the general expressions for the couplings of $W^{\pm}$ bosons in the two VL multiplets models are given by
\begin{eqnarray}
\mathcal{L}_{W^{\pm}} &=&
\frac{g}{\sqrt{2}} \left(
\begin{array}{ccccc}
\bar{u}_{L}^{1}, & \bar{u}_{L}^{2}, & \bar{u}_{L}^{3}, & \bar{U}_{1L}, & \bar{U}_{2L} \\
\end{array}
\right) \cdot \delta_{L} \cdot \gamma^{\mu}
\left(
\begin{array}{c}
d_{L}^{1} \\ d_{L}^{2} \\ d_{L}^{3} \\ D_{1L} \\ D_{2L} \\
\end{array}
\right) W_{\mu}^{+} \nonumber \\
&+& \frac{g}{\sqrt{2}} \left(
\begin{array}{ccccc}
\bar{u}_{R}^{1}, & \bar{u}_{R}^{2}, & \bar{u}_{R}^{3}, & \bar{U}_{1R}, & \bar{U}_{2R} \\
\end{array}
\right) \cdot \delta_{R} \cdot \gamma^{\mu}
\left(
\begin{array}{c}
d_{R}^{1} \\ d_{R}^{2} \\ d_{R}^{3} \\ D_{1R} \\ D_{2R} \\
\end{array}
\right) W_{\mu}^{+} + h.c. \,,
\end{eqnarray}
with
\begin{equation}
\delta_{L} =
\left(
\begin{array}{ccc}
I_{3\times 3} & & \\
& \alpha_{1} & \\
& & \alpha_{2} \\
\end{array}
\right) \,, \quad
\delta_{R} =
\left(
\begin{array}{ccc}
0_{3\times 3} & & \\
& \alpha_{1} & \\
& & \alpha_{2} \\
\end{array}
\right) \,.
\end{equation}
Note that the coefficients $\alpha_{i}$, which are listed in Table~\ref{Tab:5}, depend on the representation of the $i$-th VL quark.
In the mass basis, the left- and right-handed couplings can be written as
\begin{eqnarray}
g_{WL}^{IJ} &=& \frac{g}{\sqrt{2}}V_{CKM}^{L,IJ} = \frac{g}{\sqrt{2}}V_{L}^{u\dagger}\cdot \delta_{L} \cdot V_{L}^{d} \,, \\
g_{WR}^{IJ} &=& \frac{g}{\sqrt{2}}V_{CKM}^{R,IJ} = \frac{g}{\sqrt{2}}V_{R}^{u\dagger}\cdot \delta_{R} \cdot V_{R}^{d} \,,
\end{eqnarray}
where $V_{CKM}^{L}$ and $V_{CKM}^{R}$ are the left- and right-handed CKM matrix, respectively.
%
%
%
The Lagrangian terms for the couplings between exotic quark $X^{5/3}$/$Y^{-4/3}$ and top-type/bottom-type quarks 
can be expressed as:
\begin{equation}
\mathcal{L}_{W^{\pm}} = \frac{g}{\sqrt{2}}
\left(
\begin{array}{ccccc}
0, & 0, & 0, & \bar{X}_{1L}^{5/3}, & \bar{X}_{2L}^{5/3} \\
\end{array}
\right) \cdot
\left(
\begin{array}{ccc}
0_{3\times3} & & \\
& \alpha_{1}^{X^{5/3}} & \\
& & \alpha_{2}^{X^{5/3}} \\
\end{array}
\right) \cdot \gamma^{\mu}
\left(
\begin{array}{c}
u^1_L \\ u^2_L \\ u^3_L \\ U_{1L} \\ U_{2L} \\
\end{array}
\right)W_{\mu}^{+} + h.c. \,.
\end{equation}
The mass matrix of the $X^{5/3}$ system is diagonalised as:
\begin{equation}
M_{X^{5/3}} = V_{L}^{X^{5/3}} \cdot
\left(
\begin{array}{cc}
m_{X_L^{5/3}} & \\
 & m_{X_H^{5/3}} \\
\end{array}
\right)
\cdot V_{R}^{X^{5/3}\dagger} \,,
\end{equation}
where the mass eigenstates $X_{L}^{5/3}$ and $X_{H}^{5/3}$ are defined as:
\begin{equation}
\left(
\begin{array}{c}
X_{L}^{5/3} \\ X_{H}^{5/3} \\
\end{array}
\right)_{L/R} =
V_{L/R}^{X^{5/3}\dagger} \cdot \left(
\begin{array}{c}
X_{1}^{5/3} \\ X_{2}^{5/3} \\
\end{array}
\right)_{L/R} \,.
\end{equation}
In the mass basis, the left- and right-handed couplings of $X^{5/3}$ become:
\begin{eqnarray}
g_{WL}^{X^{5/3},IJ} &=& \frac{g}{\sqrt{2}}
\left(
\begin{array}{cc}
I_{3\times3} & \\
 & V_{L}^{X^{5/3}} \\
\end{array}
\right)^{\dagger} \cdot
\left(
\begin{array}{ccc}
0_{3\times3} & & \\
 & \alpha_{1}^{X^{5/3}} & \\
 & & \alpha_{2}^{X^{5/3}} \\
\end{array}
\right)
\cdot V_{L}^{u} \,,\\
g_{WR}^{X^{5/3},IJ} &=& \frac{g}{\sqrt{2}}
\left(
\begin{array}{cc}
I_{3\times3} & \\
 & V_{R}^{X^{5/3}} \\
\end{array}
\right)^{\dagger} \cdot
\left(
\begin{array}{ccc}
0_{3\times3} & & \\
 & \alpha_{1}^{X^{5/3}} & \\
 & & \alpha_{2}^{X^{5/3}} \\
\end{array}
\right)
\cdot V_{R}^{u} \,.
\end{eqnarray}
Similarly, the couplings of $Y^{-4/3}$ can be expressed as
\begin{eqnarray}
g_{WL}^{Y^{-4/3},IJ} &=& \frac{g}{\sqrt{2}} V_{L}^{d\dagger} \cdot
\left(
\begin{array}{ccc}
0_{3\times3} & & \\
 & \alpha_{1}^{Y^{-4/3}} & \\
 & & \alpha_{2}^{Y^{-4/3}} \\
\end{array}
\right) \cdot
\left(
\begin{array}{cc}
I_{3\times3} & \\
 & V_{L}^{Y^{-4/3}} \\
\end{array}
\right) \,,\\
g_{WR}^{Y^{-4/3},IJ} &=& \frac{g}{\sqrt{2}} V_{R}^{d\dagger} \cdot
\left(
\begin{array}{ccc}
0_{3\times3} & & \\
 & \alpha_{1}^{Y^{-4/3}} & \\
 & & \alpha_{2}^{Y^{-4/3}} \\
\end{array}
\right) \cdot
\left(
\begin{array}{cc}
I_{3\times3} & \\
 & V_{R}^{Y^{-4/3}} \\
\end{array}
\right) \,.
\end{eqnarray}
We also introduce the general expressions for the couplings between $X^{5/3}$ and $X^{8/3}$:
\begin{equation}
\mathcal{L}_{W^{\pm}} = \frac{g}{\sqrt{2}}
\left(
\begin{array}{cc}
\bar{X}_{1L}^{8/3}, & \bar{X}_{2L}^{8/3} \\
\end{array}
\right) \cdot
\left(
\begin{array}{cc}
\alpha_{1}^{X^{8/3}} & \\
 & \alpha_{2}^{X^{8/3}} \\
\end{array}
\right) \cdot \gamma^{\mu}
\left(
\begin{array}{c}
X_{1L}^{5/3} \\ X_{2L}^{5/3} \\
\end{array}
\right)W_{\mu}^{+} + h.c. \,.
\end{equation}
Note that there is no mixing between $X^{8/3}_{1}$ and $X^{8/3}_{2}$ in the two VL multiplets listed in Appendix~\ref{appendix:A}.
The left- and right-handed couplings of $X^{8/3}$ in the mass basis are given by:
\begin{eqnarray}
g_{WL}^{X^{8/3},IJ} &=& \frac{g}{\sqrt{2}}
\left(
\begin{array}{cc}
\alpha_{1}^{X^{8/3}} & \\
& \alpha_{2}^{X^{8/3}} \\
\end{array}
\right) \cdot V_{L}^{X^{5/3}} \,,\\
g_{WR}^{X^{8/3},IJ} &=& \frac{g}{\sqrt{2}}
\left(
\begin{array}{cc}
\alpha_{1}^{X^{8/3}} & \\
& \alpha_{2}^{X^{8/3}} \\
\end{array}
\right) \cdot V_{R}^{X^{5/3}} \,.
\end{eqnarray}
For exotic quark $Y^{-7/3}$, the couplings can be evaluated as:
\begin{eqnarray}
g_{WL}^{Y^{-7/3},IJ} &=& \frac{g}{\sqrt{2}} V_{L}^{Y^{-4/3}\dagger} \cdot
\left(
\begin{array}{cc}
\alpha_{1}^{Y^{-7/3}} & \\
& \alpha_{2}^{Y^{-7/3}} \\
\end{array}
\right) \,,\\
g_{WR}^{Y^{-7/3},IJ} &=& \frac{g}{\sqrt{2}} V_{R}^{Y^{-4/3}\dagger} \cdot
\left(
\begin{array}{cc}
\alpha_{1}^{Y^{-7/3}} & \\
& \alpha_{2}^{Y^{-7/3}} \\
\end{array}
\right) \,.
\end{eqnarray}

\subsection{$Z$ boson couplings}        \label{Appen:B:2}

In terms of $Z$ boson couplings to the quark sector, and for the case of two VL quarks mixing with any SM quark generation under consideration, it is possible to identify three scenarios depending on where FCNCs appear. In the Top type multiplets listed in Appendix~\ref{Appen:A:1} FCNCs appear in the up quark sector; in the Bottom type multiplets listed in Appendix~\ref{Appen:A:2} FCNCs appear in the down quark sector; finally, in the Hybrid and Mixed multiplets, Appendix~\ref{Appen:A:3},~\ref{Appen:A:4}, FCNCs appear in both sectors.




The general expression for the left-handed couplings of the $Z$ in the up quark sector can be written as:
\begin{equation}
\mathcal{L}_{Z} = \frac{g}{c_W}\, \left( \bar{u}_L^1, \bar{u}_L^2, \bar{u}_L^3, \bar{U}_{1L}, \bar{U}_{2L} \right)\cdot \left[ \left( \frac{1}{2} - 
Q_{u} s_W^2 \right) I_{5\times5}
- \Delta T_{3}^{(up)} \right] \gamma^\mu \cdot \left( \begin{array}{c}
u_L^1\\u_L^2\\u_L^3\\U_{1L}\\U_{2L} \end{array} \right) Z_\mu \,,
\end{equation}
with:
\begin{equation}
\Delta T_{3}^{(up)} = \left( \begin{array}{ccc}
0_{3\times3}&&\\ & \Delta T_{3}^{(1,u)} &\\&& \Delta T_{3}^{(2,u)}\\
\end{array} \right) \,,
\end{equation}
where $I_{5\times5}$ is the $5\times5$ unit matrix and $\Delta T_{3}^{(K,u)}=1/2-T_{3}^{(K,u)}$ is the differences between 
the SM top-type quark and $K$-th generation VL quark.
In the  mass eigenstate basis, the left-handed coupling becomes:
\begin{equation}
g_{ZL}^{u,IJ} = \frac{g}{\cw} \left[ \left( \frac{1}{2} - Q_{u} \ssw \right) \delta^{IJ} - \sum_{K=1,2} \Delta T_{3}^{(K,u)} V_L^{u*,K+3I} 
V_L^{u,K+3J} \right] \,.
\end{equation}
Analogously for the right-handed couplings we obtain:
\begin{equation}
g_{ZR}^{u,IJ} = \frac{g}{\cw} \left[ \left(- Q_{u} \ssw \right) \delta^{IJ} + \sum_{K=1,2} T_{3}^{(K,u)} V_R^{u*,K+3I} 
V_R^{u,K+3J} \right] \,.
\end{equation}
For bottom-type quark, we obtain the left- and right-handed couplings:
\begin{eqnarray}
g_{ZL}^{d,IJ} &=& \frac{g}{\cw} \left[ \left( -\frac{1}{2} - Q_{d} \ssw \right) \delta^{IJ} - \sum_{K=1,2} \Delta T_{3}^{(K,d)} 
V_L^{d*,K+3I} V_L^{d,K+3J} \right] \,,\\
g_{ZR}^{d,IJ} &=& \frac{g}{\cw} \left[ \left(- Q_{d} \ssw \right) \delta^{IJ} + \sum_{K=1,2} T_{3}^{(K,d)} V_R^{d*,K+3I} 
V_R^{d,K+3J} \right] \,,
\end{eqnarray}
where $\Delta T_{3}^{(K,d)}=-1/2 - T_{3}^{(K,d)}$.
The left-handed couplings of the $Z$ in $X^{5/3}$ quarks can be written as:
\begin{equation}
\mathcal{L}_{Z} = \frac{g}{c_W}\, \left( \bar{X}_{1L}^{5/3}, \bar{X}_{2L}^{5/3} \right)\cdot \left[ \left(\begin{array}{cc} T_{3}^{(1,X)} & \\ & T_{3}^{(2,X)} \end{array}\right)
- Q_{X}\ssw
\left(\begin{array}{cc} 1 & \\ & 1 \end{array}\right) \right] \gamma^\mu \cdot \left( \begin{array}{c}
X_{1L}^{5/3}\\X_{2L}^{5/3} \end{array} \right) Z_\mu\,,
\end{equation}
where $Q_{X}=5/3$. In the mass eigenstate, the coupling becomes:
\begin{equation}
g_{ZL}^{X^{5/3},IJ} = \frac{g}{\cw} \left[ -Q_{X}\ssw\delta^{IJ} + \sum_{K=1,2} T_{3}^{(K,X)}V_{L}^{X^{5/3}*,KI}V_{L}^{X^{5/3},KJ} \right] \,.
\end{equation}
The right-handed couplings are
\begin{equation}
g_{ZR}^{X^{5/3},IJ} = \frac{g}{\cw} \left[ -Q_{X}\ssw\delta^{IJ} + \sum_{K=1,2} T_{3}^{(K,X)}V_{R}^{X^{5/3}*,KI}V_{R}^{X^{5/3},KJ} \right] \,.
\end{equation}
For exotic quark $Y^{-4/3}$ we obtain the left- and right-handed couplings are:
\begin{eqnarray}
g_{ZL}^{Y^{-4/3},IJ} &=& \frac{g}{\cw} \left[ -Q_{Y}\ssw\delta^{IJ} + \sum_{K=1,2} T_{3}^{(K,Y)}V_{L}^{Y^{-4/3}*,KI}V_{L}^{Y^{-4/3},KJ} \right] \,,\\
g_{ZR}^{Y^{-4/3},IJ} &=& \frac{g}{\cw} \left[ -Q_{Y}\ssw\delta^{IJ} + \sum_{K=1,2} T_{3}^{(K,Y)}V_{R}^{Y^{-4/3}*,KI}V_{R}^{Y^{-4/3},KJ} \right] \,,
\end{eqnarray}
where $Q_{Y}=-4/3$.
Similarly, the left- and right-handed couplings of the exotic quarks $X^{8/3}$ and $Y^{-7/3}$ can be expressed as:
\begin{eqnarray}
g_{ZL}^{X^{8/3}} &=& g_{ZR}^{X^{8/3}} = \frac{g}{\cw} \left[ T_{3}^{(K,X^{8/3})} -Q_{X^{8/3}}\ssw \right] \,,\\
g_{ZL}^{Y^{-7/3}} &=& g_{ZR}^{Y^{-7/3}} = \frac{g}{\cw} \left[ T_{3}^{(K,Y^{-7/3})} -Q_{Y^{-7/3}}\ssw \right] \,,
\end{eqnarray}
where $Q_{X^{8/3}}=8/3$ and $Q_{Y^{-7/3}}=-7/3$.



\subsection{Higgs boson couplings}     \label{Appen:B:3}



In the interaction basis, the Yukawa interactions in top-type quarks can be written as:
\begin{equation}
\mathcal{L}_{H} = \frac{1}{v}\, \left( \bar{u}_L^1, \bar{u}_L^2, \bar{u}_L^3, \bar{U}_{1L}, \bar{U}_{2L}\right)\cdot \left[ M_{u} - M \right]  \cdot \left( \begin{array}{c}
u_R^1\\u_R^2\\u_R^3\\U_{1R}\\U_{2R} \end{array} \right) h + h.c. \,,
\end{equation}
with:
\begin{equation}
M = \left( 
\begin{array}{ccc}
0_{3\times3}&&\\&M_{1}&\\&&M_{2}
\end{array}
\right) \,.
\end{equation}
In the mass eigenstate basis the coupling of top-type quark reads :
\begin{equation}
C^{u,IJ} =\frac{M_{u}^{diag,IJ}}{v} - \sum_{K=1,2}\frac{M_{K}}{v} V_L^{u*,K+3I} V_R^{u,K+3J}\,.
\label{eq:cptp}
\end{equation}
For bottom-type quark, we obtain:
\begin{equation}
C^{d,IJ} =\frac{M_{d}^{diag,IJ}}{v} - \sum_{K=1,2}\frac{M_{K}}{v} V_L^{d*,K+3I} V_R^{d,K+3J}\,.
\end{equation}

The Higgs is also allowed to couples to exotic charged VL quarks $X^{5/3}$/$Y^{-4/3}$ if the scenario contains more than one of them: in this case, one can consider the formulas above and removing the SM quark part:
\begin{eqnarray}
C^{X,IJ} &=&\frac{M_{X}^{diag,IJ}}{v} - \sum_{K=1,2}\frac{M_{K}}{v} V_L^{X^{5/3}*,KI} V_R^{X^{5/3},KJ}\,,\\
C^{Y,IJ} &=&\frac{M_{Y}^{diag,IJ}}{v} - \sum_{K=1,2}\frac{M_{K}}{v} V_L^{Y^{-4/3}*,KI} V_R^{Y^{-4/3},KJ}\,.
\end{eqnarray}




\section{Contributions to the S,T parameters from VL quarks} 
                              \label{appendix:C}

\begin{table}[htb]
\begin{center}
\begin{tabular}{|c|l|}
\hline
Type of model & $\Pi_{x}(p^2)=$ \\ \hline
\ref{Appen:A:1:1} & $\Pi_{x}^{(A)}(p^2)$ \\ \hline
\ref{Appen:A:1:2} & $\Pi_{x}^{(A)}(p^2) + \Pi_{x}^{(B)}(p^2)$ \\ \hline
\ref{Appen:A:1:3} & $\Pi_{x}^{(A)}(p^2) + \Pi_{x}^{(B)}(p^2) + \Pi_{x}^{(D)}(p^2)$ \\ \hline
\ref{Appen:A:2:1} & $\Pi_{x}^{(A)}(p^2) + \Pi_{x}^{(C)}(p^2)$ \\ \hline
\ref{Appen:A:2:2} & $\Pi_{x}^{(A)}(p^2) + \Pi_{x}^{(C)}(p^2) + \Pi_{x}^{(E)}(p^2)$ \\ \hline
\ref{Appen:A:3:1} & $\Pi_{x}^{(A)}(p^2)$ \\ \hline
\ref{Appen:A:3:2} & $\Pi_{x}^{(A)}(p^2) + \Pi_{x}^{(B)}(p^2)$ \\ \hline
\ref{Appen:A:3:3} & $\Pi_{x}^{(A)}(p^2)$ \\ \hline
\ref{Appen:A:3:4} & $\Pi_{x}^{(A)}(p^2) + \Pi_{x}^{(C)}(p^2)$ \\ \hline
\ref{Appen:A:4:1} & $\Pi_{x}^{(A)}(p^2) + \Pi_{x}^{(B)}(p^2)$ \\ \hline
\ref{Appen:A:4:2} & $\Pi_{x}^{(A)}(p^2) + \Pi_{x}^{(C)}(p^2)$ \\ \hline
\ref{Appen:A:4:3} & $\Pi_{x}^{(A)}(p^2) + \Pi_{x}^{(C)}(p^2)$ \\ \hline
\ref{Appen:A:4:4} & $\Pi_{x}^{(A)}(p^2) + \Pi_{x}^{(B)}(p^2)$ \\ \hline
\ref{Appen:A:4:5} & $\Pi_{x}^{(A)}(p^2) + \Pi_{x}^{(B)}(p^2)$ \\ \hline
\ref{Appen:A:4:6} & $\Pi_{x}^{(A)}(p^2) + \Pi_{x}^{(B)}(p^2) + \Pi_{x}^{(C)}(p^2)$ \\ \hline
\ref{Appen:A:4:7} & $\Pi_{x}^{(A)}(p^2) + \Pi_{x}^{(C)}(p^2)$ \\ \hline
\ref{Appen:A:4:8} & $\Pi_{x}^{(A)}(p^2) + \Pi_{x}^{(B)}(p^2) + \Pi_{x}^{(C)}(p^2)$ \\ \hline
\ref{Appen:A:4:9} & $\Pi_{x}^{(A)}(p^2) + \Pi_{x}^{(C)}(p^2)$ \\ \hline
\ref{Appen:A:4:10} & $\Pi_{x}^{(A)}(p^2) + \Pi_{x}^{(C)}(p^2)$ \\ \hline
\ref{Appen:A:4:11} & $\Pi_{x}^{(A)}(p^2) + \Pi_{x}^{(B)}(p^2) + \Pi_{x}^{(C)}(p^2)$ \\ \hline
\end{tabular}
\caption{
The $\Pi_{x}(p^2)$ ($x=11,33,3Q$) in the two VL multiplets.
} \label{Tab:6}
\end{center}
\end{table}

The contributions to the $S$ and $T$ parameters (the oblique corrections) can be written in general form for the contribution of 
VL particles circulating in the loop for the one-loop two point functions in terms of the couplings of these particles. The generic couplings 
to $W$, $Z$ are given explicitly in the previous section of the Appendix.
In the VL quark model, the general formulas for $S$, $T$ and $U$ parameters are given by $\Pi_{11}(p^2)$, $\Pi_{33}(p^2)$, $\Pi_{3Q}(p^2)$ and derivative of them with respect to $p^{2}$.
The $\Pi_{x}(p^2)$ ($x=11,33,3Q$) can be decomposed into the multiple parts $\Pi_{x}^{(i)}(p^{2})$ ($i=A,B,\cdots,E$) which are based on internal particles in loop diagrams:
\begin{eqnarray}
\Pi_{x}(p^{2}) = \sum_{i}\Pi_{x}^{(i)}(p^{2}) \,.
\end{eqnarray}
The results of $\Pi_{x}(p^{2})$ in all possible models under our assumptions are listed in Table~\ref{Tab:6}.

The contributions of part A of $\Pi_{11}(p^{2})$, loops of combinations of up- and bottom-type quarks, are given by:
\begin{equation}
\Pi_{11}^{(A)}(p^{2}) = \frac{1}{g^2}\sum_{I}\sum_{J} \left[ \left( |g_{WL}^{IJ}|^{2} + |g_{WR}^{IJ}|^{2} \right) \Pi^{LL}_{T} + 2\re \left( g_{WL}^{*,IJ}g_{WR}^{IJ} \right) \Pi^{LR}_{T} \right](p^2;u_{I},d_{J}) \,,
\end{equation}
where $I,J=1,2,\cdots,5$. We define the two point functions as:
\begin{eqnarray}
\Pi_{T}^{LL}(p^2;f_{1},f_{2}) &=& -\frac{N_{c}}{16\pi^{2}} \left[ (4-2D)B_{22} - 2p^{2}(B_{1} + B_{21}) \right](p^{2},m_{f_{1}},m_{f_{2}}) \,, \\
\Pi_{T}^{LR}(p^2;f_{1},f_{2}) &=& - \frac{N_{c}}{16\pi^{2}}2m_{f_{1}}m_{f_{2}}B_{0}(p^{2},m_{f_{1}},m_{f_{2}}) \,,
\end{eqnarray}
where $N_{c}$ is the color factor and $B_{i}$ are the Passarino-Veltman functions, which are defined by \cite{Passarino:1978jh}. They satisfy the following relation at $p^2=0$:
\begin{equation}
\Pi_{T}^{LL}(0;f,f) + \Pi_{T}^{LR}(0;f,f) = 0 \,.
\end{equation}
The part B, loops of combinations of top-type quarks and $X^{5/3}_{L/H}$, can be parametrised as:
\begin{eqnarray}
\lefteqn{\Pi_{11}^{(B)}(p^{2}) =} \nonumber\\
&& \frac{1}{g^2}\sum_{I} \Biggl\{ \Bigl[ \left( |g_{WL}^{X^{5/3},4I}|^{2} + |g_{WR}^{X^{5/3},4I}|^{2} \right) \Pi^{LL}_{T} + 2\re \left( g_{WL}^{X^{5/3}*,4I}g_{WR}^{X^{5/3},4I} \right) \Pi^{LR}_{T} \Bigr](p^2;X^{5/3}_{L},u_{I})  \nonumber\\
&& + \left[ \left( |g_{WL}^{X^{5/3},5I}|^{2} + |g_{WR}^{X^{5/3},5I}|^{2} \right) \Pi^{LL}_{T} + 2\re \left( g_{WL}^{X^{5/3}*,5I}g_{WR}^{X^{5/3},5I} \right) \Pi^{LR}_{T} \right](p^2;X^{5/3}_{H},u_{I}) \Biggr\} \,.
\end{eqnarray}
The part C, loops of combinations of bottom-type quarks and $Y^{-4/3}_{L/H}$, contributes to:
\begin{eqnarray}
\lefteqn{\Pi_{11}^{(C)}(p^{2}) =} \nonumber\\
&& \frac{1}{g^2} \sum_{I} \Biggl\{ \left[ \left( |g_{WL}^{Y^{-4/3},I4}|^{2} + |g_{WR}^{Y^{-4/3},I4}|^{2} \right) \Pi^{LL}_{T} + 2\re \left( g_{WL}^{Y^{-4/3}*,I4}g_{WR}^{Y^{-4/3},I4} \right) \Pi^{LR}_{T} \right](p^2;d_{I},Y^{-4/3}_{L}) \nonumber\\
&& + \left[ \left( |g_{WL}^{Y^{-4/3},I5}|^{2} + |g_{WR}^{Y^{-4/3},I5}|^{2} \right) \Pi^{LL}_{T} + 2\re \left( g_{WL}^{Y^{-4/3}*,I5}g_{WR}^{Y^{-4/3},I5} \right) \Pi^{LR}_{T} \right](p^2;d_{I},Y^{-4/3}_{H}) \Biggl\} \,. \nonumber\\
\end{eqnarray}
The part D, loops of combinations of $X^{5/3}_{L/H}$ and $X^{8/3}$, gives:
\begin{eqnarray}
\lefteqn{\Pi_{11}^{(D)}(p^{2}) =} \nonumber\\
&& \frac{1}{g^2} \sum_{K=1,2} \Biggl\{ \left[ \left( |g_{WL}^{X^{8/3},K1}|^{2} + |g_{WR}^{X^{8/3},K1}|^{2} \right) \Pi^{LL}_{T} + 2\re \left( g_{WL}^{X^{8/3}*,K1}g_{WR}^{X^{8/3},K1} \right) \Pi^{LR}_{T} \right](p^2;X^{8/3}_{K},X^{5/3}_{L}) \nonumber\\
&& + \left[ \left( |g_{WL}^{X^{8/3},K2}|^{2} + |g_{WR}^{X^{8/3},K2}|^{2} \right) \Pi^{LL}_{T} + 2\re \left( g_{WL}^{X^{8/3}*,K2}g_{WR}^{X^{8/3},K2} \right) \Pi^{LR}_{T} \right](p^2;X^{8/3}_{K},X^{5/3}_{H}) \Biggl\} \,.
\end{eqnarray}
The part E, loops of combinations of $Y^{-4/3}_{L/H}$ and $Y^{-7/3}$, evaluates to:
\begin{eqnarray}
\lefteqn{\Pi_{11}^{(E)}(p^{2}) = \frac{1}{g^2} \sum_{K=1,2} \Biggl\{} \nonumber\\
&& \left[ \left( |g_{WL}^{Y^{-7/3},1K}|^{2} + |g_{WR}^{Y^{-7/3},1K}|^{2} \right) \Pi^{LL}_{T} + 2\re \left( g_{WL}^{Y^{-7/3}*,1K}g_{WR}^{Y^{-7/3},1K} \right) \Pi^{LR}_{T} \right](p^2;Y^{-4/3}_{L},Y_{K}^{-7/3}) \nonumber\\
&+&  \left[ \left( |g_{WL}^{Y^{-7/3},2K}|^{2} + |g_{WR}^{Y^{-7/3},2K}|^{2} \right) \Pi^{LL}_{T} + 2\re \left( g_{WL}^{Y^{-7/3}*,2K}g_{WR}^{Y^{-7/3},2K} \right) \Pi^{LR}_{T} \right](p^2;Y^{-4/3}_{H},Y_{K}^{-7/3}) \Biggl\} \,. \nonumber\\
\end{eqnarray}

The part A of $\Pi_{33}(p^{2})$, loops of combinations of top-type/bottom-type quarks, is given by:
\begin{eqnarray}
\lefteqn{\Pi_{33}^{(A)}(p^{2}) =} \nonumber\\
&& \sum_{I}\sum_{K=1,2} \Biggl\{ \left[ \left( \frac{1}{2} - \Delta T_{3}^{(K,u)} |V_{L}^{u,K+3I}|^{2} \right)^{2} + \left(T_{3}^{(K,u)}|V_{R}^{u,K+3I}|^{2} \right)^{2} \right]\Pi_{T}^{LL}(p^2;u_{I},u_{I}) \nonumber \\
&& + \biggl[ T_{3}^{(K,u)}  |V_{R}^{u,K+3I}|^{2}
-2\Delta T_{3}^{(K,u)}T_{3}^{(K,u)} |V_{L}^{u,K+3I}|^{2} |V_{R}^{u,K+3I}|^{2} \biggr] \Pi_{T}^{LR}(p^2;u_{I},u_{I}) \nonumber \\
&& + \left[ \left( -\frac{1}{2} - \Delta T_{3}^{(K,d)}|V_{L}^{d,K+3I}|^{2} \right)^{2} + \left(T_{3}^{(K,d)}|V_{R}^{d,K+3I}|^{2} \right)^{2} \right]\Pi_{T}^{LL}(p^2;d_{I},d_{I}) \nonumber \\
&& - \biggl[ T_{3}^{(K,d)} |V_{R}^{d,K+3I}|^{2}
+2\Delta T_{3}^{(K,d)}T_{3}^{(K,d)} |V_{L}^{d,K+3I}|^{2} |V_{R}^{d,K+3I}|^{2} \biggr] \Pi_{T}^{LR}(p^2;d_{I},d_{I}) \Biggr\} \nonumber \\
&+& 2\sum_{I<J}\sum_{K=1,2} \Biggl\{ \nonumber\\
&& \left[ \left( \Delta T_{3}^{(K,u)} \right)^{2} |V_{L}^{u,K+3I}|^{2} |V_{L}^{u,K+3J}|^{2}  + \left(T_{3}^{(K,u)} \right)^{2} |V_{R}^{u,K+3I}|^{2} |V_{R}^{u,K+3J}|^{2}  \right]\Pi_{T}^{LL}(p^2;u_{I},u_{J}) \nonumber \\
&& - 2\biggl[ \Delta T_{3}^{(K,u)}T_{3}^{(K,u)} \re \left( V_{L}^{u,K+3I}V_{L}^{u*,K+3J} V_{R}^{u*,K+3I}V_{R}^{u,K+3J} \right) \biggr] \Pi_{T}^{LR}(p^2;u_{I},u_{J}) \nonumber\\
&& + \left[ \left( \Delta T_{3}^{(K,d)} \right)^{2} |V_{L}^{d,K+3I}|^{2} |V_{L}^{d,K+3J}|^{2}  + \left(T_{3}^{(K,d)} \right)^{2} |V_{R}^{d,K+3I}|^{2} |V_{R}^{d,K+3J}|^{2}  \right]\Pi_{T}^{LL}(p^2;d_{I},d_{J}) \nonumber \\
&& - 2\biggl[ \Delta T_{3}^{(K,d)}T_{3}^{(K,d)} \re \left( V_{L}^{d,K+3I}V_{L}^{d*,K+3J} V_{R}^{d*,K+3I}V_{R}^{d,K+3J} \right) \biggr] \Pi_{T}^{LR}(p^2;d_{I},d_{J}) \Biggr\} \,.
\end{eqnarray}
The part B, loops of $X^{5/3}_{L}$ and $X^{5/3}_{H}$, is:
\begin{eqnarray}
\lefteqn{\Pi_{33}^{(B)}(p^{2}) =\sum_{I=1,2}\sum_{J=1,2} \Biggl\{ T_{3}^{(I,X)}T_{3}^{(J,X)} \Biggl[} \nonumber\\
&& \left( |V_{L}^{X^{5/3},I1}|^{2} |V_{L}^{X^{5/3},J1}|^{2} + |V_{R}^{X^{5/3},I1}|^{2} |V_{R}^{X^{5/3},J1}|^{2} \right) \Pi_{T}^{LL}(p^2;X^{5/3}_{L},X^{5/3}_{L}) \nonumber\\
&& + 2 |V_{L}^{X^{5/3},I1}|^{2} |V_{R}^{X^{5/3},J1}|^{2} \Pi_{T}^{LR}(p^2;X^{5/3}_{L},X^{5/3}_{L}) \nonumber\\
&& + 2 |V_{L}^{X^{5/3},I2}|^{2} |V_{R}^{X^{5/3},J2}|^{2} \Pi_{T}^{LR}(p^2;X^{5/3}_{H},X^{5/3}_{H}) \nonumber\\
&& + \left( |V_{L}^{X^{5/3},I2}|^{2} |V_{L}^{X^{5/3},J2}|^{2} + |V_{R}^{X^{5/3},I2}|^{2} |V_{R}^{X^{5/3},J2}|^{2} \right) \Pi_{T}^{LL}(p^2;X^{5/3}_{H},X^{5/3}_{H}) \nonumber\\
&& +2 \left( \re \left( V_{L}^{X^{5/3}*,I1}V_{L}^{X^{5/3},I2}V_{L}^{X^{5/3},J1}V_{L}^{X^{5/3}*,J2} \right) \right. \nonumber\\
&& \left. + \re \left( V_{R}^{X^{5/3}*,I1}V_{R}^{X^{5/3},I2}V_{R}^{X^{5/3},J1}V_{R}^{X^{5/3}*,J2} \right) \right) \Pi_{T}^{LL}(p^2;X^{5/3}_{L},X^{5/3}_{H}) \nonumber\\
&& + 2 \re \left( V_{L}^{X^{5/3}*,I1}V_{L}^{X^{5/3},I2}V_{R}^{X^{5/3},J1}V_{R}^{X^{5/3}*,J2} \right) \Pi_{T}^{LR}(p^2;X^{5/3}_{L},X^{5/3}_{H}) \Biggr] \Biggr\} \,.
\end{eqnarray}
The part C, loops of $Y^{-4/3}_{L}$ and $Y^{-4/3}_{H}$, is:
\begin{eqnarray}
\lefteqn{\Pi_{33}^{(C)}(p^{2}) =\sum_{I=1,2}\sum_{J=1,2} \Biggl\{ T_{3}^{(I,Y)}T_{3}^{(J,Y)} \Biggl[} \nonumber\\
&& \left( |V_{L}^{Y^{-4/3},I1}|^{2} |V_{L}^{Y^{-4/3},J1}|^{2} + |V_{R}^{Y^{-4/3},I1}|^{2} |V_{R}^{Y^{-4/3},J1}|^{2} \right) \Pi_{T}^{LL}(p^2;Y^{-4/3}_{L},Y^{-4/3}_{L}) \nonumber\\
&& + 2 |V_{L}^{Y^{-4/3},I1}|^{2} |V_{R}^{Y^{-4/3},J1}|^{2} \Pi_{T}^{LR}(p^2;Y^{-4/3}_{L},Y^{-4/3}_{L}) \nonumber\\
&& + 2 |V_{L}^{Y^{-4/3},I2}|^{2} |V_{R}^{Y^{-4/3},J2}|^{2} \Pi_{T}^{LR}(p^2;Y^{-4/3}_{H},Y^{-4/3}_{H}) \nonumber\\
&& + \left( |V_{L}^{Y^{-4/3},I2}|^{2} |V_{L}^{Y^{-4/3},J2}|^{2} + |V_{R}^{Y^{-4/3},I2}|^{2} |V_{R}^{Y^{-4/3},J2}|^{2} \right) \Pi_{T}^{LL}(p^2;Y^{-4/3}_{H},Y^{-4/3}_{H}) \nonumber\\
&& +2 \left( \re \left( V_{L}^{Y^{-4/3}*,I1}V_{L}^{Y^{-4/3},I2}V_{L}^{Y^{-4/3},J1}V_{L}^{Y^{-4/3}*,J2} \right) \right. \nonumber\\
&& + \left. \re \left( V_{R}^{Y^{-4/3}*,I1}V_{R}^{Y^{-4/3},I2}V_{R}^{Y^{-4/3},J1}V_{R}^{Y^{-4/3}*,J2} \right) \right) \Pi_{T}^{LL}(p^2;Y^{-4/3}_{L},Y^{-4/3}_{H}) \nonumber\\
&& + 2 \re \left( V_{L}^{Y^{-4/3}*,I1}V_{L}^{Y^{-4/3},I2}V_{R}^{Y^{-4/3},J1}V_{R}^{Y^{-4/3}*,J2} \right) \Pi_{T}^{LR}(p^2;Y^{-4/3}_{L},Y^{-4/3}_{H}) \Biggr] \Biggr\} \,.
\end{eqnarray}
The part D/E, loops of $X^{8/3}$/$Y^{-7/3}$, are:
\begin{eqnarray}
\Pi_{33}^{(D)}(p^{2}) &=& 2 \sum_{K=1,2} \left( T_{3}^{(K,X^{8/3})} \right)^{2} \left( \Pi_{T}^{LL} + \Pi_{T}^{LR} \right)(p^2;X^{8/3}_{K},X^{8/3}_{K}) \,,\\
\Pi_{33}^{(E)}(p^{2}) &=& 2 \sum_{K=1,2} \left( T_{3}^{(K,Y^{-7/3})} \right)^{2} \left( \Pi_{T}^{LL} + \Pi_{T}^{LR} \right)(p^2;Y^{-7/3}_{K},Y^{-7/3}_{K}) \,,
\end{eqnarray}
where the Part D and E are exactly cancelled by the reason of $(\Pi_{T}^{LL}+\Pi_{T}^{LR})(p^{2};m_{f},m_{f})$ at $p^2=0$:
\begin{equation}
\Pi_{33}^{(D)}(p^{2}=0) = 0 \,, \qquad \Pi_{33}^{(E)}(p^{2}=0) = 0 \,.
\end{equation}

The part A of $\Pi_{3Q}(p^{2})$, loops of combinations of top-type/bottom-type quarks, is:
\begin{eqnarray}
\Pi_{3Q}^{(A)}(p^{2}) &=& \frac{1}{2}\sum_{I} \Biggl\{ Q_{u}\left( \Pi_{T}^{LL} + \Pi_{T}^{LR} \right)(p^2;u_{I},u_{I}) - Q_{d} \left( \Pi_{T}^{LL} + \Pi_{T}^{LR} \right)(p^2;d_{I},d_{I}) \Biggr\} \nonumber \\
&+& \sum_{I}\sum_{K=1,2} \Biggl\{ Q_{u} \left[ - \Delta T_{3}^{(K,u)} |V_{L}^{u,K+3I}|^{2} + T_{3}^{(K,u)} |V_{R}^{u,K+3I}|^{2} \right] \Pi_{T}^{LL}(p^2;u_{I},u_{I}) \nonumber \\
&& + Q_{u} \biggl[ T_{3}^{(K,u)} |V_{R}^{u,K+3I}|^{2}
- \Delta T_{3}^{(K,u)} |V_{L}^{u,K+3I}|^{2} \biggr] \Pi_{T}^{LR}(p^2;u_{I},u_{I}) \nonumber \\
&& + Q_{d} \left[ - \Delta T_{3}^{(K,d)} |V_{L}^{d,K+3I}|^{2} + T_{3}^{(K,d)} |V_{R}^{d,K+3I}|^{2} \right] \Pi_{T}^{LL}(p^2;d_{I},d_{I}) \nonumber\\
&& + Q_{d} \biggl[ T_{3}^{(K,d)} |V_{R}^{d,K+3I}|^{2}
- \Delta T_{3}^{(K,d)} |V_{L}^{d,K+3I}|^{2} \biggr] \Pi_{T}^{LR}(p^2;d_{I},d_{I}) \Biggr\} \,.
\end{eqnarray}
The part B/C, loops of $X^{3/5}$/$Y^{-4/3}$, are:
\begin{eqnarray}
\lefteqn{\Pi_{3Q}^{(B)}(p^{2}) =} \nonumber\\
&& Q_{X} \sum_{K=1,2} \Biggl\{ T_{3}^{(K,X)} \biggl[ \left( |V_{L}^{X^{5/3},K1}|^{2} + |V_{R}^{X^{5/3},K1}|^{2} \right) \left( \Pi_{T}^{LL} + \Pi_{T}^{LR} \right)(p^2;X^{5/3}_{L},X^{5/3}_{L}) \nonumber\\
&& + \left( |V_{L}^{X^{5/3},K2}|^{2} + |V_{R}^{X^{5/3},K2}|^{2} \right) \left( \Pi_{T}^{LL} + \Pi_{T}^{LR} \right)(p^2;X^{5/3}_{H},X^{5/3}_{H}) \biggr] \Biggr\} \,,\\
\lefteqn{\Pi_{3Q}^{(C)}(p^{2}) =} \nonumber\\
&& Q_{Y} \sum_{K=1,2} \Biggl\{ T_{3}^{(K,Y)} \biggl[ \left( |V_{L}^{Y^{-4/3},K1}|^{2} + |V_{R}^{Y^{-4/3},K1}|^{2} \right) \left( \Pi_{T}^{LL} + \Pi_{T}^{LR} \right)(p^2;Y^{-4/3}_{L},Y^{-4/3}_{L}) \nonumber\\
&& + \left( |V_{L}^{Y^{-4/3},K2}|^{2} + |V_{R}^{Y^{-4/3},K2}|^{2} \right) \left( \Pi_{T}^{LL} + \Pi_{T}^{LR} \right)(p^2;Y^{-4/3}_{H},Y^{-4/3}_{H}) \biggr] \Biggr\} \,.
\end{eqnarray}
The part D/E, loops of $X^{8/3}$/$Y^{-7/3}$, are:
\begin{eqnarray}
\Pi_{3Q}^{(D)}(p^{2}) &=& 2Q_{X^{8/3}} \sum_{K=1,2} T_{3}^{(K,X^{8/3})} \left( \Pi_{T}^{LL} + \Pi_{T}^{LR} \right)(p^2;X^{8/3}_{K},X^{8/3}_{K}) \,,\\
\Pi_{3Q}^{(E)}(p^{2}) &=& 2Q_{Y^{-7/3}} \sum_{K=1,2} T_{3}^{(K,Y^{-7/3})} \left( \Pi_{T}^{LL} + \Pi_{T}^{LR} \right)(p^2;Y^{-7/3}_{K},Y^{-7/3}_{K}) \,,
\end{eqnarray}
where the part D and E at $p^2=0$ become
\begin{equation}
\Pi_{3Q}^{(D)}(p^{2}=0) = \Pi_{3Q}^{(E)}(p^{2}=0) = 0 \,.
\end{equation}


\section{Extra numerical results for VL multiplets} 
                              \label{appendix:D}

\begin{figure}[htb]
\begin{center}
\hspace*{-0.7cm}
\epsfig{file=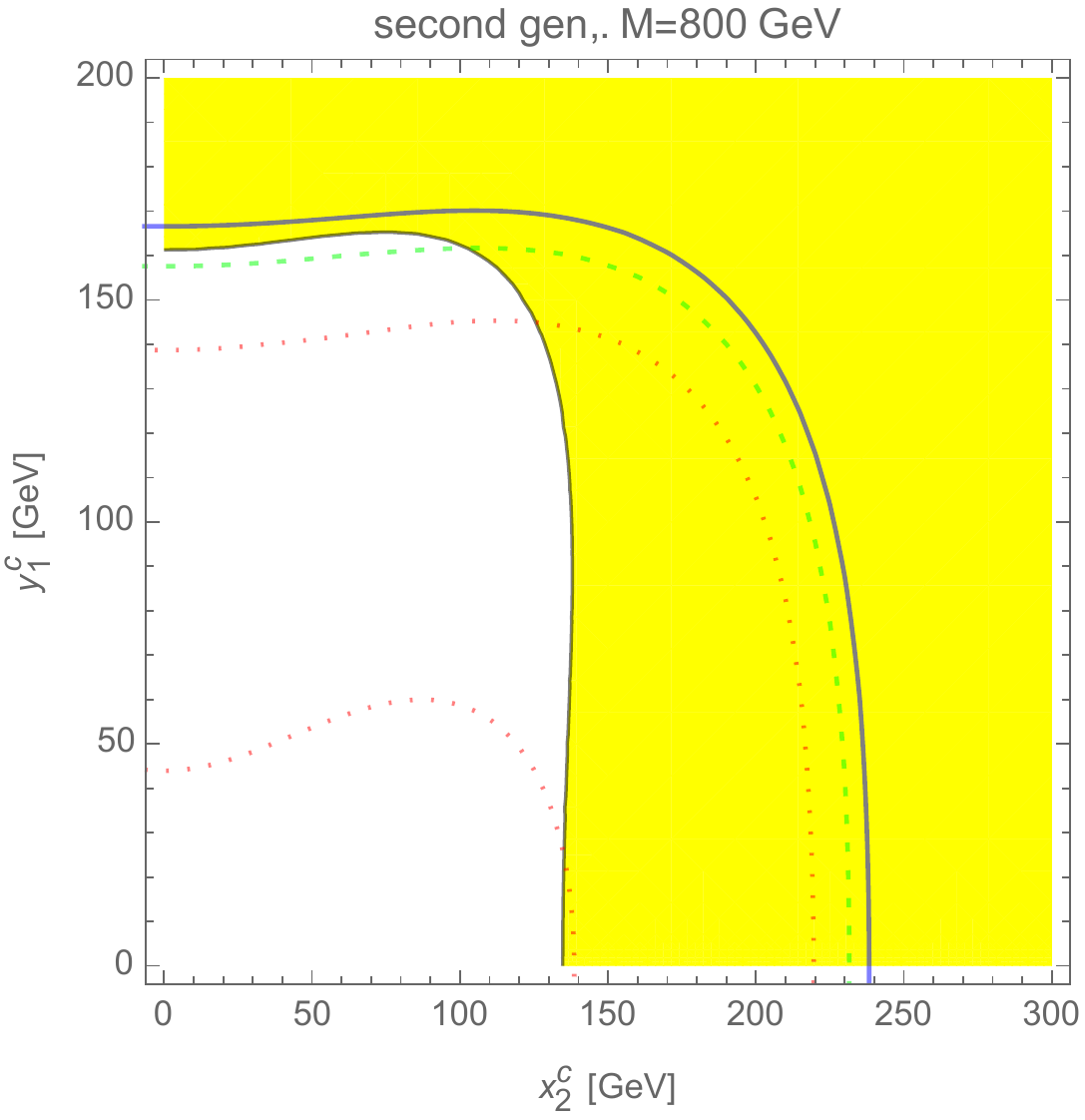,width=0.45\textwidth}
   \hspace*{0.4cm}
   \epsfig{file=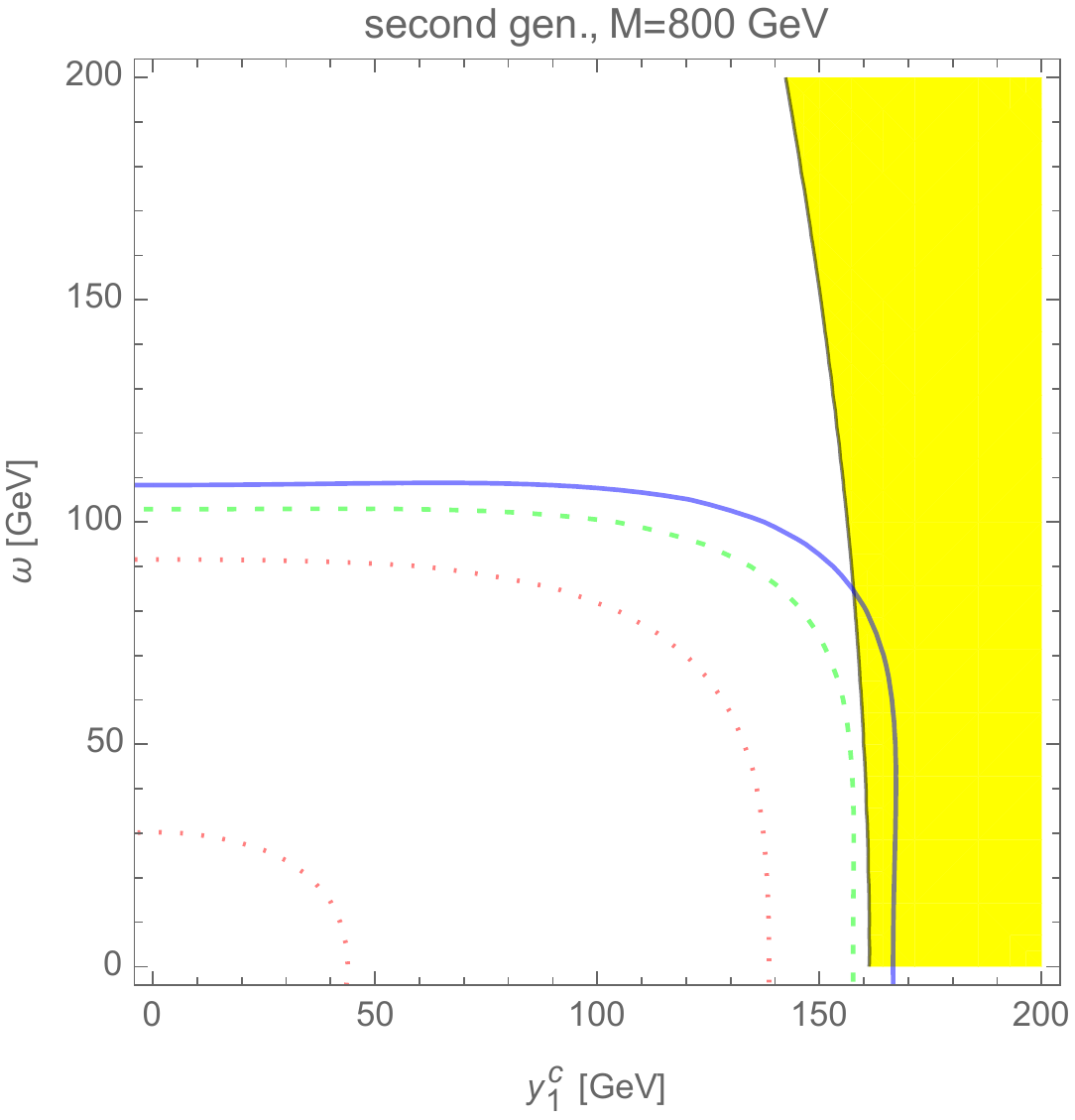,width=0.45\textwidth} 
    \\ \vspace*{0.4cm}
\hspace*{-0.7cm}
\epsfig{file=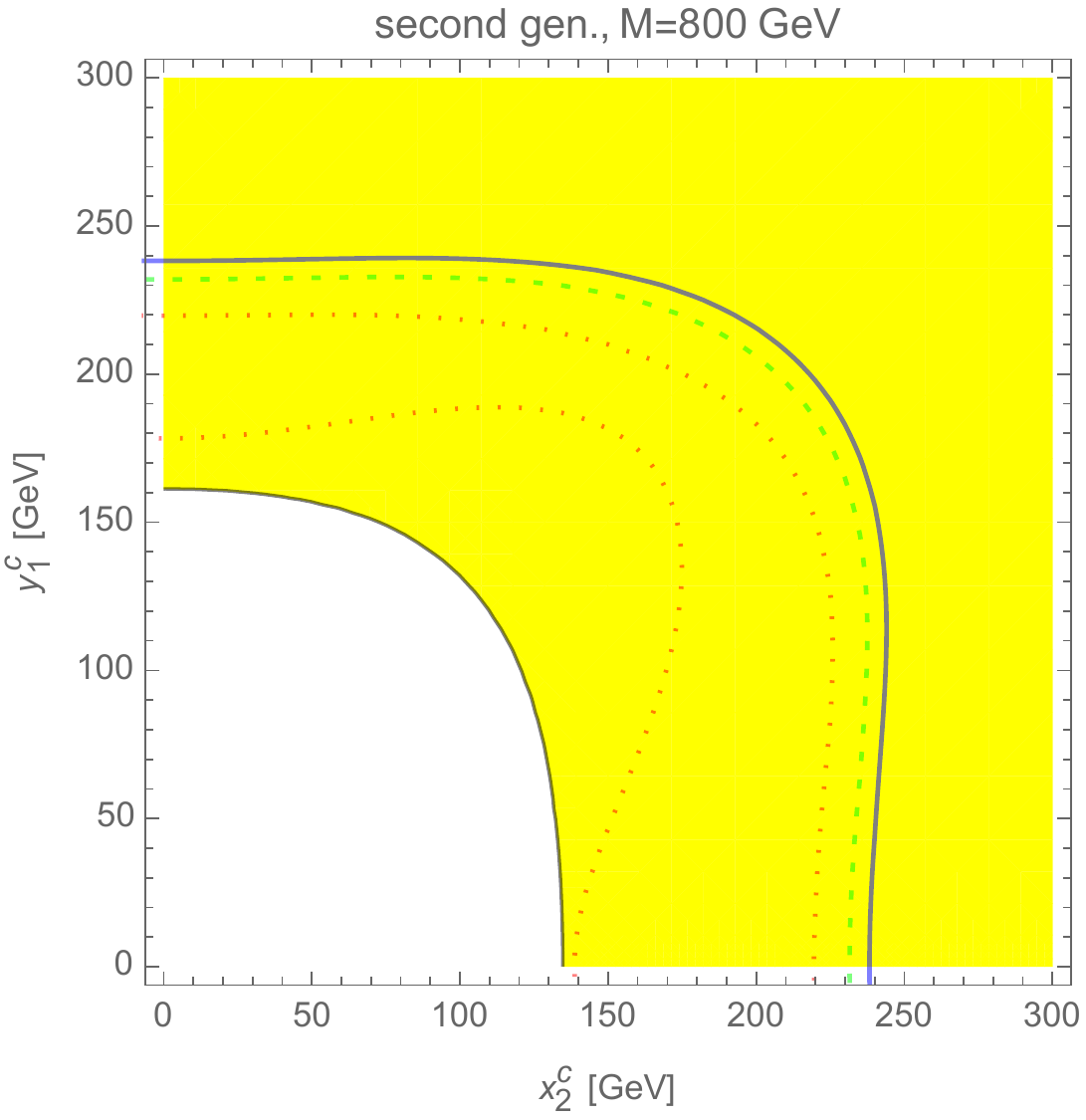,width=0.45\textwidth}
  \hspace*{0.4cm}
  \epsfig{file=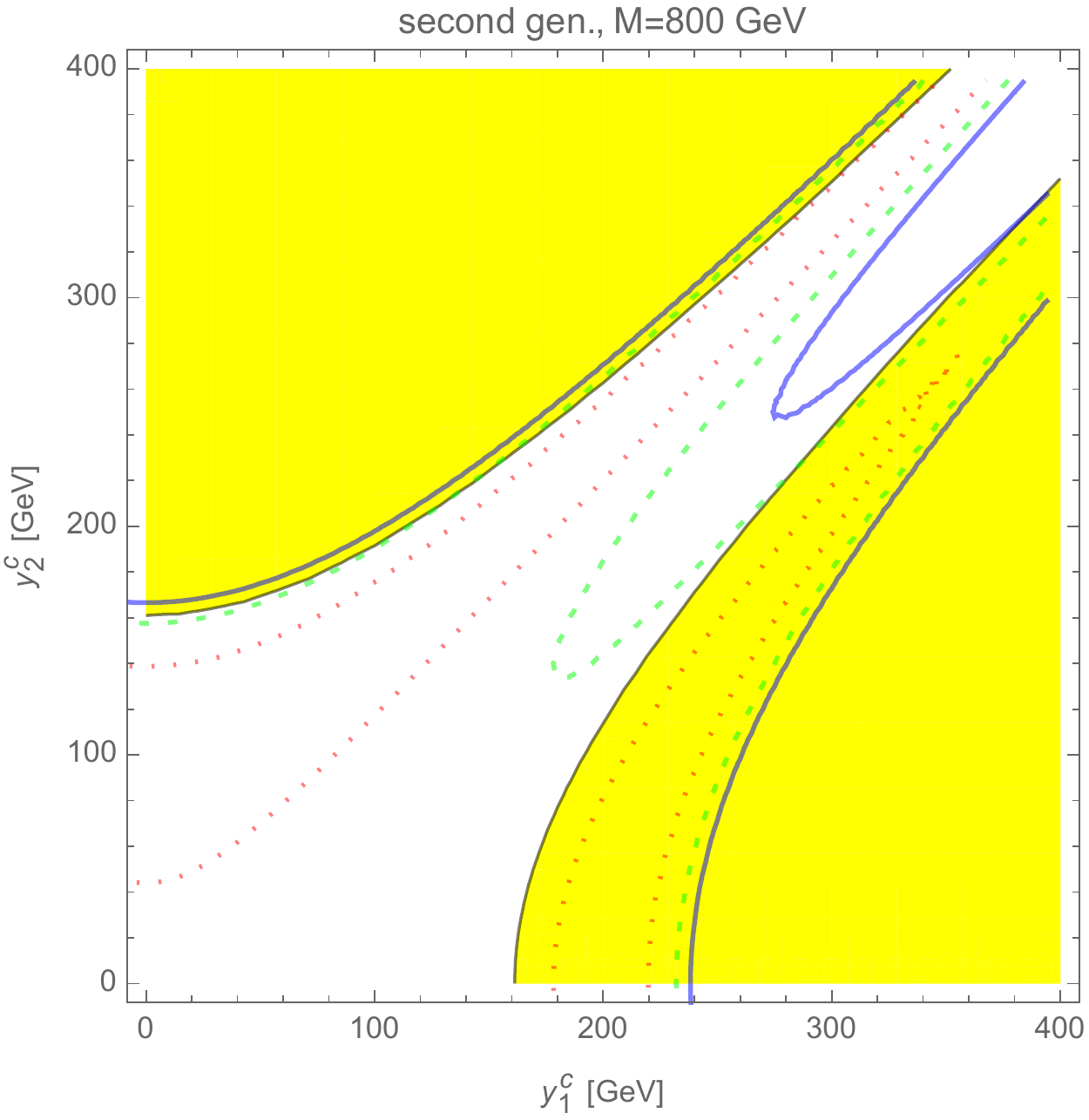,width=0.45\textwidth}
\end{center}
\caption{EWP bounds at 1 $\sigma$ (red-dashed), 2  $\sigma$ (green-dashed) and 3 $\sigma$ (blue) for VL quarks coupling with
  the second SM generations, compared with the region excluded at 3$\sigma$ by tree-level bounds (yellow
  region in the left panel). We always chose $M_1 = M_2 = M = 800$ GeV and $\omega = \omega' = 0$ (except for the upper-right plot where $\omega = \omega' \neq 0$). 
  Upper left panel: Singlet $Y=2/3$ and Doublet $Y=7/6$. Upper right panel: Doublet $Y=7/6$ and Triplet $Y=5/3$. 
  Lower left panel: Singlet $Y=2/3$ and Doublet $Y=1/6$. Lower right panel: Doublet $Y=1/6$ and Doublet $Y=7/6$.
  Only the first quadrant is shown as the figures are symmetric with respect to a sign change in the coordinates in the other three quadrants. 
  Note that in the four plots the variables and the units on the axis are not the same, so that they should not be compared directly as they refer to 
  different multiplets.}     
\label{fig:Dsnonsmd}
\end{figure}

We collect in the present appendix extra numerical results which complete those shown in the main text, in particular concerning the
limits for the case of the VL quarks coupling to the second SM generation; these are similar in form to those obtained for the coupling to the first 
SM generation, but bounds vary considerably in some cases. See figure \ref{fig:Dsnonsmd}.

\newpage


\bibliographystyle{JHEP}
\bibliography{TWOVLQ}


\end{document}